\documentclass[12pt]{article}
\usepackage{amssymb,amsmath}
\usepackage{graphicx}
\usepackage{epsfig}
\usepackage{epstopdf}
\usepackage{latexsym}
\usepackage{psfrag}
\usepackage{booktabs}
\usepackage{bbm}
\usepackage[toc]{appendix}
\usepackage{color}
\usepackage{datetime}
\usepackage[
      colorlinks=true,
      linkcolor=darkblue,  
      urlcolor=blue,    
      filecolor=blue,     
      citecolor=red,
      linktocpage=true,
      pdfstartview=FitV,
      bookmarksopen=true    
      ]{hyperref}

\definecolor{darkblue}{rgb}{0.2, 0, 0.8}

\newcommand{\be}{\begin{equation}}
\newcommand{\ee}{\end{equation}}
\newcommand{\bea}{\begin{eqnarray}}
\newcommand{\eea}{\end{eqnarray}}

\newcommand{\reef}[1]{(\ref{#1})}
\newcommand{\m}{\mu}
\newcommand{\n}{\nu}
\newcommand{\de}{\delta}
\newcommand{\pa}{\partial}
\newcommand{\Lag}{\mathcal{L}}
\newcommand{\eps}{\epsilon}
\newcommand{\s}{\sigma}
\newcommand{\cn}{\mathcal{N}}

\begin{document}  

\renewcommand*\thesection{\arabic{section}}
\renewcommand*\thesubsection{\arabic{section}.\arabic{subsection}}

\setcounter{equation}{0}
\numberwithin{equation}{section}
\setcounter{figure}{0}
\renewcommand{\thefigure}{\arabic{figure}}

\DeclareGraphicsRule{.tif}{png}{.png}{`convert #1 `dirname #1`/`basename #1 .tif`.png}


\begin{titlepage}

\begin{flushright}
MCTP-13-37
\end{flushright}
\vspace*{2.3cm}

\begin{center}
{\LARGE \bf Quantum gravity \\ via \\[3mm] supersymmetry and holography} \\

\vspace*{1.2cm}

{\bf Henriette Elvang$^{1}$ and Gary T.~Horowitz$^{2}$}
\bigskip

$^{1}$Randall Laboratory of Physics, Department of Physics,\\
University of Michigan, Ann Arbor, MI 48109, USA
\bigskip

$^{2}$Department of Physics, UCSB, Santa Barbara, CA 93106, USA

\bigskip
elvang@umich.edu, gary@physics.ucsb.edu  \\
\end{center}

\vspace*{0.1cm}

\begin{abstract}  
We review the approach to quantum gravity based on supersymmetry,
strings, and holography. This includes a survey of black holes in
higher-dimensions, supersymmetry and supergravity, as well as string
theory, black hole microstates, and the gauge/gravity duality. This
presentation will appear as a chapter in ``General Relativity and Gravitation: A Centennial Perspective", to be published by Cambridge University Press.

\end{abstract}

\end{titlepage}

\setcounter{tocdepth}{2}
{\small
\setlength\parskip{-0.5mm}
\tableofcontents
}

\newpage

\section{Introduction}

This chapter offers a survey of ideas and results in the approach to quantum gravity based on supersymmetry, strings, and holography.

Extra spatial dimensions appear naturally in this approach, 
so to set the stage we  begin in Section \ref{s:Ddim} with a discussion of general relativity in more than four spacetime dimensions.  
In higher dimensions, one encounters a richness of structure with no parallel in 4D. Even in vacuum gravity, this includes black hole solutions with non-spherical horizon topologies, black hole non-uniqueness, and regular multi-horizon black holes.  We give an overview of such solutions and their properties, both in the context of Kaluza-Klein theory and for asymptotically flat boundary conditions. 

A very interesting extension of general relativity is to include matter in such a way that the  action becomes invariant under supersymmetry transformations. 
Supersymmetry is a remarkable symmetry that relates bosons and fermions. It is the only possible extension of the Poincar\'e group for a unitary theory with non-trivial scattering processes. Supersymmetry is considered a natural extension of the standard model of particle physics; the study of how supersymmetry is broken at low energies, and its possible experimental consequences, is an important active research area in particle physics. Furthermore, independently of its potential phenomenology, supersymmetry offers strong calculational control  and that makes it a tremendously powerful tool for analyzing fundamental properties of  quantum field theories. 

When supersymmetry and general relativity are combined, the result is supergravity. The metric field is accompanied by a spin-3/2 spinor field and this gives a beautiful and enticing playground for advancing our understanding of quantum gravity. Supergravity theories exist in spacetime dimensions $D \le 11$ and they provide a natural setting for studies of charged black holes. Certain  extremal limits of charged black holes in supergravity are invariant under supersymmetry; such `supersymmetric black holes' are key for understanding the statistic mechanical nature of black hole thermodynamics, specifically the 
microstates responsible for the Hawking-Bekenstein entropy. An example of  a supersymmetric black hole is the extremal 
Reissner-Nordstrom solution. 

Section \ref{s:super} begins with a brief introduction to supersymmetry and supergravity, followed by a survey of  supersymmetric black holes and their properties. We then discuss perturbative 
quantization of gravity as a quantum field theory, an approach in which the metric field is quantized in a flat-space background and the resulting gravitons are point-like spin-2 particles. It is well-known that this approach leads to ultraviolet divergences, starting at
2-loop order in pure gravity, that --- unlike the corresponding infinities in gauge theories --- cannot be cured as gravity is a non-renormalizable theory. However, in supergravity, the symmetry between bosons and fermions results in crucial cancellations in graviton scattering amplitudes and this can delay the occurrence of the ultraviolet  divergences to higher-loop order. It has even  been suggested that with maximal supersymmetry, perturbative supergravity in 4D may be free of such ultraviolet divergences. We will offer a short description of these ideas and related results.

A  profound solution to the problem of ultraviolet divergences in quantum gravity is to treat gravitons as extended one-dimensional objects: strings. Then the short-distance behavior is regulated by the finite extent of the string and scattering processes are free of ultraviolet divergences. Thus string theory is a very promising framework for a quantum theory of gravity. 
String theory naturally incorporates supersymmetry, and general relativity --- and its extension to supergravity --- emerges as a  low-energy effective theory. String theory predicts extra spatial dimensions, so compatibility with observations requires that either these extra dimensions are compact and small (incorporating Kaluza-Klein theory) or that we live on a $3+1$ dimensional subspace of this higher-dimensional spacetime.

Section \ref{s:strings} is dedicated to an introduction to string theory. We begin with an overview of perturbative string theory and then turn to nonperturbative aspects, specifically quantum black holes.  One of the remarkable features of string theory is that it provides a precise 
microscopic description of black hole entropy and Hawking radiation for certain
black holes, specifically the supersymmetric or near-supersymmetric black holes. We describe in detail the precision-counting of black hole microstates in string theory and its match to the Bekenstein-Hawking entropy. 

At the nonperturbative level, there are arguments that quantum gravity might be holographic: this is the notion that physics in a region of space is completely described by degrees of freedom living on the boundary of this region. This idea was  originally proposed based on considerations of black hole entropy. The entropy of a black hole scales with its area, in striking contrast to most systems which have an  entropy proportional to their volume. This suggests that everything that happens inside the black hole might be encoded in degrees of freedom at the horizon.  A precise formulation of holo\-graphy emerges from string theory and is called ``gauge/gravity duality". In Section \ref{s:adscft}, we discuss holography and present a detailed account of the motivation for gauge/gravity duality and the evidence in its favor. Recent years have seen applications of 
gauge/gravity duality to a wide variety of problems in physics, including black holes, quark confinement, hydrodynamics, and
condensed matter physics. We give a brief survey of these results. 

Section \ref{s:conclusion} contains some concluding remarks. 
It is our hope that this chapter will convey the depth and richness of the subjects mentioned above and motivate the reader to pursue further information in the references provided throughout the text.


\section{Gravity in $D$ dimensions}
\label{s:Ddim}
At first sight,  general relativity in a $D$-dimensional spacetime looks much like 4D general relativity. The Einstein equation takes the same form
\be
  G_{MN} = 8\pi G_D\, T_{MN} \,,
\ee 
in which $G_D$ is the $D$-dimensional Newton's constant and 
the Einstein tensor is given 
in terms of the Ricci tensor as $G_{MN} = R_{MN} - \frac{1}{2} g_{MN} R$.  The spacetime indices $M,N$ run over $0,1,2,\dots,D-1$. The field equations can be derived using the variational principle from the $D$-dimensional Einstein-Hilbert action
\be
  S_\text{EH} = \frac{1}{16 \pi G_D} \int d^D x \,\sqrt{-g}\,R 
  ~+~ 
  S_\text{matter}\,,
\ee
where the stress-energy tensor is 
$T_{MN} = - \frac{2}{\sqrt{-g}} \frac{\de S_\text{matter}}{\de g^{MN}}$.

Given the similarities, one expects that solutions to the 4D Einstein equation have straightforward generalizations to higher dimensions. This is indeed the case, for example the Schwarzschild metric generalizes to $D > 4$ dimensions as a solution to the vacuum Einstein equations. The metric of the $D$-dimensional 
{\bf Schwarzschild-Tangherlini solution} \cite{Tangherlini:1963bw} found in 1963   is
\be
   ds^2 = -\bigg[1 - \Big(\frac{r_0}{r}\Big)^{D-3} \bigg]\, dt^2 
   +\bigg[1 - \Big(\frac{r_0}{r}\Big)^{D-3} \bigg]^{-1}dr^2
   + r^2 \,d\Omega_{D-2}^2 \,,
   \label{DSchw}
\ee
where $d\Omega_{d}^2$ is the line element for a $d$-dimensional round unit  sphere, $S^d$.
The horizon is located at $r=r_0$ and has topology $S^{D-2}$. Black hole thermodynamics works the same in higher dimensions as in 4D. 
This includes the first law and the area theorem.
For the $D$-dimensional Schwarzschild-Tangherlini 
 black holes, 
the ADM mass, temperature (calculated from the surface gravity $\kappa$), and horizon `area' are\footnote{We will refer to the horizon volume as `area' although of course it is the volume of a $(D\!-\!2)$-dimensional manifold.} 
\be
  M =\frac{(D-2) \Omega_{D-2} r_0^{D-3}}{16\pi G_D}\,,~~~~~
  T = \frac{\kappa}{2 \pi} = \frac{D-3}{4\pi r_0}\,,~~~~~
  A = \Omega_{D-2} r_0^{D-2}\,,~~
\ee
where $\Omega_{d}$ the volume of the unit $d$-sphere. The entropy is 
$S = A/(4G_D)$. These quantities satisfy $dM = T dS$, and the Smarr relation $(D-3)M = (D-2)TS$.

There are of course also important differences as $D$ varies. For example, it is well-known that there is no 3D vacuum black hole,\footnote{With a negative cosmological constant, there is a 3D black hole \cite{Banados:1992wn}.}
 so the 4D Schwarzschild solution does not generalize as we go down in dimension. Might it be that there are black hole solutions in $D>4$ that do not exist in 4D? The answer turns out to be yes.

That gravity has richer structure in higher dimensions is apparent already from solutions to the linearized Einstein equation. For example, a gravitational wave in $D$ dimensions has $D(D-3)/2$ degrees of freedom. This formula counts the well-known 2 polarizations of a gravitational wave
 in a $D=4$ dimensional spacetime. But it also tells us that in 3 dimensions there are no propagating modes of gravity. Going up in dimensions, we learn that a gravitational wave in 5D carries 5 degrees of freedom,  in 6D it is 9 degrees of freedom, and so on.
This hints that gravitational dynamics has important dimensional dependence and that going up in spacetime dimension gives `more freedom' and new phenomena may be found. In this section, we give  several examples of the rewards of studying gravity in spacetime dimensions $D>4$.

\subsection{Kaluza-Klein theory}

One motivation for studying higher-dimensional gravity is that it offers a method for unifying gravity with other forces, as first explored by Kaluza and Klein in the early 1920's \cite{Kaluza:1921tu,Klein:1926tv}. 
The idea is that pure gravity in 5-dimensions can be viewed as a Maxwell-scalar-gravity system in 4-dimensions. To see how it works, we write an ansatz for the 5D metric 
\be
  ds_\text{5D}^2 
  = 
  g_{MN} \, dX^M dX^N
  =
  e^{\phi/\sqrt{3}}  \, g_{\mu\nu}\,dx^\mu dx^\nu 
  + e^{-2\phi/\sqrt{3}} \big( dy + A_\mu dx^\mu \big)^2\,,
  \label{KKansatz}
\ee
where $M,N=0,1,2,3,4$ and $\mu=0,1,2,3$. Let us assume that $g_{\mu\nu}$, $\phi$ and $A_\mu$ are independent of $X^4= y$
and that the $y$ direction is a circle $S^1$ of radius $R$. 
Evaluating the 5D Ricci tensor with this ansatz, one finds (after partial integration) that the 5D Einstein-Hilbert action can be written 
\be
  \begin{split}
  S
  &= \frac{1}{16 \pi G_5} \int d^5X\,\sqrt{-g_5}\,  R_\text{5D}\\
  &= \frac{1}{16 \pi G_4} \int d^4x\, \sqrt{-g}
       \Big( 
         R_\text{4D} 
         - \tfrac{1}{2} 
         \partial_\mu \phi \,\partial^\mu \phi
         - \tfrac{1}{4} e^{-\sqrt{3} \phi} F_{\mu\nu} F^{\mu\nu} 
       \Big)\,,
    \end{split}
    \label{5dKK}
\ee
where $F_{\mu\nu} = \partial_\mu A_\nu -  \partial_\nu A_\mu$ is a Maxwell field strength and the 4D indices $\mu,\nu$ are lowered/raised with the 4D metric $g_{\m\n}$ and its inverse. 
The 4D Newton's constant is 
related to the 5D one by
\be
  G_4 = G_5/(2\pi R)\,.
\ee

Let us now consider the equations of motion derived from \reef{5dKK}. The 4D description gives the Einstein equation coupled to a Maxwell field and the scalar $\phi$ called the `dilaton'. In addition, we have the matter equations of motion: the Maxwell equation, with its non-minimal coupling to $\phi$, and the scalar equation of motion sourced by the Maxwell field. 
This appears to be a rather complex gravity-electromagnetic system. However, from the 5D point of view, this is nothing but vacuum gravity: the 5D equation of motion is just the vacuum Einstein equation. Thus `lifting' the 4D system to 5D unifies electromagnetism with gravity!
This is a very clean and beautiful example of unification of forces.

In the {\bf Kaluza-Klein ansatz} \reef{KKansatz}, we assumed that the metric components were independent of the $S^1$ direction $y$. Generally, we could express the dependence of $y$ in terms of the Fourier modes of the 4D fields, 
e.g.~
\be
  \phi(x,y) = \sum_n \phi_n(x) \,e^{i n y/R}\,.
\ee
 It follows from the $\phi$ equation of motion that the modes with non-zero $n$ have mass-terms of the order $|n|/R$. 
 If  the radius of the Kaluza-Klein circle $R$ is very small compared to energies we are interested in (or that otherwise appear in the system),  then in a  low-energy long-wavelength approximation these modes do not contribute.
Thus we can truncate the 
$n\!\ne\! 0$ 
modes  and focus only on the massless modes.  This means that we are effectively taking the fields to be independent of $y$, and that is exactly the Kaluza-Klein ansatz.

The $S^1$ {\bf Kaluza-Klein reduction} (also known as dimensional reduction) from 5D to 4D described above --- or, equivalently, the unifying lift from 4D to 5D --- can be generalized to $D$-dimensions with minor changes in the numerical coefficients for the dilaton dependence. It also has generalizations to reduction on other manifolds than a circle, for example on a torus $S^1 \times S^1$ or a sphere $S^d$. 
The required key property is that the lower-dimensional equations of motion are consistent truncations of the higher-dimensional ones.

 The Kaluza-Klein reduction is the proto-type for 
 {\bf compactifications} 
 of a higher-dimen\-sional system to a lower-dimensional one. Compactifications play a central role in many areas of high-energy theoretical physics, both in field theory and in string theory. For now, we focus on classical aspects of higher-dimensional gravity so we will consider various solutions to the $D$-dimensional Einstein equations, in the context of the Kaluza-Klein ansatz as well as more generally.

\subsection{Black strings}
\label{s:BS}
As a  simple, but nonetheless quite interesting, example of a Kaluza-Klein spacetime, we consider  
black strings. 
Choose the Schwarzschild solution as the 4D solution in the 
Kaluza Klein ansatz. 
The ansatz \reef{KKansatz} then has $\phi=0$ and $A_\mu = 0$ and it tells us that the metric
\be
   ds_\text{5D}^2 = -\bigg(1 - \frac{2G_4M}{r} \bigg) dt^2 
   + \bigg(1 - \frac{2G_4M}{r} \bigg)^{-1} dr^2
   + r^2 \,d\Omega_2^2 + dy^2
   \label{bs}
\ee
solves the 5D vacuum Einstein equation. At any constant $y$-slice, the geometry described by \reef{bs} looks like a 4D Schwarzschild black hole with mass $M$, so it describes a continuous uniform string of Schwarzschild black holes: it is called a  {\bf homogeneous black string}. When $y$ is a circle, the topology of the black string horizon is $S^2 \times S^1$.

Now, let us think of the black string as a uniform distribution of mass along the  circle parameterized by $y$. Suppose this distribution of mass is perturbed a little: then there will be regions of higher mass-density and regions of lower mass-density. The denser regions will tend to attract more matter and grow while the 
lower-density 
regions are depleted. 
This indicates that the black string has a classical instability. 
Indeed, when the radius of the circle $R$ is (roughly) larger than the Schwarzschild radius $2GM$, the homogeneous black string solution \reef{bs} is unstable to spherical
 linear perturbations, as demonstrated first by Gregory and Laflamme \cite{Gregory:1993vy}. The evolution of 
  the {\bf Gregory-Laflamme instability}  
 is exactly as our intuition  indicates: the black string horizon becomes non-uniform along $y$ as some parts of the string bulge while others shrink 
 when 
 the mass concentrates/depletes the corresponding regions. The $S^2$ of the constant $y$-slices are not perturbed, it remains round. 

\begin{figure}[t]
\begin{center}
\includegraphics[width=9cm]{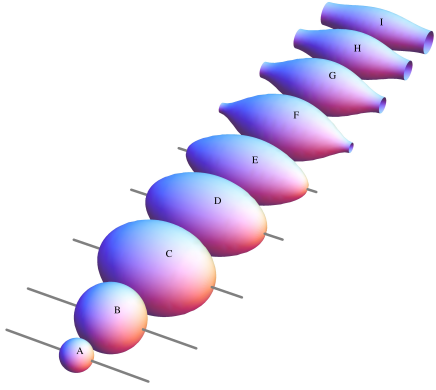}
\vspace{-10mm}
\end{center}
\caption{Embedding plots of $D=5$ Kaluza-Klein black hole horizons: A-E show the localized black holes and F-I are non-uniform black strings. (Plots from {\cite{Horowitz:2011cq}}.)} 
\label{fig:KKembed}
\end{figure}

What is the endstate of the Gregory-Laflamme instability? 
The intuitive description of the instability as mass concentrations in certain regions of the $y$ direction indicates that a localized black hole forms. Such a localized black hole would be like the $D=5$ Schwarzschild-Tangherlini 
 black hole \reef{DSchw}, but placed in a spacetime with a compact $S^1$ direction. We can easily imagine this when the black hole is much smaller than the $S^1$;  the black hole does not `know' that the $y$-direction is compact, so it is essentially unaffected. However, if the Schwarzschild radius of the black hole is comparable to the size of the $S^1$, there will be a significant backreaction that deforms the shape of the horizon (while keeping its $S^3$ topology). Such 
{\bf localized black hole} solutions have been constructed numerically; Fig.~\ref{fig:KKembed} shows the change of horizon shape as the mass of the black hole is increased for fixed size of the Kaluza-Klein circle at infinity. 

For the localized black hole to be an endstate of the Gregory-Laflamme instability requires the horizon topology to transition from the $S^2 \times S^1$ of the string to the $S^3$ of the localized black hole. Classically, the horizon cannot bifurcate without forming a naked singularity at the pinch-off \cite{Hawking:1973uf}. In classical gravity the pinch-off is not reached in finite affine time along the null generators of the horizon \cite{Horowitz:2001cz}, but a numerical analysis \cite{Lehner:2011wc} indicates that a naked singularity forms in finite asymptotic time as the horizon pinches. In fact, the numerical work \cite{Lehner:2011wc} reveals that  the horizon develops in  an approximately
 self-similar fashion at late times: the black string becomes a string of 5D black holes of various sizes connected by thin strings. These thin strings are themselves subject to the Gregory-Laflamme instability and this results in further clumping, thus giving a self-similar evolution. (This is similar to the behavior in a low-viscosity fluid stream: the Rayleigh-Plateau instability  causes a cascade of spherical beads to 
develop 
 in a self-similar manner  along the stream \cite{Cardoso:2006ks}.)

Since a naked singularity forms without fine-tuning of the initial data, this constitutes a {\it  violation of cosmic censorship}. Classical gravity can no longer be trusted near the  singularity and it is expected that quantum gravity effects must be included to understand the evolution. However, the most natural outcome is simply that the horizon bifurcates
and the endstate of the Gregory-Laflamme instability is a localized black hole.

The homogeneous black string and the localized black holes are not the only static black hole solutions to the 5D vacuum Einstein equations with Kaluza-Klein boundary conditions. As one increases the mass of a  localized black hole on a circle of fixed asymptotic size $L=2\pi R$, there is a critical mass $G_5 M/L^2 \sim 0.12$ where the horizon merges across the Kaluza-Klein circle and for larger masses one has a new {\bf inhomogeneous black string solution} whose horizon topology is $S^2 \times S^1$. It has been constructed numerically  \cite{Wiseman:2002zc,Kudoh:2003ki,Kleihaus:2006ee}; embeddings of the horizon is illustrated in Fig.~\ref{fig:KKembed}.

\begin{figure}[t]
\begin{center}
\includegraphics[width=11.5cm]{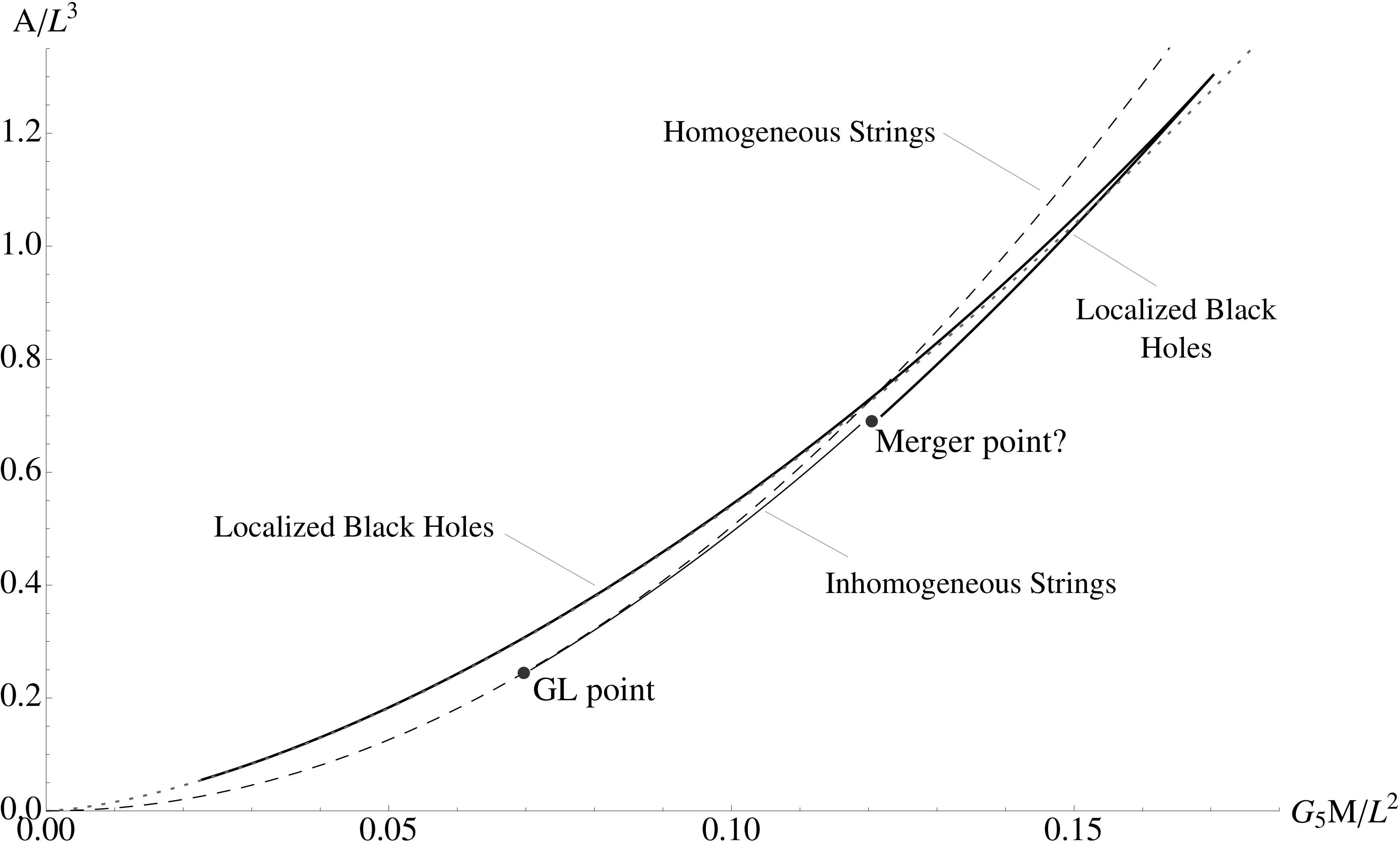}
\end{center}
\caption{Phase diagram for $D=5$ Kaluza-Klein black holes. The horizon `area' is plotted versus the black hole mass for fixed length $L=2\pi R$ of the Kaluza-Klein circle at asymptotic infinity. The localized black hole curve (solid) starts off 
as 
 $A\propto M^{2/3}$ at small mass, since for small $G_5M/L^2$ it behaves as the asymptotically flat 5D Schwarzschild-Tangherlini black hole. The dashed curve for the homogeneous black string is $A \propto M^2$ reflecting the area dependence of  the 4D Schwarzschild black hole of the string.
The inhomogeneous black string branch (also solid) begins  
at the point $G_5M \approx 0.7 L^2$ where the Gregory-Laflamme instability first sets in. Numerics make it plausible that the branch of inhomogeneous black strings merge near $G_5M \approx 1.2 L^2$ with the localized black hole branch, as indicated in the plot. (Plot from \cite{Horowitz:2011cq}.)
} 
\label{fig:KKphases}
\end{figure}

The 5D vacuum solutions discussed here --- the homogeneous and inhomogeneous black strings and the localized black hole --- all have Kaluza-Klein asymptotics: at large $r$, these 5D vacuum solutions approach 4D Minkowski spacetime times the Kaluza-Klein circle $S^1$.  Fig.\,\ref{fig:KKphases} indicates the different solution branches in a ``phase diagram" where solutions are compared for fixed size of the Kaluza-Klein circle at infinity. Note that there can be more than one solution with the same mass; so we have {\bf black hole non-uniqueness} in 5D Kaluza-Klein 
spacetimes! As shown in the phase diagram in Fig.\,\ref{fig:KKphases}, the  inhomogeneous black string joins the homogeneous black string at the onset of the 
Gregory-Laflamme 
instability. This is expected due to the existence of a static inhomogeneous perturbation at this point. It is also clear from Fig.\,\ref{fig:KKphases} that the
entropy of the inhomogeneous black string  is smaller than that of the homogeneous black string of the same mass, so 
the area theorem implies that
 it could not have been a viable endstate of the Gregory-Laflamme instability,

It interesting to note that there is no positive energy theorem for Kaluza-Klein spacetimes;  in fact there exist solutions with arbitrarily low energy\footnote{For a definition of energy in Kaluza-Klein theory, see \cite{Deser:1988fc}.} \cite{Brill:1989di,Brill:1991qe}. Moreover, the simplest Kaluza-Klein spacetime, 4D Minkowski  space times a circle, is actually unstable. It can undergo decay by nucleation of {\bf Kaluza-Klein bubbles}, which are obtained by a double analytic continuation of the 5D Schwarzschild-Tangherlini 
 black hole \cite{Witten:1981gj}. This instability can be removed, and a positive energy theorem proven,
by including fermions with periodic boundary conditions on the $S^1$ \cite{Dai}. These fermions are naturally included in the supersymmetric theories we discuss later.
The existence of Kaluza-Klein bubbles actually allow for even more classical solutions to the 5D Einstein equation with Kaluza-Klein boundary conditions: these are static, analytically known solutions that describe combinations of black strings, black holes, and Kaluza-Klein bubbles \cite{Elvang:2002br,Elvang:2004iz}. 

In our presentation  
of 5D Kaluza-Klein gravity, we have encountered a richness of structure: linearly unstable black strings, black hole non-unique\-ness, and violation of cosmic censorship. It turns out that this also carries over to black holes in asymptotically flat 5D spacetimes. This is the subject of the next section.

\subsection{Asymptotically flat black holes in $D=5$ vacuum gravity}
\label{s:D5BHs}
We have already described a black hole solution in $D$-dimensional asymptotically flat space, namely the Schwarzschild-Tangherlini 
 solution \reef{DSchw}. 
Black holes in 4D vacuum gravity are characterized 
by their mass $M$ and angular momentum $J$. The 4D rotating black hole described by the Kerr solution can be generalized to $D>4$ dimensions: the rotating black hole solutions of the $D>4$ dimensional vacuum Einstein equation were found analytically in 1986 \cite{Myers:1986un} and are called {\bf Myers-Perry black holes}. 

In 4D
spacetime, 
angular momentum is often associated with an 
axis of rotation, but this is an artifact of having three spatial directions.  
It is more general to associate angular momentum with {\em planes of rotation}; in three spatial dimensions a plane is characterized by its normal vector, but this is not true in higher dimensions. 
The 
independent 
planes of rotation 
in $D$ dimensions 
can be characterized by the 
$\lfloor (D-1)/2\rfloor$ independent generators of the Cartan subalgebra of the $D$-dimensional rotation group $SO(D-1)$. 
Thus, in addition to its mass $M$, the 5D Myers-Perry black hole is characterized by two angular momenta $J_{1}$ and $J_{2}$ associated with rotations in two independent planes, say $(x^1x^2)$ and $(x^3x^4)$. 
The 5D Myers-Perry metric is
\bea
  ds^2 &=& -dt^2 + \frac{\mu r^2}{\Delta} 
  \Big( 
    dt + a_1 \cos^2\theta\, d\phi_1+ a_2 \sin^2\theta\, d\phi_2
  \Big)^2\\ \nonumber
  &&+ \frac{\Delta}{(r^2 + a_1^2)(r^2 + a_2^2)-\mu r^2}\,dr^2\\[1.5mm] \nonumber
  &&+  (r^2 + a_1^2) (\sin^2 \theta \, d\theta^2 + \cos^2\theta \,d\phi_1^2)
    +  (r^2 + a_2^2) (\cos^2\theta \, d\theta^2 + \sin^2\theta \,d\phi_2^2)\,,
\eea
where
\be
  \Delta = (r^2 + a_1^2)(r^2 + a_2^2)
   \bigg( 1
       - \frac{a_1^2 \cos^2 \theta}{r^2+a_1^2} 
       -\frac{a_2^2 \sin^2 \theta}{r^2+a_2^2}\bigg)\,.
\ee
The  topology of the horizon is $S^3$, but with rotation turned on, its shape is no longer round, but pancaked in the planes of rotation.  The mass and angular momentum are
\be
  M = \frac{3 \pi \, \mu}{8 G_5} \,, 
  ~~~~J_i = \frac{\pi \, \mu}{4G_5} a_i \,. 
\ee
The first law of thermodynamics is now 
$dM = TdS + \Omega_{1} dJ_{1} + \Omega_{2} dJ_{2}$ with $\Omega_{1,2}$ the angular velocities of the horizon. This is a straightforward generalization of the 
first  
law for Kerr black holes. 
Just as for Kerr, there is an upper bound on the magnitude of the angular momentum for given mass: 
$M^3 \ge \big( \tilde{J}_1^2 + \tilde{J}_2^2 +2|\tilde{J}_1 \tilde{J}_2| \big)$, where 
$\tilde{J}_i = \sqrt{27\pi/32 G_5} J_i$
. When both angular momenta are non-vanishing, the 5D Myers-Perry solution approaches a smooth solution describing an extremal black hole with $T=0$, just like Kerr. 
However, when one of the angular momenta vanishes, say $J_{2} = 0$, the maximally rotating 5D Myers-Perry black hole becomes singular. It is smooth for 
$\tilde{J}_{1}^2 <M^3$, but as the angular momentum is increased from $\tilde{J}_{1} = 0$ to the maximum value, the horizon `area' decreases monotonically to zero as the horizon flattens out in the plane of rotation. 

Now it turns out that 
the 
Myers-Perry black holes are not the only regular black hole solutions to the 5D vacuum Einstein  equation: there is another class of  solutions called  {\bf black rings}. 
The black rings 
have horizon topology $S^2 \times S^1$ and
 the metrics 
 are known analytically \cite{Emparan:2001wn,Emparan:2006mm}. To get some intuition for what a black ring is, recall the black string: suppose you take a 4D Schwarzschild black hole times a line, but instead of wrapping the string on a Kaluza-Klein circle,  close it into a round ring in an asymptotically flat spacetime. Then we have a black hole with $S^2 \times S^1$ topology, i.e.~a black ring. The difference between  a black ring and a black string is that the $S^1$ of the ring is contractible, 
 whereas the $S^1$ of the string is not.  

A ring-like distribution of mass in space is going to collapse upon itself, so clearly no black ring solution can exist without something balancing its gravitational self-attraction.  In vacuum, a static ring can be constructed in 5D asymptotically flat space, but it suffers from a conical excess angle inside the plane of the ring; the excess is needed to support the ring-shaped horizon topology. However, the black ring can be balanced against self-collapse by giving it angular momentum in the plane of the ring. For a given ADM mass $M$, the minimum angular momentum needed is $\tilde{J}^2 > \frac{27}{32} M^3$ \cite{Emparan:2001wn,Elvang:2003yy} and when this bound is satisfied there are black ring solutions that are smooth everywhere outside and on the horizon. 

\begin{figure}
\begin{center}
\includegraphics[width=3cm]{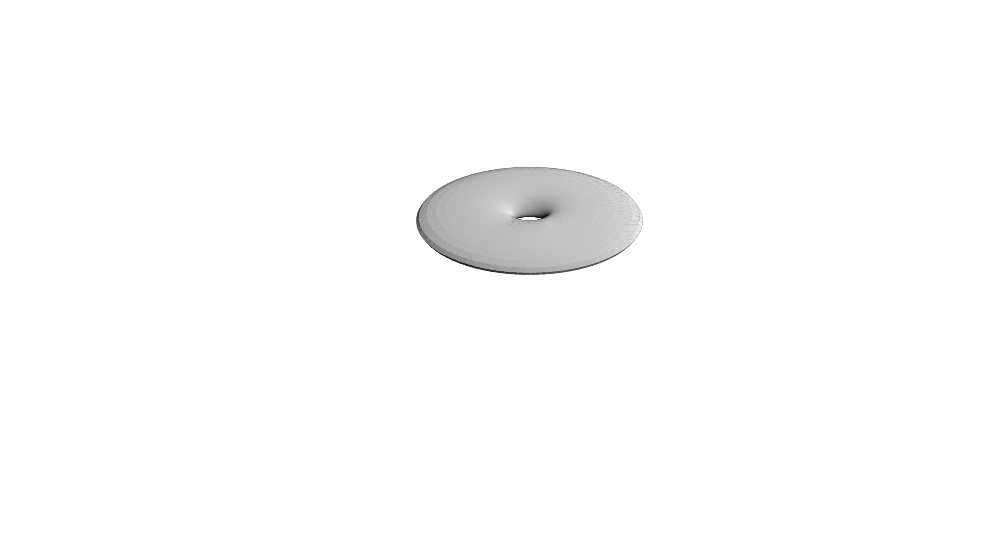}\hspace{-1.4cm}
\includegraphics[width=3cm]{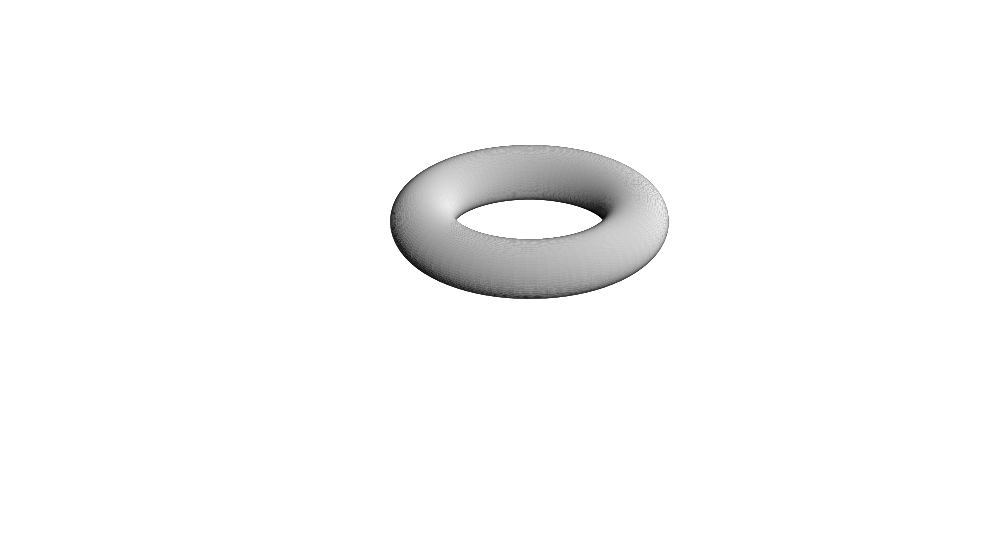}\hspace{-0.5cm}
\includegraphics[width=3cm]{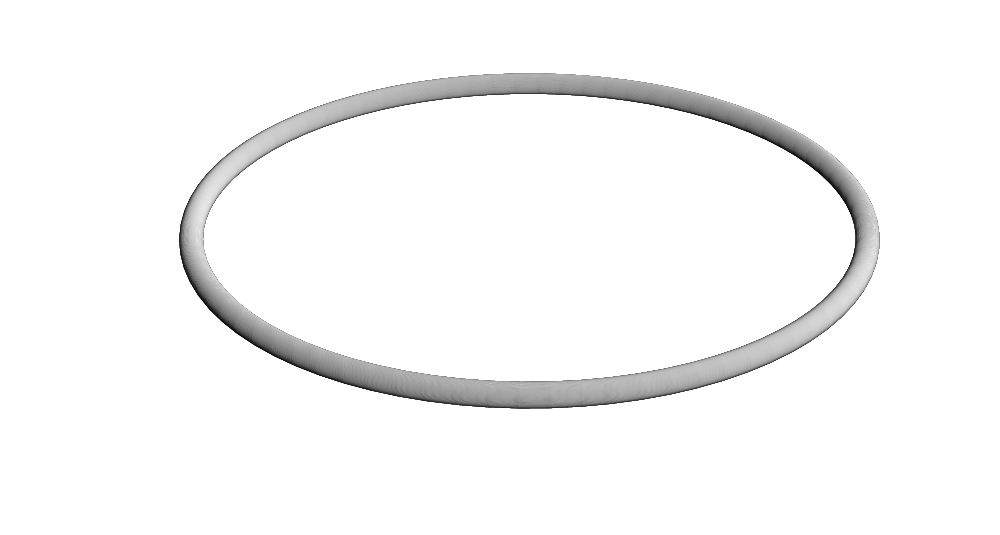}
\\[-13mm]
\includegraphics[width=10cm]{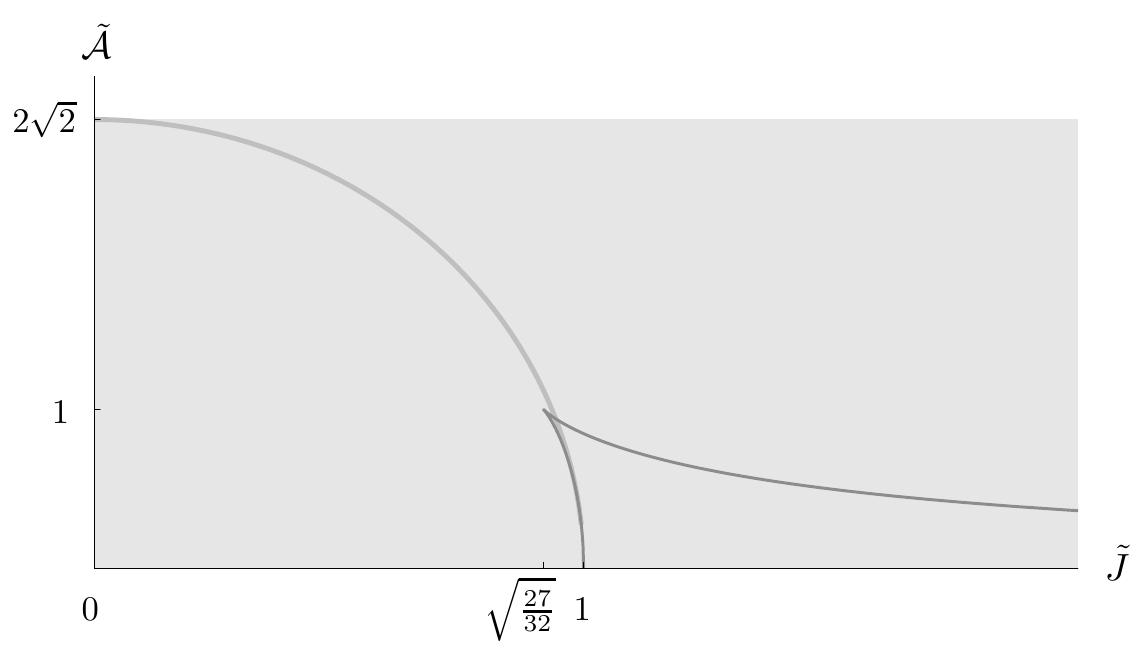}~~
\end{center}
\vspace{-5mm}
\caption{Phase diagram for asymptotically flat $D=5$ black holes with angular momentum in one plane: 5D Myers-Perry (light line),
 black rings  (dark line), and black saturns (gray shaded). The plot shows for fixed mass scale $M=1$ horizon area (entropy) $\mathcal{A}$ versus  angular momentum $J$ in rescaled units $\tilde{\mathcal{A}} = \sqrt{27/(256\pi G_5^3)}\mathcal{A}$ and 
$\tilde{J} = \sqrt{27\pi/(32G_5)} J$. (Plots from \cite{Elvang:2006dd,Elvang:2007hg}.)}
\label{fig:5dphases}
\end{figure}

As solutions to the vacuum Einstein equations, black hole thermodynamics is valid for black rings too, 
with  entropy  proportional to the three-dimensional `area' of the 
horizon.  
Figure \ref{fig:5dphases} is a plot of entropy versus angular momentum $J$ for smooth 5D Myers-Perry black holes and the black rings. The black holes and rings in this plot have rotation only in one plane, and for the ring, this is the plane of its $S^1$, as needed to balance it. 
Let us highlight some remarkable features: 
\begin{itemize}
\item There are 
{\it two branches of black rings}, 
the `lower' one consists of `fat' flattened-out black rings, while the higher entropy branch are thin black rings. The two branches meet at the cusp where the angular momentum takes its minimal possible value, $\frac{\tilde{J}^2}{M^3}=\frac{27}{32}$. 
\item In the range  $\frac{27}{32} \le  \frac{\tilde{J}^2}{M^3}< 1$, three distinct black hole solutions exists: one Myers-Perry black hole and two black rings, thin and fat. This is an exciting (and historically first) example of 
{\bf black hole non-uniqueness} 
in an asymptotically flat spacetime. Of course, this is completely different from 4D in which the Kerr black hole famously is the only smooth asymptotically flat stationary black hole vacuum solution.  
\item For the thin black rings, there is 
{\it no upper bound on the magnitude of the angular momentum}. 
As $J^2/M^{3}$ increases, the ring's $S^1$ radius  grows and the ring becomes very thin. A small part of the ring will look like a piece of a boosted black string \cite{Elvang:2006dd} --- given that it is thin, one would expect it to undergo Gregory-Laflamme instability. As the instability develops, there will be gravitational radiation from the time-varying quadruple moment of the rotating bumps on the ring, but the 
time scale 
of radiating away these bumps cannot compete \cite{Elvang:2006dd} with the 
time scale 
of the horizon pinch, so the likely endstate of the instability of an ultra-spinning black ring is a pair of black holes flying apart in such a way that the angular momentum is preserved. The pinch of the horizon would go through a naked singularity, so this would also constitute a violation of cosmic censorship.
\end{itemize}

Now just as the Myers-Perry black holes can carry angular momentum in the two independent planes of $\mathbb{R}^4$, so can 
black rings. As an intuitive picture, consider starting with a black string made from a Kerr black hole (instead of Schwarzschild) times a line. Then bend this Kerr string into a ring and set it into rotation in the plane of the $S^1$ of the ring. That gives a rotating black ring with angular momentum also on the $S^2$ of the ring cross-section. Saying the  words is easy, but  the construction of the {\bf doubly-rotating black ring} as a solution to the 5D Einstein equation is less trivial. 
An exact solution does exist
\cite{Pomeransky:2006bd}; it was constructed using the {\em `inverse scattering method'} \cite{Belinsky:1971nt,Belinsky:1979mh,Belinski:2001ph}. An analysis of the physical properties of the doubly-rotating black ring can be found in \cite{Elvang:2007hs}.

The inverse scattering method is an integrability technique that uses Lax pairs to generate new solutions to non-linear partial differential equations with input of a known solution. The method was originally used to study solitonic waves in shallow water.  In the late 1970's  it was realized \cite{Belinsky:1971nt} 
that  such techniques can also be applied to the 4D Einstein equations for co-dimension 2 systems. For example, one can use inverse scattering to generate the full Kerr solution from flat Minkowski space \cite{Belinski:2001ph}.
 Much more recently, the inverse scattering method was applied to generating solutions in 5D gravity and several new solutions were found. Let us survey them.

Consider the balance of the singly spinning black ring: for a given mass $M$ and $J$ above the lower bound, the ring adjusts its radius to achieve the needed balance. One can simulate this
 with a Newtonian model of a rotating rubber band that wants to contract to zero size: balance between the band tension and the centrifugal force is obtained for only one special radius. Now suppose the ring (or rubber band) is placed in an external central attractive potential. It then has to rotate a bit faster to be balanced at the same radius. This situation can happen too in 5D general relativity: we can imagine a rotating black ring balanced around a central 5D  Schwarzschild-Tangherlini   black hole. Remarkably, there exists an exact analytic solution to the 5D Einstein vacuum equations that realizes this {\bf Black Saturn} configuration \cite{Elvang:2007rd}. It is completely smooth everywhere outside and on the two horizons \cite{Elvang:2007rd,Chrusciel:2010ix}. The Black Saturn solution was constructed using the inverse scattering technique. 

Black Saturn displays a number of novel properties \cite{Elvang:2007rd}:
\begin{itemize}
\item It offers {\it 2-fold continuous black hole non-uniqueness}: 
for given  ADM mass $M$ (the total mass of the hole-ring system) and total ADM angular momentum $J$, there are continuously many ways of distributing the mass and angular momentum among the two black objects of the saturn system. The ring and hole can be co-rotating or counter-rotating. 
\item It shows that the 5D Schwarzschild-Tangherlini 
 solution \reef{DSchw} is {\em not} the unique solution with zero ADM angular momentum. One can arrange the ring and hole  of 
Black 
Saturn to be counter-rotating in such a way that the total system has zero angular momentum at infinity. This leaves the freedom of the mass distribution between the hole and the ring, thus leaving a continuous 1-parameter family of $J=0$ solutions which are degenerate (as far as asymptotic data goes) with the 5D Schwarzschild-Tangherlini 
 black hole (but have smaller entropy). 
\item The Black Saturn system illustrates frame-dragging effects as can be seen by studying the effect of, say, the rotating ring on the central black hole. 
\item No Black Saturn configuration has higher entropy than that of the 5D static
 Schwarz\-schild-Tangherlini 
  black hole, but it can come arbitrarily close. In fact the whole gray shaded region of phase space with $S < S_\text{Schw}$ and $\tilde{J}\ge 0$ in Fig. \ref{fig:5dphases} is filled out by a continuum of Black Saturn configurations \cite{Elvang:2007hg}.
\end{itemize}

In our solar system, the planet Saturn has more than one ring and this is also possible for Black Saturn. One can use the inverse scattering method  to construct exact Black Saturn solutions with any number of  rings rotating in the same plane. And why keep the `planet'? Drop the black hole at the center and simply just have a {\bf multi-ring system}; the simplest case with two rings in the same plane is called a {\bf di-ring} solution 
\cite{Iguchi:2007is,Evslin:2007fv}. 
And we can take this even further: why should the rings  be in the same plane? After all, we have two independent planes of rotation, so how about arranging two rotating rings in the two orthogonal planes? Such a system is known as a {\bf bi-ring} (or bicycling rings) and the exact solution has been constructed with
the inverse scattering method 
\cite{Izumi:2007qx,Elvang:2007hs}.

The solutions we have discussed above are  ``special" in the sense that they have more symmetry than is strictly needed: they have three commuting Killing vectors $\pa_t$, $\pa_{\phi_1}$, and $\pa_{\phi_2}$. The black hole rigidity theorem \cite{Hollands:2006rj,Moncrief:2008mr} requires only one rotational isometry, and even before these theorems were established it was conjectured that such less-symmetric non-static stationary black hole solutions exist \cite{Reall:2002bh,Moncrief:2008mr}.  This has been demonstrated  by the construction \cite{Emparan:2009vd} of so-called {\bf helical black rings} 
using 
asymptotic matching methods in the limit of large angular momenta. Helical black rings have the same horizon topology as the ring, $S^2 \times S^1$, but are shaped as a ``slinky" bent into a ring.

The richness of black holes in 5D Einstein vacuum gravity is clearly remarkable and unparalleled in 4D. Let us now briefly discuss what is known (and not known) about black holes in asymptotically flat spacetimes with $D>5$.

\subsection{Asymptotically flat black holes in $D>5$ vacuum gravity}
\label{s:Dge5}
Examples of black holes in higher-dimensional Kaluza-Klein theory are easy to obtain as direct products of lower-dimensional vacuum solutions times circles. For instance, we get a 6D black string from 5D Schwarzschild-Tangherlini 
 times $S^1$. Or taking the product of 4D Schwarzschild and 
a torus 
$S^1 \times S^1$, we get a 6D {\bf black membrane}. 
Of course this generalizes: 
the product of $n$-dimensional Schwarzschild-Tangherlini 
 times a $p$-torus gives a {\bf black $p$-brane} in a spacetime with $D=(n+p)$ dimensions. One can 
also construct stationary solutions such as a 6D rotating `black cylinder' as the product of a black ring and a circle. 

The asymptotically flat black holes we met in 5D generalize to higher dimensions. This includes the Myers-Perry rotating black holes. Consider a $D$-dimensional  Myers-Perry black hole with rotation just in one plane. The metric is \cite{Emparan:2003sy}
\be
 \begin{split}
 ds^2 \,=\,
 &
  -dt^2 
 + \frac{\m}{r^{D-5} \rho^2} \Big( dt + a\, \sin^2 \theta \, d\phi\Big)^2
 + \big( r^2 + a^2 \big) \sin^2\theta \, d\phi^2 \\
 &\,
  + \rho^2 \, d\theta^2 
 + r^2  \cos^2\theta\, d\Omega_{D-4}^2
 +\frac{\rho^2}{\Delta} \, dr^2\,,
 \end{split}
\ee
where
\be
  \rho^2 = r^2 + a^2 \,\cos^2\theta
  ~~~~~\text{and}~~~~~
  \Delta = r^2 + a^2 - \frac{\mu}{r^{D-5}} \,.
\ee
The mass is $M = (D-2) \Omega_{D-2} \mu/(16\pi G_D)$ and the angular momentum is $J=2 M a/(D-2)$. While the  angular momentum of the $D=5$ Myers-Perry black hole is bounded from above, it turns out that there is no such bound on the angular momentum for $D>5$: for given mass of the black hole, the angular momentum can be arbitrarily large. In this ultra-spinning limit, the black hole flattens out in the plane of rotation. At some stage it becomes very similar to a thin black membrane and is then expected \cite{Emparan:2003sy} to undergo an instability much like the Gregory-Laflamme instability. In several cases, this instability has been seen numerically. If the instability mode preserves the isometry of the generators of the rotation, then one can imagine that it causes a pinch that splits the rotating spherical black hole into a black ring! Or a black saturn. Or a multi-ring saturn-like system. (Generally, the fewer number of disconnected horizons, the higher the entropy of the system.) One can also imagine that there exist `lumpy' black holes whose horizons are topologically spherical analogous of the inhomogeneous black strings discussed in Section \ref{s:BS}. The possibilities and discussion of the phase diagram can be found in \cite{Emparan:2008eg}.

In $D \ge 5$, there are also black rings and di-rings and bi-rings and saturns; but beyond 5D, no exact solutions are currently known for these. 
Black rings (and helical black rings) in $D > 5$ 
have been constructed using the {\bf blackfold method} 
\cite{Emparan:2007wm,Emparan:2009at,Emparan:2009cs,Camps:2012hw}. 
(There have also been numerical constructions \cite{Kleihaus:2012xh} of $D > 5$ black rings.) 
 The blackfold method exploits the fact that black holes in higher dimensions can have more than one characteristic scale. For example, for a black ring, there is one scale associated with the size $R_{S^1}$ of the $S^1$ of the ring and another $R_{S^2}$ with the $S^2$. In the ultra-rotating regime where the ring is thin, the scales are separated $R_{S^2} \ll R_{S^1}$. The blackfold methods exploit such separation of scales to solve the Einstein equations in a matched asymptotic expansion. The method has also been used to construct black holes with more exotic horizon topologies, for example products of odd-spheres. For an overview of possibilities for the horizon topologies in 
$4 \le D \le 11$, see table 1 in \cite{Emparan:2009vd}.

So far, we have discussed only black holes in higher-dimensional Einstein vacuum gravity. It is natural to introduce matter fields and also consider charged black holes. This can be done in a general context, but  with an aim toward constructing the quantum theory, in the following we will focus on supergravity theory.

\section{Supergravity}
\label{s:super}
Supersymmetry is a symmetry that mixes bosons and fermions. It is the only possible extension of Poincar\'e spacetime symmetry for a unitary theory with non-trivial scattering processes \cite{Coleman:1967ad,Haag:1974qh}. Not only does this make supersymmetry a natural candidate for physics beyond the standard model of particle physics and a beautiful path to the unification of forces, it also provides an extremely powerful tool for understanding gauge theories --- and black holes!  The combination of general relativity with supersymmetry 
is   {\bf supergravity}. 
We will briefly introduce the ideas of supersymmetry and supergravity and then
discuss their  
impact on our understanding of charged black holes and how it improves the perturbative quantum theory.

\subsection{Supersymmetry}
\label{s:susy}
Let us begin with a simple, but concrete, example of supersymmetry. Consider in 4D flat space the Lagrangian for a 2-component Weyl fermion $\chi$ and 
a 
complex scalar field $\phi$ interacting via Yukawa terms and a quartic scalar interaction:
\be
  \label{Lsusy}
  \Lag = 
  i \chi^\dagger \bar{\sigma}^\mu \pa_\mu \chi 
   - \pa_\mu \bar\phi \,\pa^\mu \phi
  +\tfrac{1}{2} g \, \phi\, \chi \chi + \tfrac{1}{2} g^* \, \bar\phi \, \chi^\dagger \chi^\dagger
      - \tfrac{1}{4}|g|^2\,  |\phi|^4  \,.
\ee
The bar on $\phi$ denotes the complex conjugate and we have introduced the $2\times 2$ matrices $\sigma^\m = (1,\s^i)$ and $\bar{\sigma}^\m = (1,-\s^i)$, where $\s^i$ are the Pauli matrices.
In addition to the usual Poincar\'e symmetry, $\Lag$ also has a symmetry that mixes the fermions and bosons: 
\be
  \label{susytrInt}
  \begin{array}{rclcrcl}
  \delta_\eps \phi &\!\!\!=\!\!\!& \eps^\alpha \chi_\alpha \,,
  && 
  \delta_\eps \bar\phi &\!\!\!=\!\!\!& \eps^\dagger_{\dot{\alpha}} \chi^{\dagger{\dot{\alpha}}} \,,\\[1mm]
  \delta_\eps \chi_\alpha &\!\!\!=\!\!\!& -i  \s^\m_{\alpha \dot{\beta}} \, \eps^{\dagger \dot{\beta}} \pa_\m \phi
  + \tfrac{1}{2} g^* {\bar\phi}^2 \eps_\alpha\,,
  &&
  \delta_\eps \chi^\dagger_{\dot{\alpha}} &\!\!\!=\!\!\!& 
  i  \pa_\m \bar\phi\,\eps^{\beta} \s^\m_{\beta \dot{\alpha}} 
  + \tfrac{1}{2} g \phi^2\,\eps^\dagger_{\dot{\alpha}}
  \, \,.
\end{array}
\ee
This is an example of a {\bf supersymmetry transformation}.
The anti-com\-muting constant spinor $\eps$ is an infinitesimal  supersymmetry parameter (a fermionic analogue of the infinitesimal angle $\theta$ of a rotation transformation).

The anti-commuting conserved Noether supercharges $Q$ and $Q^\dagger$ resulting from supersymmetry give symmetry generators that extend the 
Poincar\'e algebra to a graded Lie algebra. The commutator of two supersymmetry transformations is a translation, $[\delta_{\eps_1},\delta_{\eps_2}] \sim (\eps_1^\dagger \sigma^\mu \eps_2)\, \pa_\mu$, so this induces the algebra 
\be
   \{ Q^\dagger , Q\} \sim P^\mu\,,~~~~~
   \{ Q , Q\}  = 0\,,~~~~~
   \{ Q^\dagger , Q^\dagger\}  = 0\,.
\ee
In addition, one has $[ Q^{(\dagger)}, P^\mu ]= 0$ and 
$[ Q, M^{\mu\nu } ] \sim  Q$.

In the quantum theory, the fields $\phi$ and $\chi$ in \reef{Lsusy} create a spin-0 or spin-$1/2$ particle from the vacuum, respectively. Since the fields are related by supersymmetry, so are the corresponding particles. The algebra outlined above implies that $P^2 = P_\m P^\m$ commutes with the supersymmetry generators, so this means that particles related by supersymmetry --- i.e.~in the same {\bf supermultiplet} --- must have the same mass. In our example \reef{Lsusy},  the boson and fermion are both massless. The supercharges act on a particle with spin $s$ by relating it to a particle with spin $s \pm \tfrac{1}{2}$. 

Supersymmetry implies that the number of on-shell bosonic and fermonic degrees of freedom are equal.\footnote{Off-shell counting of bosonic and fermonic degrees of freedom also match by inclusion of auxiliary fields. See for example textbooks such as \cite{Wess:1992cp} and \cite{Freedman:2012zz}.} 
 In our example \reef{Lsusy},
the complex scalar encodes two real degrees of freedom and the on-shell Weyl fermion 
similarly gives 
two real degrees of freedom (the positive and negative helicity states of the massless spin-1/2 fermion). Similarly, the two helicity states $\pm1$ of  
a 
massless vector boson, as a photon or gluon, are matched by the two $\pm1/2$ helicity states of its supersymmetric partner fermion, a photino or gluino. 

{\bf Extended supersymmetry} means that one has $\mathcal{N}$ pairs of supersymmetry charges $Q^A, Q_A^\dagger$ with $A=1,2,\dots,\mathcal{N}$. In that case, the supersymmetry algebra allows for the possibility of a central charge extension  
$\{ Q_\alpha^A, Q_\beta^B\} \sim \epsilon_{\alpha\beta} Z^{AB}$, where   
$Z^{AB}$ is antisymmetric. 
The model we described above in \reef{Lsusy} has $\cn=1$ supersymmetry. 
In a supersymmetric theory, we distinguish   
the internal `flavor' symmetries that commute with the supersymmetry generators 
 from 
the {\bf R-symmetries} that do not. An $\cn=1$ supersymmetric theory may have a $U(1)$ R-symmetry, while theories with extended supersymmetry can have non-abelian R-symmetry, 
typically $SU(\cn)$, 
 that rotates the supercharges among each other. 

In a 4D theory with spin no 
greater
than 1, the maximal admissible amount of supersymmetry is $\cn=4$. The 4D $\cn=4$ supersymmetric theory turns out to be unique: it is the maximally supersymmetric extension of Yang-Mills theory and it is known as 
``{\bf $\cn=4$ super Yang-Mills theory}" (SYM).  
Its spectrum of particles consists of 
the gluon,   
$(\cn\!=)4$ 
spin-1/2 gluinos, and 6 scalars. With the two helicity states of the gluon and the 6 real scalars, this amounts to 8 bosonic degrees of freedom and $4 \times 2= 8$ fermionic degrees of freedom. 
$\cn=4$ SYM  
has truly remarkable properties, for example the beta-function vanishes at all orders in perturbation theory, so there is no running of the gauge coupling. The theory is conformal, meaning that its super-Poincar\'e symmetry is enhanced to the superconformal group $SU(2,2|4)$. 
The bosonic part of this group is the 4D conformal group $SU(2,2) \sim SO(4,2)$  and 
$SU(4) \sim SO(6)$ R-symmetry. The fermionic part is generated by  16 supersymmetry generators $Q^A$ and $Q_A^\dagger$ and 16 superconformal generators $S^A$ and $S_A^\dagger$. 
The $\cn=4$ SYM theory plays a key role in many modern developments in high energy physics. The theory can also be obtained by keeping only the massless modes of 10-dimensional $\cn=1$ super Yang-Mills theory after Kaluza-Klein reduction on a 6-torus.

Above we introduced supersymmetry in the context of flat Minkowski space and used a constant spinor, $\pa_\m \eps = 0$, as the parameter in the supersymmetry transformations. This is called {\bf global} or {\bf rigid supersymmetry}.
The next possibility to consider is `gauged' supersymmetry, i.e.~{\bf local supersymmetry}, where the supersymmetry parameter depends generally on the local  spacetime coordinates, $\eps = \eps(x)$: the result is {\bf supergravity}. 

\subsection{Supergravity}
\label{s:sg}
Supergravity is the wonderful combination of supersymmetry and general relativity. A general feature of supergravity is that the gravitational field $g_{\mu\nu}$ is partnered with a Rarita-Schwinger field $\psi_\mu$; thus the spin-2 graviton is paired with a spin-$3/2$ gravitino. In spacetime dimensions $D=2,3,4$ (mod 8), the gravitino field  $\psi_\mu$ can be Majorana 
(real),\footnote{In other dimensions, one uses a Dirac or symplectic Majorana spinor; the super\-symmetry trans\-formations \reef{SGtransf} are then modified accordingly to ensure that $\delta_\eps e_\mu^a$ is real. For an overview of spinor representations in $D$-dimensions, see Table 3.2 in \cite{Freedman:2012zz}.} 
and in these cases the most fundamental structure of supergravity can be described in terms of the following action \cite{Freedman:2012zz}\footnote{Here and henceforth, we use Greek letters $\mu,\nu\,\dots$ for the $D$-dimensional
coordinate frame indices.}
\be
  \label{SGaction}
  S= \frac{1}{16 \pi G_D} \int d^D x \, \sqrt{-g} \, \Big[ R 
  - \overline{\psi}_\mu \gamma^{\mu\nu\rho} D_\nu \psi_\rho \Big]\,  .
\ee
Here,
$\gamma^{\mu\nu\rho} = \gamma^{[\mu} \gamma^\nu\gamma^{\rho]}$ is the fully antisymmetric product of 
three 
$\gamma$-matrices of the $D$-dimensional Clifford algebra. 
The gravitino covariant derivative,
\be
  D_\nu \psi_\rho =  \pa_\nu \psi_\rho + \frac{1}{4} \omega_{\nu ab} \gamma^{ab} \psi_\rho\,,
\ee
is given in terms of the torsion-free spin-connection $\omega_{\nu ab}$. 
(The Christoffel connection is 
not needed 
because of the contraction with the antisymmetric gamma-matrix.)
Using the vielbein $e^\nu_a$, we have
\be
   \omega_{\mu}^{ab} = 2 e^{\n[a} \pa_{[\mu} e_{\nu]}{}^{b]} 
   - e^{\nu [a} e^{b]\rho} e_{\m c} \pa_\n e_\rho{}^c\,.
\ee

Consider the local supersymmetry transformation 
\be
  \label{SGtransf}
  \delta_\eps e_\mu^a = \frac{1}{2} \bar{\eps} \gamma^a \psi_\m\,,
  \hspace{1cm}
  \delta_\eps \psi_\mu =  D_\m \eps \,.
\ee
It is instructive to outline how \reef{SGtransf} acts on the action \reef{SGaction}; for full detail, see \cite{Freedman:2012zz}. First, recalling the derivation of Einstein's equation from the action principle, we are familiar with the result of varying the Einstein-Hilbert part of the action:
\be
   \delta_\eps \Big( \sqrt{-g} \, R\Big) ~\to~ 
    \sqrt{-g}  \,\Big(R _{\mu\nu} - \frac{1}{2} g_{\m\n} R\Big) 
    \big(-\bar{\eps}\, \gamma^\mu \psi^\nu\big)\,.
    \label{delEH}
\ee
Next, the  variation of the spinors in the spin-3/2 kinetic term 
gives  
---  after partial integration and careful tracking of the order of the fermion fields --- a term proportional to $\gamma^{\m\n\rho} D_\m D_\n \psi_\rho$. Antisymmetrization of the Lorentz-indices on the covariant derivatives 
allows  
us to replace $[D_\m, D_\n]$ with the Riemann curvature tensor; explicitly we have, at linear order in the gravitino field,
\be
   \delta_\eps \Big(
      -\sqrt{-g}\,\overline{\psi}_\mu \gamma^{\mu\nu\rho} D_\nu \psi_\rho \Big) 
      \Big|_{\text{lin.}\,\psi}
   ~\to~ 
   \frac{1}{4}\sqrt{-g} \,\bar{\eps} \gamma^{\m\n\rho} \gamma^{ab} 
   R_{\m\n ab} \psi_\rho\,.
\ee
Now the product of gamma-matrices can be expanded on a basis of rank $r=1,3, 5$ antisymmetric products of gamma-matrices $\gamma^{\rho_1 \dots \rho_r}=\gamma^{[\rho_1} \cdots \gamma^{\rho_r]}$.
Upon contraction with the Riemann tensor, the rank-5 term  
$\gamma^{\rho \m\n ab}R_{\m\n ab}$ vanishes thanks to the Bianchi identity. The rank-3 terms vanish as a result of application of the Bianchi identity and the symmetry properties of the Ricci tensor. Finally, one is left with two rank-1 contributions:
\be
   \delta_\eps \Big(
      -\sqrt{-g}\,\overline{\psi}_\mu \gamma^{\mu\nu\rho} D_\nu \psi_\rho \Big) 
        \Big|_{\text{lin.}\,\psi}
   ~\to~ 
   \sqrt{-g}  \,\Big(R _{\mu\nu} - \frac{1}{2} g_{\m\n} R\Big) 
    \big(\bar{\eps}\, \gamma^\mu \psi^\nu\big)\,.
   \label{SGvarF}
\ee
This cancels the variation of the Einstein-Hilbert term \reef{delEH} and we therefore see that the action \reef{SGaction} is invariant under the supersymmetry transformation \reef{SGtransf} to 
{\em linear order in the fermions}. The cancellation of the variations \reef{delEH}  and \reef{SGvarF} was first demonstrated in \cite{Freedman:1976xh}. 
It is remarkable how the proof of linearized supersymmetry relies on a delicate  interplay between fundamental identities:  the commutator of covariant derivatives in Riemannian geometry, spin and the Clifford algebra, and Fermi-statistics and its connection to the anti-commutation of the fields without which the Majorana gravitino kinetic term would be a total derivative. 
Invariance at non-linear order is dimension dependent and can require additional fields and other terms in the action. 
For minimal $\cn\!=\!1$ supergravity in 4D, 
no other fields are needed, but the action \reef{SGaction} must be supplemented by terms quartic in the gravitino field \cite{Freedman:1976xh}. 
Local supersymmetry  was also demonstrated in \cite{Deser:1976eh}. 

As with global supersymmetry, there is an equal number of fermionic and bosonic on-shell degrees of freedom in supergravity theories. 
Massless particles in $D$-dimensional spacetime are characterized by the irreducible representations of the `little group' $SO(D-2)$ (the part of the Lorentz group that leaves the 
null 
 momentum vector invariant). The graviton is symmetric and traceless, so that amounts to $D(D-3)/2$ bosonic degrees of freedom; this is 
 the same counting as the number of independent polarizations of a gravitational wave in $D$-dimensions, as discussed early in Section \ref{s:Ddim}. A Majorana gravitino in the vector-spinor representation of $SO(D-2)$ has $(D-3)2^{\lfloor (D-2)/2 \rfloor}$ degrees of freedom.
So for $D=4$, the graviton and the Majorana gravitino each have 2 degrees of freedom and hence the 
\mbox{$\cn\!=\!1$}
supergravity multiplet in 4D consists precisely of the graviton and the gravitino. 

One can couple other fields to supergravity as `matter' supermultiplets. For example in 4D we can add to the $\cn\!=\!1$ supergravity action
$N_v$ copies of 
 $\cn\!=\!1$ vector multiplets (consisting of a gauge boson and its gaugino partner) or 
$N_\chi$ copies of 
 $\cn\!=\!1$ chiral multiplets (1 spin-1/2 fermion and 1 complex scalar) while preserving $\cn\!=\!1$ supersymmetry. If a model has only the supergravity multiplet and no matter multiplets, we call it {\bf pure supergravity}.

Next, consider extended supergravity, 
i.e.~supergravity theories with more than one gravitino field. 
The $D=4$, $\cn\!=\!2$ pure supergravity theory \cite{Ferrara:1976fu} has 
 2 bosonic degrees of freedom for the graviton, $2 \times 2$ fermionic degrees of freedom from the two gravitinos, and finally 2 more bosonic  degrees of freedom from a spin-1 graviphoton. One can couple 
to it 
 extra $\cn\!=\!2$ vector supermultiplets (1 gauge boson, 2 gauginos, 1 complex scalar) and still preserve $\cn\!=\!2$ supersymmetry of the full action.

We will be describing black holes in higher dimensions, so let us next consider supergravity in five dimensions. The on-shell 5D graviton has 5 degrees of freedom. There is no 
spinor 
representation that can match this, so in 5D we cannot have a simple $\cn\!=\!1$ supergravity multiplet consisting of just a graviton and a gravitino. Instead, we can take the gravitino to be a symplectic Majorana spinor with $2\times 4$ degrees of freedom and include a graviphoton with 3 degrees of freedom in the on-shell supermultiplet. This is the field content of minimal supergravity theory in 5D and it has $\cn\!=\!2$ supersymmetry \cite{Hawking:1981bu}. 
The bosonic action for  minimal supergravity in 5D is not just Einstein-Maxwell theory, but also has a Chern-Simons term $A \wedge F \wedge F$.

For a theory whose highest spin particle is the spin-2 graviton,\footnote{Theories with states of spin higher than 2 have been constructed in anti-de-Sitter space (AdS), see \cite{Fronsdal:1978rb}, \cite{Fradkin:1986qy} and the newer review \cite{Vasiliev:1999ba}. This is particular interesting in connection with the gauge-gravity duality, see for example \cite{Klebanov:2002ja} and  \cite{Gaberdiel:2010pz}.} the maximal amount of supersymmetry allowed in 4D is $\cn=8$.
This is easy to see by working down from the highest helicity state of $+2$ and reducing the helicity by 1/2 at each application of the supersymmetry charge. 
After application of $\cn=8$ supercharges, one reaches the helicity $-2$ state of the graviton. Thus having more than 8-fold supersymmetry would give states with spin higher than 2 in 4D. 

$\cn\!=\!8$ supersymmetry in 4D gives a uniquely determined supergravity theory, simply called 
 `{\bf $\cn\!=\!8$ supergravity}'. 
 Its spectrum 
of $2^8=256$ massless states is organized into fully antisymmetric rank $r$ representations of the global $SU(8)$ R-symmetry: the two states of the graviton, 8 pairs of gravitinos, 28 pairs of graviphotons, 56 pairs of spin-1/2 gravi-photinos, and 70 scalars. 

Supergravity with 
a 
spin-2 graviton as the highest spin state exists in dimensions $D \le 11$. To see how the bound 
on the spacetime dimension 
arises, start in $D=11$ where the minimal spinor is a 32-component Majorana spinor. Upon dimensional reduction on a 7-torus, an 11D Majorana gravitino gives eight Majorana gravitinos in 4D. Indeed, the dimensional reduction of 11D supergravity on a 7-torus is 
$\cn\!=\!8$ supergravity theory in 4D. 
The minimal spinor representation in $D>11$ has more than 32 components (e.g.~for $D=12$  it is 64), so starting with   a gravitino in $D>11$ and reducing toroidally to 4D gives $\cn>8$ gravitinos in 4D. If this were a 4D supergravity theory, it would have states with spin greater than 2. Thus we conclude that we cannot have supergravity in $D>11$. 

In $D=11$, the gravitational field $g_{\m\n}$ encodes  44  
on-shell 
degrees of freedom. The 11D gravitino is a Majorana spinor in the vector-spinor representation, so it has 128 degrees of freedom. Matching the fermionic and bosonic degrees of freedom requires an antisymmetric 3-form field $A^{(3)}_{\m\n\rho}$: it precisely contains the needed 84 bosonic degrees of freedom. The 11D supergravity theory is unique and the bosonic part of the action is \cite{Cremmer:1978km}
\be\label{sugra11D}
 S = \frac{1}{16 \pi G_{11}}
 \int d^{11}x\, \bigg[\sqrt{-g}\Big( R 
 - \frac{1}{4!} F^{(4)}_{\m\n\rho\sigma}F^{(4)\,\m\n\rho\sigma} \Big)
 - \frac{\sqrt{2}}{3}  A^{(3)} \wedge F^{(4)}  \wedge F^{(4)} \bigg]\,,
\ee
where $F^{(4)}_{\m\n\rho\sigma}$ are the components of the 4-form field strength $F^{(4)}=dA^{(3)}$. The Chern-Simons term is needed for supersymmetry. The fermionic terms include the standard kinetic gravitino 
term 
of \reef{SGaction}, but also terms coupling the gravitinos to $F^{(4)}$. The 3-form potential $A^{(3)}$ naturally encodes the electric charge of a membrane in 11 dimensions; its electric charge is captured by Gauss' law $Q_\text{E} \propto \int_{S^7} \star F^{(4)}$, where $\star$ indicate the 11D Hodge dual and the $S^7$ is transverse to the membrane. Similarly, 
$Q_\text{M} \propto\int_{S^4}  F^{(4)}$ calculates the magnetic charge of an object extended in 5-spatial directions in 11D. Thus, the fundamental objects carrying electric and magnetic charges in 11D supergravity are 2-branes and 5-branes: they are called 
M2-  
and M5-branes and we will meet them again later in our discussion of 
string theory and M-theory in Section \ref{s:strings}.

There are two distinct $\cn\!=\!2$ supergravity theories in 10D: they are called Type IIA and Type IIB and differ by  whether the supersymmetry generators have  different chirality (Type IIA) or the same chirality (Type IIB). Type IIA  
can be  
obtained as the Kaluza-Klein reduction of 11D supergravity on a circle. Both Type IIA and Type IIB supergravity contain an antisymmetric 2-form potential $B_{\mu\nu}$. The objects that are electrically charged, $Q_\text{E} \propto \int_{S^7} \star H$,  under the corresponding 3-form flux $H=dB$ are 1-dimensional: they are strings! 
Indeed, it turns out that the Type II supergravity theories are low-energy limits of superstring theories with $\cn\!=\!2$ supersymmetry. 

Upon Kaluza-Klein compactification of the 11D supergravity theory to lower dimensions, one obtains many other interesting $D$-dimensional supergravity theories. For example, the 5D minimal $\cn\!=\!2$ supergravity theory described above is a certain truncation of  11D supergravity  on a 6-torus. 
And,
as noted earlier, 
  $\cn\!=\!8$ supergravity in 4D arises from 11D supergravity by reduction on a 7-torus.

In contemporary applications, compactifications of 11D supergravity, or 10D Type IIA/IIB supergravity, on curved manifolds 
are  
very important. When a Kaluza-Klein reduction of supergravity is performed on a manifold with positive curvature, such as a $p$-sphere $S^p$, the resulting lower-dimensional theory is  `gauged' supergravity.
 One can think of the `gauging' as having the gravitinos  charged under the gauge fields. Gauged supergravity typically comes with a  non-trivial scalar potential --- or in the simplest cases a negative cosmological constant.  Whereas Minkowski space is the simplest `vacuum' solution for ungauged supergravity,  anti-de-Sitter space (AdS) is a simplest solution in gauged supergravity.  As an example, Type IIB supergravity on an $S^5$ gives a 5D gauged supergravity theory \cite{Gunaydin:1984qu} that plays a central role in studies of the gauge-gravity  duality. We discuss this further in Section \ref{s:adscft}.

\subsection{Charged black holes, BPS bounds, and Killing spinors}
\label{s:bps}
In Section \ref{s:Ddim} we discussed higher-dimensional black holes as solutions to the vacuum Einstein equations.  It is very interesting to study classical solutions in supergravity, in particularly those 
with 
special supersymmetric properties, as we now describe.

Denoting generic bosonic and fermonic fields by $B$ and $F$,  supersymmetry transformations generically take the schematic form \cite{Freedman:2012zz}
\be
 \delta_\eps B = \bar{\eps}\, f(B)\, F  + O(F^3)
 \,,
 ~~~~~\text{and}~~~~~
 \delta_\eps F = g(B)\,\eps + O(F^2) \,,
 \label{susyFB}
\ee
where $f$ and $g$ are functions of the bosonic fields
and their derivatives. 
We are interested in classical solutions (of the supergravity equations of motion)  that  the supersymmetry transformations \reef{susyFB} leave invariant.
Typically, we consider solutions that only have non-trivial bosonic fields, i.e.~all the fermion fields are set to zero, $F=0$. 
Since the supersymmetry variation of bosons \reef{susyFB} are  proportional to the fermion fields, they automatically vanish, $\delta_\eps B=0$, on a purely bosonic solution. On the other hand, we get non-trivial constraints from the condition 
that fermion variations vanish, $\delta_\eps F=0$. In the simplest form \reef{SGtransf}, the constraint is $0 = \delta_\eps \psi_\mu =  D_\m \eps$, so it requires the existence of a covariantly constant spinor $\eps$. More generally, there will be other fields involved in the condition $\delta_\eps F = 0$;
we will see examples shortly. 
 The spinors that solve the constraints arising from setting the supersymmetry variations of the fermion fields to zero are called {\bf Killing spinors}. If a classical  solution has $n$ parameters characterizing its Killing spinors, 
 it is said to  preserve $n$ 
supersymmetries. 

The existence of Killing spinors 
has  
important implications. For example, the bispinor products of Killing spinors $\bar \eps_1 \gamma^\m \eps_2$ are Killing vectors associated with the ordinary (bosonic) symmetries of the spacetime solution. 
And very importantly, the Killing spinor equations imply a set of first order equations consistent with the equations of a motion, making it easier to find exact solutions.

Another important implication is that the existence of Killing spinors implies that certain energy bounds are saturated. These are called BPS bounds after Bogomol'nyi, Prasad, and Sommerfield \cite{Bogomolny:1975de,Prasad:1975kr}.
Thus solutions with Killing spinors are often called {\bf BPS solutions}. 

As a simple example, Witten's proof of the positive energy theorem \cite{Witten:1981mf} shows that the ADM mass is positive $M \ge 0$ with equality precisely when there exists a covariantly constant spinor, $D_\m \eps = 0$. In particular,  Minkowski space has a covariantly constant Killing spinor
and obviously it has $M=0$. It is a BPS solution
in pure $\cn=1$ supergravity.

Pure $\cn=1$  {\em gauged}  supergravity in 4D has a negative cosmological constant $\Lambda = -3/L^2$ and the Killing spinor equation is
\be
   0 = \delta_\eps \psi_\mu =  D_\m \eps - \frac{1}{2L} \,\gamma_\mu  \eps \,.
\ee
Four-dimensional anti-de-Sitter space (AdS$_4$) 
admits such a Killing spinor, so AdS 
with radius $L$ 
is a BPS solution in gauged supergravity.

Recall from Section \ref{s:sg} that the bosonic sector in pure $\cn=2$ supergravity in 4D consists of the gravitational 
field $g_{\mu\nu}$ and the graviphoton field $A_\m$. The purely bosonic part of the action turns out to be Einstein-Maxwell theory. The vanishing of the supersymmetry transformation of the gravitino fields in this theory gives a  Killing spinor equation of the form\footnote{Here we are setting $G_4 =1$ for simplicity.}
\be
  \hat{D}_\mu \eps 
  \,\equiv\,  D_\mu \eps  
  - \tfrac{1}{4}  F_{\n\rho} \gamma^\n \gamma^\rho \gamma_\mu \eps 
  \,= \,0\,.
\ee
An argument similar to Witten's \cite{Witten:1981mf} shows  that the mass $M$ and electric and magnetic charges, $Q$  and $P$, of 
regular 
solutions to the equations of motion of $\cn=2$ supergravity in 4D satisfy the bound  \cite{Gibbons:1982fy}
\be
  M \ge \big( Q^2 + P^2\big)^{1/2} \,. 
  \label{MQP}
\ee
Equality holds precisely when the solution admits a Killing spinor $\hat{D}_\mu \eps =0$.

The bound \reef{MQP}  looks very familiar: it is precisely the bound the Reissner-Nordstrom black hole must satisfy in order to have a smooth horizon! Thus, the {\em extremal Reissner-Nordstrom black hole is a BPS solution of $\cn\!=\!2$ supergravity in 4D}.

The temperature of the extremal Reissner-Nordstrom black hole is zero. However, extremality in the sense  of zero temperature --- or coinciding inner and outer horizons --- does not necessarily mean that the solution is BPS. For example, the extremal Kerr black hole is not BPS: without an electromagnetic charge, 
a 
solution with $M>0$ cannot saturate the  BPS bound \reef{MQP} and hence it does not admit a Killing spinor. Similarly, no Kerr-Newman black hole with $J \ne 0$ saturates the bound \reef{MQP}. 
Hence, there are no asymptotically flat rotating BPS black holes in 4D ungauged supergravity.

Given our discussion in Section \ref{s:D5BHs} of vacuum solutions describing black holes and black rings in 5D, it is natural to ask if they have charged cousins. For simplicity, we first  answer this question  in the context of  minimal 5D supergravity (described briefly in Section \ref{s:sg}) and then generalize. In 5D, the equivalent of the Reissner-Nordstrom black hole is a static, charged black hole with a round $S^3$ horizon. In its simplest form, it is an electrically charged solution of the equations of motion in minimal 5D supergravity. It has an extremal limit in which the inner and outer horizon coincide; in this limit, the solution is supersymmetric and saturates the appropriate 5D BPS bound $M \ge \tfrac{\sqrt{3}}{2} Q$ \cite{Gibbons:1993xt}.

The 5D version of the Kerr-Newman solution is a charged version of the Myers-Perry black hole described in Section \ref{s:D5BHs}. It can have angular momenta $J_1$ and $J_2$ in both the two independent planes of 5D spacetime. The solution was first constructed in \cite{Cvetic:1996xz} and, in its simplest version, it is a solution to minimal 5D supergravity. The BPS limit of this charged rotating black hole is called the BMPV black hole \cite{Breckenridge:1996is}. The BMPV black hole has $M = \tfrac{\sqrt{3}}{2} Q$ and 
--- unlike in 4D ---  it can still carry angular momentum provided that the magnitudes are the same in the two planes of rotation, $|J_1|=|J_2|$.

Black rings can also carry charges \cite{Elvang:2003yy} and they have limits in which they are  BPS solutions. BPS black rings were first constructed as exact solutions in minimal 5D supergravity \cite{Elvang:2004rt}. These 
BPS rings have $M = \tfrac{\sqrt{3}}{2} Q$, but --- contrary to the BMPV 
black holes --- the angular momentum $J_1$ in the plane of the ring must be strictly greater than the angular momentum of the $S^2$ (the orthogonal plane): $|J_1|>|J_2|$.  
Charged 
black rings  
have a new feature: they carry a non-conserved `dipole' charge \cite{Elvang:2003yy,Elvang:2004rt} associated with application of Gauss' law with an $S^2$ surrounding a piece of the ring. This measures a string-like charge density along the $S^1$ of the black ring; since this ring is a contractible circle, the `dipole' charge is not conserved, but it impacts the solution non-trivially and is required for smoothness of the horizon.

The  BMPV black hole and the charged black rings described above have a natural generalization \cite{Breckenridge:1996is,Bena:2004de,Elvang:2004ds,Gauntlett:2004qy} in which they carry conserved charges of 3 distinct gauge fields of a 5D supergravity theory obtained from reduction of Type IIB supergravity in 10D on a 5-torus. The `minimal' solutions are recovered in the limit where the three charges are equal. 
The BPS black holes with  three different charges play a key role in Section \ref{s:strings} when we discuss how string theory offers a precise microscopic account of black hole entropy. 

\subsection{Perturbative quantum gravity}
\label{s:perturbative}
The focus of 
this section is 
on 
the application of standard quantum field theory in flat spacetime to scattering of gravitons, the spin-2 particles associated with the quantization of the gravitational field $g_{\m\n}$. More precisely, we expand the gravitational field around a flat space background: $g_{\m\n} = \eta_{\m\n} + \kappa h_{\m\n}$, where $\kappa^2 = 8\pi G_D$. 
The fluctuating field $h_{\m\n}$ is the {\bf graviton field}. 
Consider pure gravity without matter and expand the Einstein-Hilbert action in powers of $\kappa h_{\m\n}$:
\be
  \begin{split}
    S_\text{EH}
    &= \frac{1}{2\kappa^2} \int d^D x \,\sqrt{-g}\, R 
    \\
    &=  \int d^D x \,
    \Big[ 
      h \pa^2 h + \kappa  \,h^2 \pa^2 h
      + \kappa^2  \,h^3 \pa^2 h
      + \kappa^3  \,h^4 \pa^2 h + \dots
    \Big]\,.
  \end{split}
    \label{EHaction2}
\ee
Since the Ricci-scalar $R$ involves two derivatives, every term in the expansion has two derivatives. There are infinitely many terms, with increasingly delightful  assortments of index-structures; in \reef{EHaction2} we have written them schematically  as $h^{n-1}\partial^2 h$. There are no mass terms in \reef{EHaction2}, so the particles associated with quantization of the gravitational field $h_{\m\n}$  are massless: they have spin-2 and are the {\bf gravitons}. 

It is interesting to study graviton scattering processes, but we have to gauge fix the action \reef{EHaction2} before extracting the Feynman rules. A standard choice is {\bf de Donder gauge}, $\pa^\m h_{\m\n} = \frac{1}{2} \pa_\n h_\m{}^\m$, which brings the quadratic terms in the action to the form
\be
   h \pa^2 h 
   ~\to~
    -\frac{1}{2} h_{\m\n} \Box h^{\m\n}
    + \frac{1}{4} h_{\m}{}^\m \Box  h_\n{}^\n \,.
\ee
The propagator resulting from these quadratic terms is
\be
  P_{\m_1 \n_1, \m_2 \n_2}
   = -\frac{i}{2} 
   \Big(
     \eta_{\m_1\m_2}  \,\eta_{\n_1\n_2}
   +\eta_{\m_1\n_2}  \,\eta_{\n_1\m_2}
   - \frac{2}{D-2} \, \eta_{\m_1 \n_1}\, \eta_{\m_2 \n_2}
   \Big)
   \frac{1}{\,\,k^2}\,.
   \label{dedonder}
\ee
The external lines in graviton Feynman diagrams have two Lorentz-indices that must be contracted with graviton polarization vectors. In 4D, the polarizations encode the two helicity $h=\pm 2$ physical graviton states. They can be constructed as products of 
spin-1 photon polarization vectors 
$\eps^\m_\pm(p_i)$. Picking a basis where $\eps^\m_\pm(p_i)^2 = 0$, the graviton polarizations 
\be
  e_-^{\m\n}(p_i) = \eps^\m_-(p_i) \eps^\n_-(p_i) \,,
  \hspace{1cm}
  e_+^{\m\n}(p_i) = \eps^\m_+(p_i) \eps^\n_+(p_i) \,.
  \label{gravpol}
\ee
are automatically symmetric and traceless.

The infinite set of 2-derivative interaction terms $h^{n-1} \pa^2 h$ yield complicated  Feynman rules for $n$-graviton vertices for {\em any} $n=3,4,5,\dots$.  Together with the 3-term de Donder propagator \reef{dedonder}, this is a clear indication that calculation of tree-level graviton scattering amplitudes from Feynman diagrams is highly non-trivial. 
Nonetheless, it turns out that the final result for the on-shell amplitudes can be written in a relatively simple form (for an overview, see \cite{Elvang:2013cua}). In fact, there is an interesting relationship between tree-level graviton amplitudes and gluon amplitudes in Yang-Mills theory. For the case of 4-particle amplitudes this relationship is
\be
  M^\text{tree}_4(1234) 
  =~ - 
  s \,A^\text{tree}_4[1234] \,A^\text{tree}_4[1243] \,,
  \label{KLT4}
\ee
where $M^\text{tree}_n$ denotes a tree 
$n$-graviton
amplitude and $A_n^\text{tree}$ a (color-ordered) tree 
$n$-gluon
amplitude. 
The prefactor is the kinematic invariant Mandelstam variable
$s= - (p_1 + p_2)_\mu(p_1 + p_2)^\mu$. 
In four dimensions, the relation between the scattering states in \reef{KLT4} is 
\be
  \text{graviton}^{\pm 2}(p_i) 
  =  \text{gluon}^{\pm 1}(p_i) \otimes   \text{gluon}^{\pm 1}(p_i) \,,
  \label{gravglue}
\ee 
where $\pm$ indicates the helicity state.

There are similar (although somewhat more involved, see Appendix A of \cite{Bern:1998sv}) expressions for $M^\text{tree}_n$ in terms of sums of products of two $A_n^\text{tree}$ for all $n$. These are called the {\bf KLT relations} after Kawai, Lewellen and Tye \cite{Kawai:1985xq} who first derived such relations between closed and open string amplitudes; the field theory relations, such as \reef{KLT4}, are obtained in the limit where the string tension goes to infinity and the string behaves as a point particle. We discuss string theory in Section \ref{s:strings}.

From the point of view of the 
Lagrangian \reef{EHaction2}, the KLT relations are 
very  
surprising. Field redefinitions and clever gauge choices can bring the Feynman rules into a KLT-like form; see \cite{Bern:1999ji,Bern:2000mf,Siegel:1993xq} and the review \cite{Bern:2002kj}.  More recently, another form of the relation between gravity  and gauge theory amplitudes has been found: it is known as {\bf BCJ duality relations}, named after Bern, Carrasco, and Johansson \cite{BCJ}. Contrary to the KLT relations, the BCJ relations can also be applied at loop-level; not only do they offer a powerful alternative to the gravitational Feynman rules, they also hint at a possible deeper structure in perturbative gravity. The study of the surprisingly rich and enticing mathematical structure of scattering amplitudes in both Yang-Mills theory and in gravity is currently an exciting area of research (see for example \cite{ArkaniHamed:2012nw,Elvang:2013cua}). 

Let us now discuss the behavior of graviton loop amplitudes in the high-energy (ultraviolet, UV) limit. Consider for example a 1-loop diagram with $m$ external gravitons and only cubic interactions. The numerator of the loop-integrand can have up to $2m$ powers of momenta, since each graviton interaction vertex has two derivatives, and with $m$ propagators, this naively gives
\bea
 \text{gravity 1-loop diagram}  
  \sim
  \int^\Lambda d^4 k \,\frac{(k^2)^m}{(k^2)^m} 
  \sim 
  \Lambda^4 \,.
  \label{mgon}
\eea
This is power-divergent as the UV cutoff $\Lambda$ is taken to $\infty$ for all $m$. On the other hand, for Yang-Mills theory, the interactions are at most 1-derivative, so the 
integral 
\reef{mgon}
 (now with $m$ external gluons) 
has at most $k^m$ in the numerator, and hence it is manifestly UV finite for $m>4$.

However, the power-counting is too naive. There can be cancellations  within each diagram. Moreover, individual Feynman diagrams should not necessarily be taken too seriously since they are not gauge invariant. So cancellations of 
 {\bf UV divergences}  can take place in the sum of diagrams, rendering the on-shell amplitude better behaved than naive power-counting indicates.

Actually, pure gravity in 4D is 
finite at 1-loop order \cite{'tHooft:1974bx}: all the 1-loop UV divergences cancel!  
This is can be seen from the fact that the only viable 1-loop counterterm in pure gravity in 4D must be quadratic in the Ricci tensor, but by a field redefinition such a term can be completed to the Gauss-Bonnet term which is a total derivative. 
At 2-loop order, pure gravity indeed has a divergence \cite{Goroff:1985sz,vandeVen:1991gw}. In Yang-Mills theory, divergences are treated with the procedure of renormalization. However, in gravity, it would take an infinite set of local counterterms to absorb the divergences and hence the result is unpredictable: pure gravity is a non-renormalizable theory.

From the point of view of renormalization, the theory described by the Einstein-Hilbert action is naturally regarded as an {\bf effective field theory} that cannot be extrapolated to arbitrarily high energy. 
To see this, note that the 4D gravitational coupling $\kappa \sim G_4^{1/2}$ has  dimension of (mass)$^{-1}$. Perturbative calculations rely on an expansion in the small dimensionless coupling $E \kappa$, where $E$ is the energy scale of the scattering process.  Thus, perturbation theory is only valid at energies much smaller than $G_4^{-1/2} \sim M_\text{Planck} \sim 10^{19}$\,GeV. In other words, perturbation breaks down at high energies and from this point of view Einstein gravity is an effective field theory. 
As a classical effective field theory, general relativity is hugely successful and captures classical gravitational phenomena stunningly as shown by experimental tests.

Viewing gravity as an effective theory, we can study the perturbative amplitudes. The tree-level amplitudes capture the classical physics and there are no UV divergences to worry about. Could we imagine adding matter fields to cure the 2-loop divergence in pure gravity? Gravity with generic matter is 1-loop divergent \cite{'tHooft:1974bx,Deser:1974cz}, but it turns out that  any 4D theory of pure ungauged supergravity is finite at 1- and 2-loop order \cite{Grisaru:1976ua,Grisaru:1976nn,Tomboulis:1977wd,Deser:1977nt}. {\it Supersymmetry helps to tame the UV divergences.} This is 
in part 
due to cancellations between the boson and fermion loops. 
To date, only one explicit example of a UV divergence has been calculated in an amplitude in a pure ungauged supergravity theory in 4D, namely at 4-loop order in $\cn=4$ supergravity \cite{Bern:2013uka}.

It has been proposed \cite{Bern:2006kd} that maximal supergravity, $\cn=8$, in 4D could perhaps be ultraviolet finite. Explicit calculations \cite{Bern:2006kd,Bern:2007hh,Bern:2008pv,Bern:2009kd,Bern:2010tq} have demonstrated finiteness of 4-graviton amplitudes up to and including 
4-loop order, 
while symmetry arguments have established that no divergences can occur until 7-loop order \cite{Elvang:2010jv,Drummond:2010fp,Elvang:2010kc,Beisert:2010jx,Green:2008bf}. 
This can also be analyzed using superspace methods, see \cite{Bossard:2011tq} and references therein. 
The known symmetries do not constrain the divergences past 7-loop order, so it seems that UV finiteness would require the theory to have some hitherto hidden structure. Perhaps the relationships with Yang-Mills theory can clarify this.

From a field theory perspective, gravity needs a suitable UV completion, i.e.~another theory that reduces to general relativity in the low-energy limit. This is true for several reasons.
First, as we have just discussed, perturbation theory breaks down at high energies. Second, even if each order in a supergravity perturbation theory were UV finite, the perturbation series is not likely convergent; hence non-perturbative information is needed for a complete theory. Finally, the UV completion is  also needed in order to make sense of microscopic quantum properties of non-perturbative objects such as black holes.  A very successful candidate for such a UV complete theory of quantum gravity is string theory, which is the subject of our next section.

\section{String theory}
\label{s:strings}
String theory  combines the ideas of the previous two sections (higher dimensions and supersymmetry) and reduces to supergravity in a 
low-energy limit. In this section we give an overview of string theory, focussing on gravitational aspects of this large subject.\footnote{For an introduction to string theory, see \cite{Zwiebach:2004tj}. More complete  references include \cite{Polchinski:1998rq,Becker:2007zj}.} We will see that  it has many remarkable properties, including providing a theory of black hole microstates which reproduce both the Hawking-Bekenstein entropy and Hawking radiation.

 \subsection{Perturbative string theory}

 String theory starts with the idea that 
 particles are not really point-like, but excitations of a one-dimensional string. The string can be closed (topologically a circle) or open (a line segment). For now, we will focus on closed strings.  
As a string travels through spacetime, 
it traces out a  two-dimensional worldsheet. 
 To describe the dynamics of this worldsheet, we introduce local coordinates $(\sigma,\tau)$ on the worldsheet and $X^\mu$ on spacetime, so the position of the worldsheet is given by $X^\mu(\sigma,\tau)$.   If we introduce a nondynamical worldsheet metric $\gamma_{ab}$, then the string action can be written as follows:
\be\label{polyakov}
S[X^\mu,\gamma_{ab}] = {1\over 4\pi \ell^2_s} \int d\tau d\sigma \sqrt{-\gamma} \gamma^{ab}\partial_a X^\mu \partial_b X^\nu g_{\mu\nu}\,.
\ee
Here, 
$\ell_s$ is a new length scale in string theory (called the ``string length") which determines the string tension $T = 1/(2\pi \ell_s^2)$. The worldsheet metric $\gamma_{ab}$ is essentially ``pure gauge" since $S$ is invariant under both worldsheet diffeomorphisms and Weyl rescalings. One can remove it by solving its equation of motion $\delta S /\delta  \gamma_{ab} = 0$ to find that  $\gamma_{ab} $ is proportional to the induced metric on the worldsheet. Substituting this back into $S$ reduces the action to just the area of the induced metric. In other words, the worldsheet must be an extremal surface. This is a useful way to picture the motion of classical strings, but to quantize the string in flat spacetime,  it is much more convenient to work with (\ref{polyakov}), since  it is quadratic in the dynamical fields $X^\mu$.

There are several different approaches to quantize this string (e.g. light cone gauge or covariant quantization) which differ in how one treats the gauge freedom,  but they all agree on the physical results. The first thing one discovers is that the theory is only consistent in 26 spacetime dimensions. This surprising result arises in different ways depending on the approach one uses to quantize the string. If one completely fixes the gauge by going to a light cone gauge, one breaks manifest Lorentz invariance. Lorentz invariance is recovered only in $D=26$. In a covariant quantization, there are negative norm states created by the operators associated with $X^0(\sigma,\tau)$. These states are removed by constraints (associated with the gauge invariance) in $D=26$. 

 The physical spectrum of this bosonic string includes a scalar tachyon with $m^2 < 0$. The existence of this tachyon  
shows 
 that 26-dimensional Minkowski space is unstable in this theory. Although it is possible that this theory has a stable ground state, it has never been found. Instead, one proceeds by adding fermions, $\psi^\mu$, to the worldsheet and makes the two-dimensional theory supersymmetric. Quantizing  this superstring in flat spacetime, one now finds  that it is consistent in ten spacetime dimensions. 

The $D=10$ spectrum is now tachyon-free and consists of the following massless bosonic modes: a symmetric traceless ``graviton" $h_{\mu\nu}$, a scalar ``dilaton" $\phi$,  an antisymmetric ``Kalb-Ramond" field $B_{\mu\nu}$, and some ``Ramond-Ramond" fields $F_{\mu \cdots \nu}$ which are higher rank generalizations of a Maxwell field. In addition,  there 
is  
an infinite tower of higher mass and higher spin states. Finally, there are  fermionic partners for each of these bosonic states so that the complete spectrum is invariant under spacetime supersymmetry. Note that we only impose worldsheet supersymmetry, but nevertheless, the spacetime spectrum turns out to be supersymmetric.\footnote{This involves imposing a consistency condition known as the GSO projection \cite{GSO}.} At large mass, the spectrum is highly degenerate, with an entropy proportional to the mass. This can be understood by  thinking of a highly excited string as a random walk on a discrete grid, made up of segments of length $\ell_s$. Due to the tension, a string with $n$ segments has mass $M \sim n/\ell_s$. If the string can move in $p$ directions at each step, then the total number of configurations is\footnote{It is actually slightly less than this if we require the string to return to its starting point and form a closed loop, but this correction is subleading at large mass.} $p^n$, so the entropy is $S = n \log p \sim M\ell_s$.

The existence of the graviton is the first indication that a theory of strings has something to do with gravity. Much stronger evidence comes from quantizing the string in a static curved spacetime.  As we have said, the action  (\ref{polyakov}) is classically invariant under rescaling $\gamma_{ab}$, but quantum mechanically there is a conformal anomaly. To calculate this anomaly, one analytically continues the spacetime and worldsheet metrics to Euclidean 
signature\footnote{To evaluate a path integral, one often analytically continues to Euclidean space to convert the oscillating integrand into a convergent integral.}, 
and expands  $g_{\mu\nu}(X)$ in Riemann normal coordinates about a point $X_0$.  The anomaly can then be computed perturbatively in powers of  $\ell_s/L$ where $L$ is a typical length scale of the curvature. To leading order, the conformal anomaly vanishes if the  spacetime  satisfies Einstein's equation: $R_{\mu\nu} = 0$. If one couples the string to other background fields corresponding to other massless bosonic modes of the string, one recovers the equations of supergravity \cite{Callan:1985ia}. 
This is an important point: {\it in string theory, the full classical equations of motion for the spacetime fields come from demanding conformal invariance of the quantized worldsheet theory.} It is a remarkable and deep fact about string theory that Einstein's equation arises as a condition on the background fields (which are like coupling constants) in a 
two-dimensional 
quantum field theory. 
In general, these equations receive higher order corrections involving higher derivative terms, but they are usually negligible unless the curvature is of order the string scale. Thus general relativity (or supergravity) arises as the leading order classical equations of motion  in string theory. 

\begin{figure}[t]
\begin{center}
\includegraphics[width=8cm]{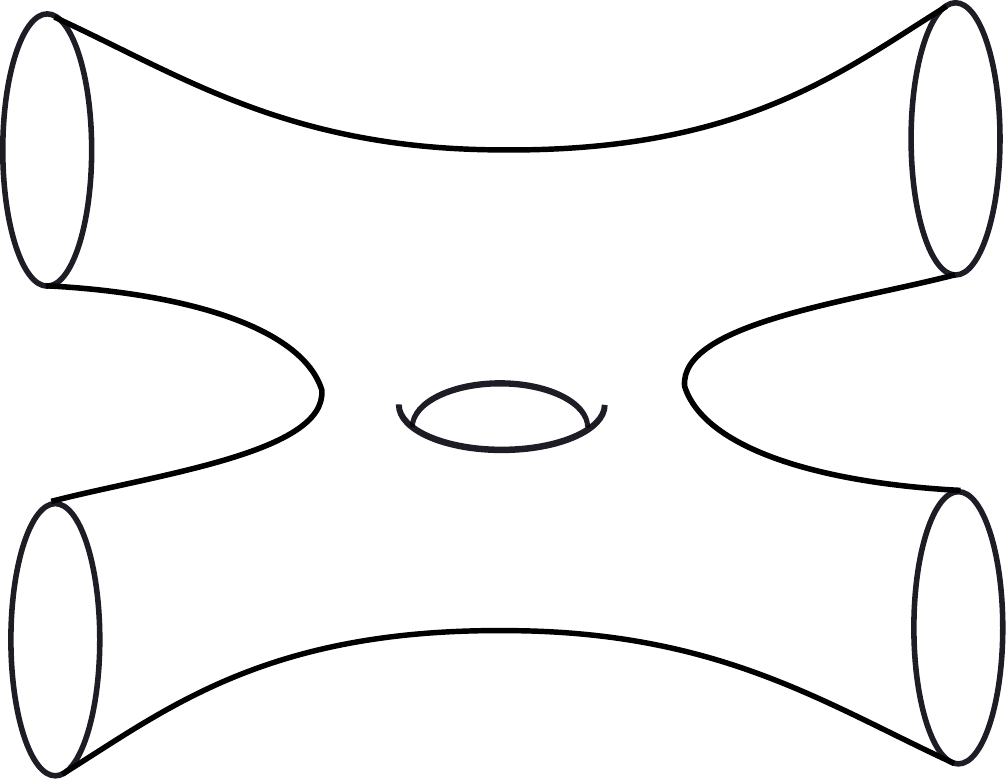}
\end{center}
\caption{A one-loop string diagram.} 
\label{fig:pants}
\end{figure}

One of the early successes of string theory was that it provided a perturbatively finite quantum theory of gravity. To describe this we have to introduce string interactions.  The basic assumption of string theory is that strings interact via a simple splitting and joining interaction. By quantizing a single string above, we have described a first quantized string. One might have thought that the next step would be to construct a string field theory in which the first quantized states $\Phi[X^\mu(\sigma), \psi^\mu(\sigma)]$ are promoted into field operators and one introduces interactions by adding cubic terms to the action. Progress has been made in this direction (see, e.g., \cite{Witten:1985cc,Sen:1999xm,Moeller:2000xv}), but perturbative scattering amplitudes in string theory are usually computed in a first quantized framework using analogs of Feynman diagrams, see Fig. \ref{fig:pants}

This is most well developed starting with Minkowski spacetime, but it can be extended to other static backgrounds. One  again analytically continues both the spacetime and worldsheet metrics to Euclidean signature. The scattering amplitudes are obtained by summing over worldsheet topology, and computing the path integral
\be\label{pathint}
\int DX^\mu D \psi^\mu D\gamma_{ab}\  e^{-S[X^\mu, \psi^\mu, \gamma_{ab}] }
\ee
for each topology. 
One usually takes the external  strings 
to infinity, so the amplitude corresponds to an S-matrix element. By a conformal transformation, the external strings can then be mapped to points on a compact Riemann surface, with the state of the string represented by an operator inserted at that point.
After fixing the $\gamma_{ab}$ gauge freedom  of diffeomorphisms and Weyl rescalings, there remains a 
finite-dimensional 
``moduli space" of metrics to integrate over for each worldsheet topology. 

The analog of the loop expansion in ordinary quantum field theory is the genus expansion of the string worldsheet. The reason for this is that the dilaton couples to the string via the scalar curvature ${\cal R}$ of the string worldsheet. In other words, one adds to the string action (\ref{polyakov}) the term ${1\over 4\pi}\int \phi {\cal R}[\gamma]$. By the Gauss-Bonnet theorem, if the dilaton is a constant $\phi_0$, this contributes $\phi_0 \chi$ to the action where $\chi$ is the Euler number of the string worldsheet.  
This is related to the genus $g$ by $\chi = 2(1-g)$. 
In the path integral \eqref{pathint}
the net effect is to 
weigh 
different worldsheet topologies by $g_s^{-\chi}$ where the string coupling $g_s$ is determined by the dilaton: $g_s \equiv e^{\phi_0}$. 

As an example, when the external states are all gravitons, the tree level  amplitude for two strings to scatter into two strings  reproduces the $2 \rightarrow 2$ graviton scattering of general relativity in a limit where strings become point-like.
Newton's constant is not an independent parameter in string theory, but determined in terms of $g_s$ and $\ell_s$. The ten-dimensional Newton's constant  is  $G_{10} = 8\pi^6 g_s^2 \ell_s^8$.

Unlike quantum field theory in which the number of Feynman diagrams grows rapidly  with the order of perturbation theory, string theory has only one diagram at each loop order (for a fixed number of external legs). The statement that string theory is UV finite is the observation that for each loop order, the integral \eqref{pathint} has no UV divergences. 
Intuitively, this is because the string is an extended object and there is no special point where the interaction takes place. The extent of the string regulates the short distance divergence in graviton scattering amplitudes that was discussed in 
Section \ref{s:perturbative}.
When these amplitudes were first 
computed 
in the 1980's, there were some remaining subtleties associated with the fermions and the integration over ``supermoduli" space. However,  
these issues have all been resolved \cite{Witten:2012bh,Witten:2013cia}. 

While it is remarkable that string theory picks out a unique spacetime dimension, 
it is clearly too large compared to our everyday observations of a 4D spacetime.   
As discussed earlier, a standard way to make contact with the real world is to compactify six of the dimensions. To preserve spacetime supersymmetry in the remaining four noncompact dimensions, the internal space is highly constrained. It turns out that it must be a complex manifold with Ricci flat Kahler metric \cite{Candelas:1985en}. Such spaces are called {\bf Calabi-Yau spaces}. Simple examples include the six torus, $T^6$, $T^2 \times K3$, and a quintic hypersurface in 
$\mathbb{CP}_4$. 
The realization of the importance of Calabi-Yau spaces in string theory started a very fruitful collaboration between algebraic geometers and string theorists which has continued for over twenty five years. 

 A common problem in earlier studies of Kaluza-Klein compactification was to obtain chiral fermions (as observed in the standard model
of particle physics) in the lower-dimensional spacetime. Even if one starts with chiral fermions in the higher-dimensional spacetime, the reduced spacetime always had an equal number of left-handed and right-handed fermions. String theory has several ways to cure this problem. The simplest is to use the fact that the higher-dimensional theory is not just pure gravity. If one starts with nonzero gauge  fields in the higher-dimensional spacetime, the lower-dimensional theory can have chiral fermions. In Calabi-Yau compactifications, various properties of fermions are simply related to the topology of the internal manifold (e.g., the number of generations is  related to the Euler number).

Strings sense spacetime very differently from point particles. In particular, two geometrically different spacetimes can be indistinguishable in string theory. A simple example of this is flat spacetime with one direction compactified to a circle of radius $R$. The part of the string spectrum that depends on $R$ includes  the usual momentum modes with energy $O(1/R)$,
 but there are also winding modes with energy $O(R/\ell_s^2)$. If we change the radius to $\ell_s^2/R$, this spectrum is completely invariant. The winding and momentum modes simply trade places. In fact, one can show that all interactions are also invariant, and strings cannot tell the difference between these two spacetimes.  This is the simplest example of {\bf T-duality} \cite{Giveon:1994fu}: whenever the solution is independent of a periodic spacelike direction, one can  change variables in the path integral (\ref{pathint}) and rewrite the action in terms of a different set of background fields (which include inverting the radius of the periodic spacelike direction). Since a change of variables does not change the physics, the two backgrounds are equivalent in string theory. 

Another example of different geometries being equivalent in string theory is {\bf mirror symmetry} \cite{Lerche:1989uy,Greene:1990ud}: 
for solutions of the form $M_4 \times K$ where $M_4$ is 
four-dimensional Minkowski space and $K$ is a Calabi-Yau manifold, one can change a sign of a certain charge in the worldsheet theory and the interpretation changes from strings moving on $M_4 \times K$ to strings moving on $M_4 \times \tilde K$ where $\tilde K$ is a geometrically and topologically different Calabi-Yau space. Since the sign of the charge is arbitrary from the worldsheet standpoint, these two compactifications are equivalent in string theory. As one striking application, mirror symmetry was used to  count the number of holomorphic curves of various degrees in a given Calabi-Yau manifold \cite{Candelas:1990rm}, reproducing and greatly generalizing results that had been obtained by mathematicians.

Mirror symmetry has been used to show that spacetime topology change is possible in string theory.
A given Calabi-Yau space usually admits a whole family of 
Ricci flat metrics, so one can construct a solution in which the four large
dimensions stay approximately flat and the geometry of the Calabi-Yau 
space changes slowly from one Ricci flat metric to another. In this process
the  Calabi-Yau space can develop a curvature singularity resulting from
a topologically nontrivial $S^2$  being shrunk down
to zero area. In the  mirror geometry, there is no singularity and the evolution can be continued. In the original description, 
 the evolution corresponds to continuing through the geometrical
singularity to a nonsingular Calabi-Yau space on the other side with different topology \cite{Aspinwall:1993nu}. It should perhaps be emphasized that examples like this show that area is not quantized in string theory. In many supersymmetric examples, the area of certain surfaces in the internal space can vary continuously. They give rise to massless scalar fields  in the noncompact directions that are known as ``moduli".

In the mid-1980's, {\it five} different perturbative string theories were constructed
 that were all consistent in ten spacetime dimensions. There was a theory of open strings called Type I with $\cn=1$ supersymmetry, and two theories of closed strings with $\cn=2$ supersymmetry \cite{Schwarz:1982jn}. The latter two differed in whether the two supersymmetry generators had different chirality (Type IIA) or the same chirality (Type IIB). As the names suggest, the 
 low-energy 
 limit of these two string theories are the  $\cn=2$ supergravity theories in  
10D  
mentioned in Section \ref{s:sg}. In addition, there were two theories of closed strings with $\cn=1$ supersymmetry which required either $E_8\times E_8$ or $SO(32)$ gauge groups in ten dimensions (called {\bf heterotic strings}) \cite{Gross:1984dd}. These gauge groups were required by anomaly cancellation. 
In fact, 
it was this discovery \cite{Green:1984sg} that sparked an explosion of interest in string theory in 1984. It seemed  remarkable that string theory not only picked a unique spacetime dimension, but also an essentially unique gauge group. A decade later it was realized that  the five perturbative string theories are all related by a series of ``dualities" (which include T-duality), so there was really only one theory with different weak coupling limits. This new insight was possible due to an improved understanding of some nonperturbative aspects of string theory which we now discuss.

\subsection{Nonperturbative aspects of string theory and quantum black holes}

In the mid-1990's,
 it was discovered that string theory is not just a theory of strings. 
There are other extended objects called branes. The name comes from membranes which are two-dimensional, but branes
exist in any dimension: $p$-branes are $p$-dimensional extended objects. Branes are nonperturbative objects with a tension that is
inversely related to a power of the coupling $g_s$. 

The most common type
of brane is called a {\bf D-brane} and it has a tension $T\propto 1/g_s$. So
one could never
see these objects in perturbation theory in $g_s$. Even though they are very
heavy,  the gravitational field they produce is governed by $G_{10} T\sim g_s$
so as $g_s\rightarrow 0$, there should be a flat space description of these
objects and it was found by Polchinski \cite{Polchinski:1995mt}.
At weak coupling, a D-brane is a 
surface in spacetime on which open strings can end. The D 
stands for ``Dirichlet" and refers to the boundary conditions on the  ends
of the open 
strings.  In fact, D-branes were discovered by applying T-duality to a theory of open strings. Open string worldsheets have boundaries, and by looking at how the Euler number changes when open strings interact, one finds that the open string
coupling constant, $g_o$, satisfies $g_o^2 = g_s$. The tension of a D-brane can be understood by viewing it as a soliton of the open string theory: $T \propto 1/g_o^2 \propto 1/g_s$.
The endpoints of the open strings move freely along the brane but cannot leave the brane unless  they join and form a closed string. The
massless states of an open string include a spin-1 excitation, so
every D-brane comes with a $U(1)$ gauge field. When $N$ D-branes
coincide in spacetime, the open strings stretching from one to another also
become massless. This enhances the resulting gauge group from $U(1)^N$
to $U(N)$. These
D-branes are also sources for the $p$-form ``Ramond-Ramond" fields $F_p$.

D-branes have found applications in string phenomenology, i.e., the attempt to connect string theory with standard four-dimensional particle physics. String theory clearly unifies the graviton with many matter degrees of freedom.  
One difficulty 
 is that one often has too many light degrees of freedom. In particular, the moduli associated with the size of certain surfaces in the internal directions correspond to 
4D  
massless scalars which we do not see.  One way to make  the models more realistic  is, roughly speaking,  to wrap a brane around the surface in the internal space (which will try to contract it) and also add flux associated with $F_p$
 (which will try to expand it). Under certain circumstances these forces balance at one size of the surface \cite{Kachru:2003aw}. In terms of the lower-dimensional theory, the scalar field now has a large mass
and  has no 
low-energy 
dynamical effects.

Soon after D-branes were discovered, evidence was  found for a strong-weak coupling duality called {\bf S-duality}.  The evidence included comparing the masses of certain BPS states whose masses are fixed by the charges. Using these dualities, it was argued that all the perturbative string theories are related. In addition, it was proposed that there is an 
eleven-dimensional 
theory called {\bf M-theory}, whose dimensional reduction on a circle of radius $R = g_s \ell_s$ yields Type IIA string theory \cite{Witten:1995ex}. In particular, D0-branes have just the right properties to represent the momentum modes around this circle. M-theory can thus be viewed as a strong coupling limit of Type IIA string theory. 
Its  
low-energy 
limit is 
eleven-dimensional 
 supergravity. As discussed in Section \ref{s:sg}, this theory has only a metric, four-form field strength and spin-$3/2$ field. The bosonic action is given in 
(\ref{sugra11D}). The 
four-form 
can carry electric charges of 2-branes and magnetic charges of 5-branes, so we deduce that
 M-theory is not a theory of strings, but a theory including 
  2-branes and 5-branes, which are called 
M2- and M5-branes. 
 
 \begin{figure}[t]
 \begin{center}
\includegraphics[width=8cm]{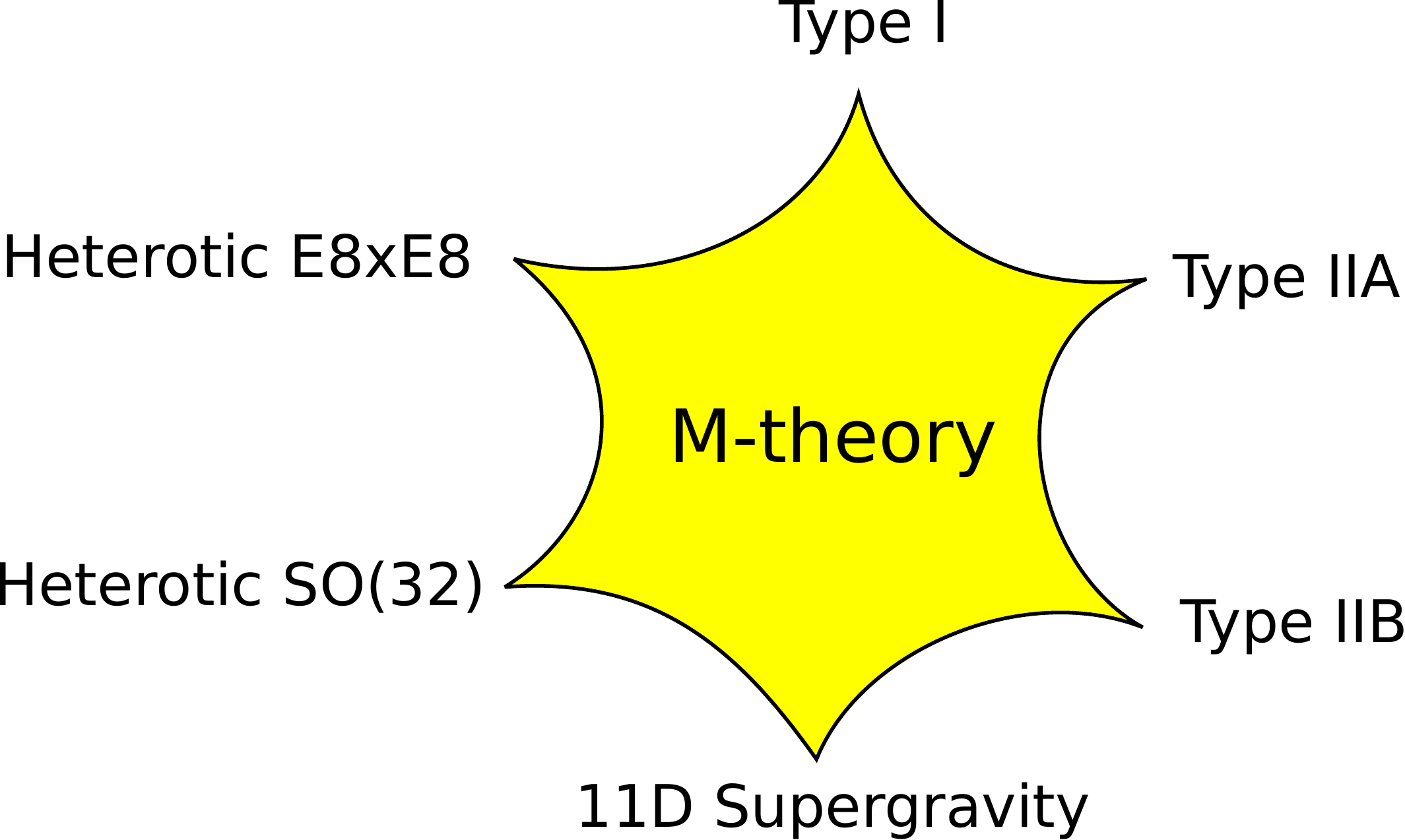}
\end{center}
\caption{A schematic view of different weak coupling limits of M-theory.} 
\label{fig:Mstar}
\end{figure}
 
Putting all the dualities together, M-theory can be viewed as having six different weak coupling limits corresponding to the five different 10D perturbative string theories, and 11D supergravity (see Fig. \ref{fig:Mstar}). We do not yet understand the fundamental principles underlying M-theory. The best description we have of this theory is in terms of a matrix model, i.e., a quantum theory describing a collection of matrices depending only on time \cite{Banks:1996vh}. In this description, space emerges from the properties of the matrices.

One of the main successes of string theory is its ability to reproduce the Hawking-Bekenstein entropy of certain black holes as well as Hawking radiation, from a microscopic quantum theory.  For many years it was thought that strings could not explain the entropy of Schwarzschild black holes since the string entropy is proportional to the mass, whereas the entropy of Schwarzschild (in $D=4$) goes like the mass squared. However, this is misleading. The string entropy is $S_\text{string} \sim M \ell_s$  whereas the black hole entropy is $S_\text{BH} \sim G_4 M^2$ and the 
four-dimensional 
Newton's constant is $G_4 \sim g_s^2 \ell_s^2$. So before one can ask if strings can explain black hole entropy, one must first specify the string coupling. The natural point to compare them is when the Schwarzschild radius is of order the string scale, since this is when the Schwarzschild solution starts to receive stringy corrections. At this point, the black hole entropy is \cite{Susskind:1993ws}
\be
S_\text{BH} \sim  M r_0 \sim M \ell_s \sim S_\text{string}\,,
\ee
so the entropies agree. Since $r_0=2 G_4 M \sim \ell_s $ implies $g_s \sim (M \ell_s)^{-1/2}$,  the coupling remains small for a large mass black hole. 
This argument continues to hold in $D>4$,  and can be generalized to include charges and angular momentum. It leads to a simple {\bf  correspondence principle} between black holes and strings \cite{Horowitz:1996nw}: when the curvature at the horizon becomes of order the string scale, the typical black hole state becomes an excited string state with the same charges and angular momentum. This provides a simple picture for the endpoint of Hawking evaporation: when black holes evaporate down to the string scale, they turn into highly excited strings which then continue to decay down to an unexcited string which is just another elementary particle.

For certain black holes, the entropy can be reproduced exactly in string theory. This was first shown for nonrotating, 
 extremal 5D
 black holes \cite{Strominger:1996sh} and soon generalized to 
near-extremal
 \cite{Callan:1996dv,Horowitz:1996fn} and rotating 
 \cite{Breckenridge:1996is} 5D black 
holes.
  Similar results hold for 
extremal \cite{Maldacena:1996gb,Johnson:1996ga} and near-extremal \cite{Horowitz:1996ac} 4D black holes.
Due to the importance of this result, we now describe one example in some detail.
(The following discussion is based on \cite{Maldacena:1996ix}.) The Type IIB supergravity action includes the following terms:
\be\label{IIBaction}
S = {1\over 16\pi G_{10}} \int d^{10} x \sqrt{-g}\left [ e^{-2\phi}[ R + 4 (\nabla \phi)^2] - \frac{1}{12} F_3^2 \right ] ,
\ee
where $F_3$ is a Ramond-Ramond three-form field-strength and $\phi$ is the dilaton. This action is written  in terms of the so-called string metric which is the metric that appears in (\ref{polyakov}), i.e., the metric that the strings directly couple to. 
We will compactify four directions $x_i$ ($i = 6,7,8,9$) on circles of length\footnote{To simplify the presentation, for the remainder of this Section we will set $\ell_s = 1$.} $2\pi$. In the resulting six-dimensional spacetime, $F_3$ has both electric and magnetic type charges, each carried by one-dimensional extended objects:
\be
Q_1 = {1\over 4\pi^2 g_s} \int e^{2\phi} *F_3\,,~~~~~~~
Q_5 = {1\over 4\pi^2 g_s} \int F_3\,.
\ee
The integrals are over an $S^3 $ surrounding the object, and $*$ denotes the 
six-dimensional 
Hodge  dual. 
The factor of $e^{2\phi}$ is needed in the first integral since $ *F_3$ by itself is not  a closed form after dimensional reduction to six dimensions. 
The labels come from the fact that $Q_1$ is the charge carried by D1-branes and $Q_5$ is the charge carried by D5-branes (which wrap the $T^4$). The charges are normalized so that $Q_i$ are integers which simply count the number of branes of each type. 

One can show that the following is a black brane solution to the equations of motion coming from the action (\ref{IIBaction}): 
\be\label{bhmetric} 
ds^2 = f(r)^{-1} \left [-dt^2 + dx^2_5 + {r_0^2\over r^2} \big(\cosh \sigma dt + \sinh \sigma dx_5\big)^2 + \left( 1 +{g_sQ_1\over r^2}\right) dx_idx^i\right ] 
\ee
$$+ f(r) \left [ \left( 1-{r_0^2\over r^2}\right)^{-1} dr^2 + r^2 d\Omega_3^2\right],$$
where 
\be
f(r) =  \left( 1 +{g_sQ_1\over r^2}\right)^{1/2} \left( 1 +{g_sQ_5\over r^2}\right)^{1/2}.
\ee
The matter fields take the form:
\be 
e^{-2\phi}  = \left( 1 +{g_sQ_5\over r^2}\right) \left( 1 +{g_sQ_1\over r^2}\right)^{-1},
\ee
\be
F_3 = 2g_sQ_5 \epsilon_3 + 2g_sQ_1 e^{-2\phi} *\epsilon_3\,,
\ee
where $\epsilon_3$ is the volume form on a unit $S^3$ and $*\epsilon_3$ is its six-dimensional dual.

If $x_5$ is periodically identified with period $2\pi R$, then momentum in this direction should be quantized: $P = n/R$. From (\ref{bhmetric}) one finds
\be\label{bhmomentum}
n = {r_0^2 R^2 \sinh 2\sigma \over 2g_s^2}\,.
\ee
If one Kaluza-Klein reduces to five dimensions, one has a spherical black hole with three charges $Q_1, Q_5, n$.

The total energy of this solution is\footnote{To compute the ADM energy, one should first conformally rescale the metric by a function  of $\phi$ so the action (\ref{IIBaction}) contains the standard Einstein-Hilbert term. The result is called the Einstein metric.}
\be\label{bhenergy}
E = {RQ_1\over g_s}  +  {RQ_5\over g_s} + {n\over R} + {Rr_0^2 e^{-2\sigma}\over 2g_s^2}\,.
\ee
The 
extremal 
limit corresponds to $r_0 \rightarrow 0, \sigma \rightarrow \infty$ with $n$ fixed. In this limit the last term vanishes. 
The remaining three terms are exactly what one would expect from $Q_1$ D1-branes wrapping the $S^1$ with radius $R$, $Q_5$ D5-branes wrapping the $T^4$  with unit radii and the $S^1$ with radius $R$, and momentum $P = n/R$. This is a consequence of the fact that the extremal solution is BPS,  so the energy is  uniquely determined by the charges.

The Bekenstein-Hawking entropy is
\be\label{bhentropy}
S_\text{BH} = {A\over 4G_{10}} = 2\pi \sqrt{Q_1 Q_5}\,{r_0 R\cosh\sigma\over g_s}.
\ee
String theory can reproduce this entropy in both the 
extremal  and near-extremal limits 
by counting states  at weak coupling, i.e., in flat spacetime. 
In the extremal case, one expects the number of states to be independent of coupling since the solution is supersymmetric and the states are BPS. (It was a surprise to find that this agreement continued to hold for slightly near extremal solutions also.) 
So  we want to count the number of bound states of D1-branes and D5-branes with given momentum in a flat spacetime compactified on a $T^4$ with unit radii 
 and an $S^1$ with radius $R$. Since the D5-branes wrap the small $T^4$,  
low-energy 
excitations only move along the $S^1$, and can be described by  an effective $1+1$ dimensional theory.

 In the 
extremal 
limit, the entropy reduces to 
\be\label{extbhentropy}
S_\text{BH} = 2\pi \sqrt{Q_1 Q_5 n}\,.
\ee
 Note that the dependence on all continuous parameters such as $R$ and 
$g_s$  
has 
 dropped out. 
An examination of the  low-energy excitations of $Q_1$ D1-branes and $Q_5$ D5-branes shows that there are $4Q_1Q_5$ massless bosons. These can be viewed as open strings  connecting one of the D1-branes with one D5-brane. The factor of 4 arises since there are two possible orientations for the strings and for each orientation, the ground state is two-fold degenerate. When these D1-D5 strings are excited, the D1-D1 and D5-D5 open strings become massive and do not contribute to the entropy. This is reflection of the fact that branes are bound together.

Free fields in $1+1$ dimensions have
independent right and left moving modes.
BPS states with nonzero momentum correspond
to exciting only the right moving modes of these massless fields. So 
one can simply count
the number of states of $4Q_1Q_5$ bosonic fields (plus an equal number of
fermionic fields required by supersymmetry) on a circle of radius $R$
with total right moving
momentum $P=n/R$. In the limit
of large $n$, the answer is $e^S$ where
\be\label{stent}
S = 2\pi \sqrt{Q_1 Q_5 n}\,,
\ee
in perfect agreement with the Bekenstein-Hawking entropy (\ref{extbhentropy}).

Now suppose we add a small amount of energy to the system keeping the charges fixed, so it is no longer 
extremal. 
To maximize the entropy, the energy will excite the lightest modes.  If $R$ is large, the lightest modes are just the momentum modes.  So the energy excites some additional left and right moving modes. At weak coupling there is very little interaction between the modes, but it is
 not zero. However, if $r_0^2 \ll g_s Q_1, g_sQ_5$ and $g_s^2 n/R^2 \ll g_sQ_1, g_sQ_5$, we are in a ``dilute gas" regime where interactions are negligible. This corresponds to a 
near-extremal  
 black hole. In this case, the entropy computed at weak coupling is just the sum of the entropies of the left and right moving modes:
\be\label{stringentropy} 
S = 2\pi \sqrt{Q_1 Q_5}\big(\sqrt {n_R} + \sqrt {n_L}\big)\,,
\ee
where $n_R$ and $n_L$ are defined by setting $P = (n_R -n_L)/R$ and requiring that $(n_R + n_L)/R$ be the contribution to the energy from the momentum modes.
From (\ref{bhmomentum}) and (\ref{bhenergy}) we get
\be 
n_R = {r_0^2 R^2 e^{2\sigma} \over 4g_s^2}\,, 
\qquad n_L = {r_0^2 R^2 e^{-2\sigma} \over 4g_s^2}\,.
\ee
Substituting into (\ref{stringentropy}) we see that the counting at weak coupling precisely reproduces the entropy of the 
near-extremal 
black hole (\ref{bhentropy}). 

We now include some interactions between the left and right moving modes.  Occasionally, a left-moving mode
can combine with a right-moving mode to form a closed string which
can leave the brane. This corresponds to the decay of an excited configuration
of D-branes and is the 
weak-coupling 
analog of Hawking radiation.
Given the remarkable agreement between the entropy of the black hole
and the counting of states of the D-branes, the next step is 
to ask how the radiation emitted by the D-brane compares to Hawking
radiation.  In both cases, the radiation is
approximately thermal with the same temperature. 
This is expected since the entropy as a function of energy
agrees in the two cases. What
is surprising is that the overall rate of radiation agrees \cite{Das:1996wn}.
What is even more remarkable is that the deviations from the
blackbody spectrum also agree \cite{Maldacena:1996ix}. 
On the black hole side, these deviations
arise since the radiation has to propagate through the curved spacetime
outside the black hole. This contains potential barriers which give
rise to frequency-dependent greybody factors. On the D-brane side
there are deviations since the modes come from separate 
left and right moving
sectors on the  D-branes, with different effective temperatures. The calculations of these deviations
could not look more
different.
On the black hole side, one solves a wave equation in a black hole 
background. The solutions involve hypergeometric functions. On the
D-brane side, one does a calculation in D-brane perturbation theory.
Remarkably, the answers are identical.
 It is worth emphasizing that it is not just the dependence on a few parameters which agree. One is comparing the decay rate as a function of frequency and the entire
functional form agrees on both sides.  It is as if the black hole 
knows that its states are described by an effective $1+1$ dimensional
field theory with left and right moving modes. 

As mentioned earlier,  this precise agreement between  D-branes and black holes holds for black holes in 4D as well as 5D,\footnote{For extensions to black rings, see for example \cite{Cyrier:2004hj,Bena:2004tk,Bena:2005ay}.} but always in the 
near-extremal limit. Far from extremality, one still expects string theory 
to provide a microscopic description of black holes, but one can no longer  rely on the weak-coupling D-brane picture. The extrapolation to strong coupling now produces significant changes in the properties of the quantum states. One approach to this problem is discussed in the next section.


\section{Holography and gauge/gravity duality} 
\label{s:adscft}

 We begin in Section \ref{s:holo} by briefly reviewing  some general arguments suggesting that quantum gravity might be holographic.  In Section \ref{s:ggc}, we formulate gauge/gravity duality, a precise form of holography that emerges from string theory. We also discuss the evidence for this duality and its consequences. Section \ref{s:applications} offers an overview of some applications of gauge/gravity duality. 

\subsection{Holography}
\label{s:holo} 

The suggestion that quantum gravity might be holographic was originally motivated by black hole entropy \cite{'tHooft:1993gx,Susskind:1994vu}. 
The idea was simply that the fact that a black hole has an entropy that scales with its area is in striking contrast to most systems which have an entropy that is proportional to their volume. This suggests that everything that happens inside the black hole could be somehow encoded in degrees of freedom at the horizon. 

In trying to find a more precise formulation of holography, it is natural to consider spacetimes  which asymptotically approach anti-de Sitter (AdS) 
spacetime\footnote{See \cite{Avis:1977yn} for an early study of quantum field theory in AdS that set the stage for many later developments.}.  There are several reasons for this. First, a static slice in AdS is a constant negative curvature hyperboloid,  so the area of a sphere at proper radius $R$ grows exponentially with $R$. Thus, compared to flat space, there is ``more room at infinity" for the holographic description to live.
Second,  the conformal boundary at infinity is  timelike, so a holographic description at infinity could live on an ordinary spacetime. Finally, black hole thermodynamics is  better behaved in AdS since the negative curvature acts like a confining box. In particular, large black holes in AdS have positive specific heat and can be in thermal equilibrium with their Hawking radiation \cite{Hawking:1982dh}. 

A general argument for holography in quantum gravity was given recently by Marolf \cite{Marolf:2008mf}. His argument can be made for either asymptotically flat or asymptotically AdS boundary conditions, but we focus on the AdS case here.  Consider first perturbative quantum gravity about AdS. At linear order, any field at a point in the interior can be evolved back and expressed as an integral over operators along a surface at large radius. Thus, the set of boundary operators defined at large radius form a  complete set of operators at linear order. The same argument can be made at  each order in perturbation theory.  This is not yet a statement about holography, but more analogous to expressing the value of a field in terms of initial data at an earlier time. However the Hamiltonian itself is a boundary operator. This is a unique feature of diffeomorphism invariant theories coming from the fact that the Hamiltonian can be expressed as a surface integral at infinity. Since the Hamiltonian generates time evolution, we can express any boundary operator ${\cal O}$ at one time in terms of a boundary operator at a different time by solving
\be\label{evolve}
{d\over dt} {\cal O}(t) = i[H,{\cal O}(t)].
\ee
It follows that, at least in  perturbation theory, any observable in the bulk can be expressed in terms of a boundary observable at one fixed time. This is a statement of holography. 

At the full nonperturbative level, the argument for the completeness of boundary observables is  less clear, not least because of the difficulty in defining observables in full quantum gravity.  However, the Hamiltonian is still presumably a boundary observable, so given a set of boundary observables at one time, one can always evolve them using (\ref{evolve}) and relate them to observables at a later time. Thus any information available at the boundary at one time  is also available at any later time. In particular, this is true for an evaporating black hole. This result can be called ``boundary unitarity". A similar argument can be made even for asymptotically flat spacetimes using observables defined on spacelike and null  infinity.

\subsection{Gauge/gravity duality}
\label{s:ggc} 
A much more precise formulation of holography was found  by Maldacena by studying 
extremal and near-extremal 
black holes in string theory \cite{Maldacena:1997re}. The holographic theory turns out to be an ordinary (supersymmetric) gauge theory. The equivalence between the gravitational and gauge theories is often called gauge/gravity duality.
 The motivation for this duality is the following. Consider a stack of $N$ D3-branes in Type IIB string theory.
  At weak coupling, 
  $g_s N\ll1$, 
  the excitations are described by open strings on the brane and closed strings off the brane. The massless states consist of\
     ${\cal N} =4 $ supersymmetric $U(N)$ gauge theory 
  (described in Section \ref{s:susy}) on the brane and Type IIB supergravity off the brane. In a  
 low-energy 
 limit, one has only long wavelength supergravity modes, but 
keeps  
 all modes of the gauge theory since this theory is conformally invariant.  These two sectors decouple at low energy since the dimensionless coupling to gravity is $G_{10}E^8$.

When 
$g_s N\gg1$, 
the backreaction of the branes becomes important and 
produces 
the following metric:
\be\label{D3brane}
ds^2 = H^{-1/2}\Big[-dt^2 + dx_1^2 +dx_2^2 +dx_3^2 \Big] 
+ H^{1/2}\Big[dr^2 + r^2 d\Omega_5^2\Big]\,,
\ee
where 
\be\label{AdSradius}
H(r) = 1 + {L^4\over r^4}\,,\qquad L^4 = 4\pi g_s N \ell_s^4\,.
\ee
The branes no longer appear explicitly, but are represented by a nonzero flux of a 5-form field strength $F_5$. This metric describes an extremal black brane. There is a smooth degenerate horizon at $r=0$. The excitations are closed strings in the above spacetime. 
Low-energy 
excitations consist of either arbitrary closed strings which are very close to the horizon (and hence have a large redshift) or massless closed strings far from the horizon with very low frequency.  The latter sector is again long-wavelength modes of IIB supergravity.
In a  low-energy 
limit, these two sectors decouple, since the absorption cross section of the black hole goes to zero as $\omega \rightarrow 0$. The 
near-horizon 
limit of (\ref{D3brane}) is obtained by dropping the  ``1"
in $H$, i.e., setting $H=L^4/r^4$. The resulting spacetime is the product of $S^5$ with radius $L$ and five-dimensional anti-de Sitter spacetime ($AdS_5$) with radius of curvature $L$:
 \be\label{Poincare} 
  ds^2 = {r^2\over L^2}\Big[-dt^2 + dx_1^2 +dx_2^2 +dx_3^2 \Big] +{ L^2 dr^2\over r^2}.
 \ee

We see that at both weak and strong coupling, the 
 low-energy 
 description of a stack of D3-branes has two decoupled sectors. In each case, one sector is 
 low-energy 
 supergravity modes. It is thus natural to identify the other two sectors. This means that we have a system which at weak coupling looks like supersymmetric Yang-Mills (SYM) and at strong coupling looks like strings in $AdS_5\times S^5$. But SYM also exists at strong coupling, and string theory exists at weak coupling, so these two descriptions must be equivalent. Thus we are led to the following remarkable conjecture: 
 \vskip .5 cm 

\noindent  {\it Four-dimensional ${\cal N} =4 $ supersymmetric $U(N)$ gauge theory is equivalent to IIB string theory with  $AdS_5\times S^5$ boundary conditions.}
  \vskip .5 cm
 \noindent This is the simplest (and most well studied) example of  {\bf gauge/gravity duality},
 the equivalence between a theory of (quantum) gravity and a nongravitational gauge theory. We will discuss some generalizations later. 
  String theory with  boundary conditions $AdS_5\times S^5$ has two dimensionless parameters $ (g_s, L/\ell_s)$, and the gauge theory has two dimensionless parameters, $(g_\text{YM}, N)$. They are related by $  4\pi g_s = g_\text{YM}^2$, and $ (L/\ell_s)^4 = g_\text{YM}^2 N$. 
 
The coordinates (\ref{Poincare}) do not cover all of $AdS_5$ but only the so-called ``Poincare patch". As it stands, the duality is between string theory on spacetimes asymptotic to (\ref{Poincare}) and SYM on Minkowski spacetime, where the Minkowski space can be viewed as the conformal boundary of (\ref{Poincare}). However, it is easy to extend the duality to all of $AdS_5$ where the conformal boundary is now the Einstein static universe, $S^3\times R$, and the SYM lives on this spatially compact space. The interior of $AdS_5$ is often called the ``bulk".
 
 Recall that Newton's constant in ten dimensions is given in string theory by $ G_{10} \sim g_s^2 \ell_s^8$. It follows from (\ref{AdSradius}) that $L^4 \sim N \ell_p^4$,
  where $\ell_p$ is the Planck length. Thus if $N$ is $O(1)$, the curvature in the bulk is of order the Planck scale everywhere. This is an interesting regime from the standpoint of quantum gravity. The gauge theory is relatively simple, but it is hard to give a physical interpretation of  any SYM observables in terms of the dual gravitational theory.
 
 In the opposite limit when $N$ is large, we have $L\gg \ell_p$, so
 typical curvatures are much smaller than the Planck scale and quantum gravity effects are suppressed. In the gauge theory it is convenient to consider the
 't Hooft limit: $N\rightarrow \infty$, $g_\text{YM} \rightarrow 0$ with $\lambda \equiv g_\text{YM}^2 N$ held fixed.  
The  't Hooft coupling 
 $\lambda$ then acts as the natural  coupling constant in this limit and only Feynman diagrams that can be drawn on a plane contribute. 
 On the string theory side,  
 when $\lambda \gg 1$ 
we have  
 $L \gg \ell_s$, so stringy excitations in the bulk are suppressed and one can work with just supergravity modes.  Note that in this limit where the gravity side is simple, the gauge theory is strongly coupled and poorly understood. Conversely, when  $\lambda \ll 1$, the gauge theory is weakly coupled and well understood, but the gravity side is very stringy. It is this strong/weak coupling aspect of the duality which allows two very different sounding theories to be equivalent.

Since $G_{10} \sim g_s^2 \ell_s^8\sim L^8/N^2$,  and $L$ is held fixed in the  't Hooft limit, it follows that
 the gravitational backreaction of all states becomes negligible unless the energy of the state grows at least as fast as $N^2$. So all states with $E< O(N^2)$ can be described by fields propagating on AdS. Since there are $O(N^2)$ degrees of freedom in the gauge theory, states with energy $E \sim  O(N^2)$ are natural to consider since they correspond to  exciting each degree of freedom by an amount independent of $N$.  
 
 Note that a four-dimensional gauge theory is describing a ten-dimensional theory of gravity. So it is an extreme type of hologram which encodes six extra spatial dimensions. For perturbations of $AdS_5\times S^5$ one can show (by comparing representations of $SO(6)$) that information about position on $S^5$ is encoded in products of the six scalar operators $\phi^i$ in  $\cn = 4$ SYM. The radial dimension in $AdS_5$ is related to an energy scale in the dual gauge theory. This is suggested by the fact that in Poincare coordinates (\ref{Poincare}) the metric is invariant under $r \rightarrow ar, \ (t,x_i) \rightarrow (t,x_i)/a$. So small radius corresponds to large distance or low energy in the gauge theory.

 The claim that a four-dimensional gauge theory could describe all of 
 ten-dimensional string theory sounds so crazy that one might think that one could disprove it quite easily. Let us try.  Having extra spatial dimensions usually leads to more quantum states (since, e.g., one can have momentum modes in more directions). So we will compare the entropy in the two theories at a high temperature $T$. In the gauge theory, the entropy is given by the usual formula for a thermal gas with $N^2$ degrees of freedom
 \be
 S_\text{SYM} \sim N^2 T^3 V_3\,,
 \ee
 where $V_3$ is the volume of the three-dimensional space.
 On the gravity side, it would seem that one could exceed this including only the massless modes of the string. A thermal gas  of massless particles
  in ten dimensions has energy density proportional to $T^{10}$, so from the first law, its entropy density is proportional to $T^9$. It would appear that at sufficiently high temperature,  the entropy in the bulk vastly exceeds that on the boundary. 
 
 Of course this estimate is incorrect since it ignores the fact that the gas will collapse to form a black hole.   In addition to the usual spherical black holes, AdS has black holes in which the horizon geometry is flat. Since we have estimated the entropy of SYM on flat space, these ``planar black holes" are the appropriate comparison. They take the form
 \be
 ds^2 = {r^2\over L^2}\left [ \left(1-{r_0^4\over r^4}\right ) dt^2 + dx_i dx^i \right] + \left(1-{r_0^4\over r^4}\right )^{-1} {L^2 dr^2\over r^2}  + L^2 d\Omega_5^2\,,
 \ee
 and have a temperature $T = 3r_0/4\pi L^2$. Their entropy is
 \be
 S_\text{BH} = {A\over 4G_{10}} \sim  {L^8 T^3V_3\over G_{10}} \sim N^2 T^3 V_3\,,
 \ee
which agrees with the estimate from the gauge theory. In fact, this shows that the gauge theory has enough microstates to reproduce the black hole entropy. Gauge/gravity duality relates the black hole to a thermal state in the dual field theory. 
 
 It is difficult to compare the numerical coefficient in the above comparison of the entropy. 
 This would require an exact calculation of the number of states in the gauge theory at strong coupling. If one instead does the calculation at weak coupling, one gets an answer which differs from the black hole entropy by  a factor of $3/4$: $S_\text{BH} = 3/4 S_\text{SYM}$  \cite{Gubser:1996de}. 
 It should be emphasized that this difference is not a problem for gauge/gravity duality:  $e^{S_\text{SYM}}$ is the number of SYM states at weak coupling,
 and the number of states of a given energy is expected to decrease with coupling. This is because   the form of the Yang-Mills Hamiltonian implies that increasing the coupling will increase the potential energy of each SYM state, and hence lower the total number of states at fixed energy. The duality predicts that at strong coupling, the entropy will agree with the black hole.
 It is surprising that what appears to be a complicated QFT calculation only changes the answer by a factor of $3/4$.  This  prediction  is waiting to be verified by a direct calculation.
 
 This situation should be contrasted with that at the end of the previous subsection.  In that case, we again compared a black hole entropy with a calculation at weak coupling and found precise agreement. The key difference is that that  black hole was supersymmetric  (or nearly supersymmetric),
 so the mass  of each state is fixed by the charge. The entropy is then independent of the coupling constant, and a precise comparison is possible.  
  
 Only gauge invariant observables can be compared on both sides of the duality. A large class of such observables can be compared as follows  
  \cite{Gubser:1998bc,Witten:1998qj}.  For every field $\Phi$ in the bulk there is a corresponding operator 
$\mathcal{O}$
  in the dual gauge theory. 
Supergravity fields correspond to simple gauge invariant operators constructed as a single trace of a local product of the super Yang-Mills fields. (An early check of the duality was that there was indeed a one-to-one correspondence between supergravity fields and suitable SYM operators.)
 Asymptotic values of the bulk fields act as sources for the dual operator in the following sense:  the string theory partition function with boundary condition\footnote{More precisely,  $\Phi$ typically vanishes asymptotically, and  $\Phi\rightarrow \Phi_0/ r^{\Delta}$ where $\Delta$ is related to the mass of the bulk field.}  $\Phi\rightarrow \Phi_0$ should equal the field theory partition function with action
\be
 S[\Phi_0] = S_\text{SYM} + \int \Phi_0 
 \,\mathcal{O}\,.
 \ee
In other words
\be
Z_\text{string\ theory}(\Phi \rightarrow \Phi_0)  = \int DA \ D\phi \ e^{iS[\Phi_0]} \equiv Z_\text{SYM}[\Phi_0].
\ee
In the 't Hooft limit with large $\lambda$, the left hand side can be approximated by just the supergravity fields, and further approximated by the exponential of the action of the classical solution with the boundary condition $\Phi \rightarrow \Phi_0$. By taking a derivative with respect to the source $\Phi_0$, one can show that the expectation value of $\mathcal{O}$ is related to a subleading term in  the asymptotic behavior of the classical solution.

Gauge/gravity duality is a conjecture. It has not yet been proven. Since SYM is a complete nonperturbative quantum theory and we do not have an independent complete nonperturbative description of string theory, one might wonder what a proof would consist of. In fact, one might be tempted to define nonperturbative string theory in terms of the dual SYM and claim the duality is true by definition. But this is much too quick. A proof is necessary and would consist of showing that everything we know about string theory, including the space of classical solutions (with AdS boundary conditions) and perturbation theory about them, is reproduced in the SYM theory.  

Although there is no proof, there is by now overwhelming evidence that gauge/gravity duality is correct. The early evidence included the fact that the symmetries on the two sides agree: ${\cal N} =4 $ supersymmetric gauge theory in $D=4$ is conformally invariant, so it is invariant under $SO(4,2)$.  
As described in Section \ref{s:susy}, the theory includes 6 scalars that transform 
in the fundamental representation of the R-symmetry $SO(6)$. 
$AdS_5\times S^5$ has an isometry group which is precisely $SO(4,2) \times SO(6)$. The supergroups also agree. More nontrivial checks came  later and include a vast number of calculations in which a physical quantity is computed on the two sides of the duality. Although the two calculations often look very different,  
the final answers agree.  We mention a few examples below:

\begin{itemize}

\item Wilson loops in the gauge theory are natural (nonlocal) gauge invariant operators. Given a curve ${\cal C}$ one considers\footnote{The objects that can be computed holographically are actually slight generalizations of the usual Wilson loop which include the six scalars in the gauge theory.}
 $W = {\rm Tr}\ P \exp[\int_{\cal C} A]$,
 where $P$ denotes path ordering and Tr denotes trace in the adjoint representation of the gauge group. The expectation value 
$\langle W\rangle $  
 of these Wilson loops can be calculated on the gravity side by considering string worldsheets in spacetime that end on the loop ${\cal C}$ at infinity. The area of the string worldsheet is then related to $\langle W\rangle $. In certain cases, one can compute $\langle W\rangle $ exactly in the gauge theory and find complete agreement with the gravity calculation \cite{Drukker:2000rr}.

\item  Renormalization group (RG) flow is the  quantum field theory process of  integrating out the high energy modes to obtain a new effective theory at lower energy. $\cn = 4$ SYM is conformally invariant, so there is no scale, and the RG flow is trivial. However, 
one can add relevant operators, for example mass-terms, 
to 
this theory and find that RG flow leads to a different conformal field theory at low energy with fewer degrees of freedom. On the gravity side, this corresponds to modifying the boundary conditions at infinity and finding a new static solution to Einstein's equation. One finds that at small radius, this new solution approaches \eqref{Poincare} but now with $L$ replaced by a new AdS radius $\tilde L$. There is detailed agreement between the CFT one gets at low energy and the new AdS \cite{Freedman:1999gp}. In particular the new AdS length scale $\tilde L$, is related to the number of degrees of freedom in the  new low-energy dual theory in just the same way as in the asymptotic region.

\item All the states of supergravity in the bulk have precise descriptions in the dual gauge theory. What about the excited string states? In general, it is hard to identify the dual of these states, but this has been done in a certain limit \cite{Berenstein:2002jq}. If one starts with a null geodesic wrapping the $S^5$, one can take a Penrose limit 
and 
obtain a 10D plane wave. In the gauge theory, this corresponds to considering states of the form  
$\text{Tr}\,[Z^J]|0\rangle$, 
where $J$ is a large angular momentum and $Z =  \phi_5 +i\phi_6$  ($\phi_i$ are the six scalars of $\cn = 4$ SYM). The complete spectrum of the string in the bulk plane wave  
background 
 is exactly reproduced in the gauge theory by replacing some of the 
$Z$'s
with $\phi_i$, $i = 1,2,3,4$, or 
$D_\mu Z = \partial_\mu Z + [A_\mu,Z]$. 
It is as if the 
$Z$'s
in the gauge theory create a string with transverse oscillations generated by $\phi_i$ and $D_\mu Z$.

\item If one restricts to a class of states preserving 1/2 of the supersymmetry, one can make the correspondence between the gravity and gauge theory much more explicit \cite{Lin:2004nb}.  Let us work with global AdS so the field theory lives on $S^3 \times R$. We will  actually restrict to fields that are independent of $S^3$ and hence
reduce to $N\times N$ matrices. In fact, we only consider states created by a
single complex matrix, so they can be described by a one-matrix model. (In terms of the six scalars in the gauge theory, this is again $Z =\phi_5 +i\phi_6$.) This
theory can be quantized exactly in terms of free fermions, and the states can be labeled by arbitrary closed curves  on a plane. (The plane represents phase space and the closed curves denote the boundary of regions that are occupied.) The states are all invariant under $SO(4)\times SO(4)$ where the first factor corresponds to rotations on the $S^3$ and the second factor corresponds to rotations of the remaining four scalars $\phi_1, \cdots \phi_4$ in the gauge theory.

On the gravity side, one
considers solutions to ten-dimensional supergravity involving just the metric
and self-dual five-form $F_5$. The field equations are  simply $dF_5=0$ and
\be\label{fieldeq}
R_{\mu\nu} = F_{\mu\alpha\beta\gamma\delta}{F_\nu}^{\alpha\beta\gamma\delta}.
\ee
There exists a large class
of stationary solutions to (\ref{fieldeq}), which have an
$SO(4)\times SO(4)$ symmetry and can be obtained by solving a linear equation \cite{Lin:2004nb}.
These solutions are nonsingular, have no event horizons, but can have 
complicated topology. They are also labeled by arbitrary closed curves
on a plane. This provides a precise way to map states in the field
theory into bulk geometries.  Only for some ``semi-classical" states
is the curvature below the Planck scale everywhere, but the 
matrix/free-fermion 
description readily describes all the states in this class, of all topologies, within a single Hilbert space.

\item 
The above examples check  gauge/gravity duality in the large $N$ limit where the bulk is described by supergravity. There is also evidence that the duality remains true at finite values of $N$.   A striking example is the `string exclusion principle' \cite{Maldacena:1998bw}. Graviton states on $S^5$ arise in the gauge theory from acting on the vacuum with an operator involving traces of products of the $\phi^i$.  However, these fields are $N \times N$ matrices, so the traces cease to be independent for products of more than $N$ fields. This leads to 
an 
upper bound on the angular momentum $J$ on $S^5$:
\begin{equation}
J / N \leq 1.
\end{equation}
  From the point of view of supergravity this is mysterious, because the graviton states exist for  arbitrary $J$.  However, there is an elegant resolution in string theory \cite{McGreevy:2000cw}   using something called the ``Myers effect" \cite{Myers:1999ps}.  This is the fact that a stack of D-branes in an external field can become polarized and take the shape of a sphere. In a certain limit, the graviton can be viewed as a stack of D0-branes, and one can show that when
it moves sufficiently rapidly on the sphere, it will blow up into a spherical D3-brane, and $J = N$ turns out to be the largest D3-brane that will fit in the spacetime.  Thus the same bound is found on both sides of the duality, and this is a nonperturbative statement in $N$: it would be trivial in a power series expansion in $1/N$.

\end{itemize}

So far we have 
discussed gauge/gravity duality with asymptotic $AdS_5\times S^5$ boundary conditions. This is the most well studied example of gauge/gravity duality, but many other examples exist. Applying a similar argument to a stack of M2-branes shows that 
M-theory 
on asymptotically $AdS_4\times S^7$ is equivalent to a $2+1$ dimensional field theory of gauge fields and matter with a Chern-Simons action for the gauge fields, called ABJM theory
\cite{Aharony:2008ug}. Applying the argument to a stack of M5-branes shows that 
M-theory 
 on asymptotically $AdS_7\times S^4$ 
spacetimes 
is equivalent to a 
still-mysterious 
$5+1$ dimensional field theory describing low-energy 
excitations of the M5-branes.
These dualities can also be extended in other ways, such as replacing the $S^n$ with other Einstein spaces. The corresponding change in the gauge theory is known in many cases \cite{Klebanov:1998hh}. One can also put the gauge theory on any spacetime, not just Minkowski space and  the Einstein static universe. In these cases, the boundary condition in the bulk is a space-time which is only locally asymptotically AdS.

Let us mention an example 
of a nontrivial check of the $AdS_4$ duality. For certain $2+1$ supersymmetric gauge theories, one can calculate exactly the Euclidean partition function of the theory on a squashed $S^3$ \cite{Martelli:2011fw}. This is possible using a technique known as  localization. One can then find the dual euclidean gravitational solution which asymptotically approaches the squashed $S^3$. If one computes the gravitational action and compares with the gauge theory partition function one finds 
$Z_\text{gauge} = e^{-S_\text{grav}}$. 
Each side is a  function of the squashing parameter, and the two functions agree exactly. Once again, the  calculations  on each side look completely different, but the final answers agree.  
Similarly, the maximization of the free energy (`$f$-maximization') of the 3D ABJM field theory \cite{Aharony:2008ug} on $S^3$ has a matching gravitational dual description \cite{Freedman:2013oja}.

\subsection{Applications}
\label{s:applications}

Given the overwhelming evidence for gauge/gravity duality, we now assume its validity and ask what it can teach us. The duality can be used in both directions to learn about quantum gravity and also 
about aspects of nongravitational 
strongly coupled 
physics. The following applications all represent fields of active research.
 
 {\bf Quantum black holes:} An immediate consequence of gauge/gravity duality is that the process of forming and evaporating a small black hole must be unitary. This is because it can be mapped to a process in the dual gauge theory in which states evolve by a standard Hamiltonian. However, the details of this map are not yet clear. In particular, it is still unknown how the information gets out of the black hole, and what is wrong with Hawking's original semiclassical argument \cite{Hawking:1976ra} that black hole evaporation would lead pure states to evolve to mixed states. 

For over a decade after the discovery of gauge/gravity duality, many people believed that the information could be restored by keeping  
track 
of subtle correlations in the Hawking radiation that were missed in a semi-classical treatment. However, it was shown in  \cite{Mathur:2009hf} that this can never work: small corrections to Hawking's calculations are not sufficient to get the information out. One alternative that has been suggested is ``fuzzballs" \cite{Mathur:2008nj}. This is the idea that the standard black hole solutions describe ensemble averages, and individual pure states do not have horizons. They are instead described by classical solutions (or more generally quantum states) that extend out to the 
would-be 
horizon. In support of this idea, a large class of supergravity solutions have been constructed which are stationary, nonsingular, and have the same mass and charge as an extremal black hole. It remains to be seen whether all black hole microstates can be realized in this way.

Another alternative has been proposed recently.  
A key assumption in Hawking's argument that black hole evaporation would not be unitary was that the horizon locally looks like flat space to an in-falling observer. 
Since we now believe that the evolution is unitary, people have started to question this assumption.
It has been suggested \cite{Almheiri:2012rt}  that someone falling into an evaporating black hole would 
hit a ``firewall" at the horizon and burn up. It was argued that this would be true even 
for a large black hole as long as it has evaporated to at least half its original mass.  
This has caused considerable controversy.  
The only thing that is clear is that the following three plausible sounding statements are inconsistent: (1) Information is not lost in black hole evaporation, i.e., pure states evolve to pure states. (2) Observers falling into a large black hole pass through the horizon unaffected. (3) Quantum field theory in curved spacetime is a good approximation outside a large black hole. 
At the moment there is no resolution in sight. The fundamental questions raised by Hawking 40 years ago are still unanswered.

{\bf Quark confinement:}  String theory began in the late 1960's as a model of hadrons. As a result of quark confinement, quarks often act like they live at the ends of a string. (It was only in the 1970's that string theory was reinterpreted as a theory of quantum gravity.) It light of this, it is interesting that gauge/gravity duality provides a simple geometric picture of quark confinement. The idea is that the potential between two quarks on the boundary can be computed in terms of a string in the bulk which ends on the quarks.  Since strings have a tension, they want to minimize their length.  Given the geometry of $AdS$,  such strings do not stay in the asymptotic region, but extend into the bulk. Suppose the bulk geometry smoothly caps off at some radius, e.g., because a circle pinches off there\footnote{It is easy to construct bulk solutions with this property, see e.g. \cite{Horowitz:1998ha}.}.
 Then the string  connecting two quarks   separated by a large distance ${\cal L}$ on the boundary drops quickly to this minimum radius, moves a distance  ${\cal L}$ at that radius, and then returns to the boundary. This means the length of the string increases linearly with ${\cal L}$ resulting in a linearly growing potential between the quarks, i.e., the quarks are confined. It is remarkable that a complicated strongly coupled quantum field theory effect such as quark confinement can be given such a simple geometric description. Since we do not currently have a gravitational dual of pure QCD, one cannot yet use holography to argue for quark confinement in the standard model.

{\bf Hydrodynamics:} The long wavelength limit of any strongly coupled field theory is expected to be described by hydrodynamics. It has been shown that general relativity indeed reproduces standard (relativistic) hydrodynamics in the boundary theory.  To see this, one uses the boundary stress tensor which can be defined for any asymptotically 
AdS
spacetime \cite{Balasubramanian:1999re,de Haro:2000xn}. Under gauge/gravity  duality, this is equal to the expectation value of the stress tensor in the dual field theory. 
One then starts with the planar black hole representing a system in equilibrium at temperature $T$, and adds long wavelength perturbations. The boundary stress tensor is conserved and takes the form of a perfect fluid plus corrections involving derivatives of the four-velocity. The first derivative term represents viscosity. Dissipation in the dual theory simply corresponds to energy flowing into the black hole.

One can show that the ratio of the shear viscosity $\eta$ to entropy density $s$ is a universal constant for any theory with a gravity dual \cite{Kovtun:2004de}:
\be
{\eta\over s} = {1\over 4\pi}.
\ee
This number is very low compared to ordinary fluids. Remarkably, when experiments at RHIC and LHC collide heavy ions together, they produce a quark/gluon plasma which has very low viscosity. The measured viscosity is in fact close to the value predicted from the gravity dual. This is difficult to explain using traditional methods.

The connection between gravity and fluid dynamics raises an interesting question. Turbulence is a common feature of fluids, but perturbations of black holes are expected to decay and not show turbulent behavior. In recent work \cite{Adams:2013vsa} it has been shown that under certain conditions, black holes in AdS do show turbulent behavior.

{\bf Condensed matter:} Given the success with heavy ion collisions, people became more ambitious and started to apply gauge/gravity duality to study properties of finite density quantum matter, i.e., the subject of condensed matter. Despite the fact that there is no obvious analog of the large $N$ limit in this case, classical gravity analogs of several condensed matter phenomena have been found. The advantage of this duality is that it allows one to calculate transport properties of strongly correlated systems at finite temperature. This is difficult to do using standard condensed matter techniques, but is easy to do holographically. One starts with a black hole in $AdS$ which represents the equilibrium system at temperature $T$. To compute transport using linear response, one simply perturbs the black hole.

We focus on one example: superconductivity.
  In standard superconductors, pairs of electrons with opposite spin can combine to form a charged boson called a Cooper pair. Below a critical temperature, these bosons condense and the DC conductivity becomes infinite. To construct a ``holographic superconductor", i.e., the gravitational dual to a superconductor,  we need just gravity coupled to a Maxwell field and 
a   
  charged scalar field.  A charged black hole corresponds to a system at temperature equal to the Hawking temperature, $T$, and nonzero charge density (or chemical potential). To represent a nonzero condensate, one needs 
  a static charged scalar field outside the black hole. This is like black hole ``hair".  So to describe a superconductor, we need to find a black hole that has hair at low temperatures, but no hair at high temperatures. More precisely, we need the usual Reissner-Nordstrom AdS black hole (which exists for all temperatures) to be unstable to forming  hair at low temperature. At first sight, this is not an easy task, since it contradicts our usual intuition that black holes have no hair. 
  
A surprisingly simple solution to this problem was found by Gubser \cite{Gubser:2008px}. He argued that a charged scalar field around a  charged black hole in AdS would have the desired property. Consider
\be\label{action}
 S=\int d^{4}x \sqrt{-g}\left(R + {6\over L^2} -\frac{1}{4}F_{\mu \nu}F^{\mu \nu} - |\nabla\Psi-i qA\Psi|^2 - m^2|\Psi|^2\right). 
\ee
This is just general relativity with a negative cosmological constant $\Lambda = -3/L^2$, coupled to a Maxwell field and charged scalar with mass $m$ and charge $q$. It is easy to see why black holes in this theory might be unstable to forming scalar hair:
for 
 an electrically charged black hole, the effective mass of $\Psi$ is 
 $m^2_\text{eff} = m^2 + q^2 g^{tt} A_t^2$. 
 But the last term is negative, so there is a chance that 
 $m^2_\text{eff}$  
 becomes sufficiently negative near the horizon to destabilize the scalar field. Detailed calculations confirm that scalar hair does indeed form at low temperature \cite{Hartnoll:2008vx}.  Why wasn't such a simple type of hair  noticed earlier? One reason is that this does not work for asymptotically flat black holes. In that case,
 the scalar field simply radiates away some of the mass and charge  of the black hole in a form of superradiance.
  
The conductivity can be calculated by perturbing this black hole with boundary conditions on the Maxwell field at infinity that correspond to adding a uniform electric field. The induced current is read off from a subleading term in the perturbation. One finds that at low temperature when the scalar hair is present, the DC conductivity is infinite, showing one really does have a superconductor \cite{Hartnoll:2008vx}. 
A similar calculation can be done in five dimensions, corresponding to a $3+1$-dimensional superconductor. However the four-dimensional bulk calculation is appropriate for some ``high temperature" superconductors such as the cuprates, in which the superconductivity is associated with two-dimensional CuO planes.
 
To make the model more realistic, we can add the effects of a lattice by requiring the chemical potential be a periodic function. This corresponds to a periodic asymptotic boundary condition on $A_t$. One then numerically finds the rippled charged black holes with this boundary condition. One can then perturb this solution and compute the conductivity as a function of frequency. At high temperature (when the scalar field is zero), the result shows a finite DC conductivity followed by a power law fall-off $|\sigma(\omega)| = B/\omega^{2/3} +C$ \cite{Horowitz:2012ky}. Exactly this same type of power law fall-off (but without the constant off-set $C$) is seen in certain cuprates in their normal phase before they become superconducting.
This behavior is not understood from standard condensed matter arguments. It is believed to be a result of strong correlations. Experiments show that the coefficient B and the exponent 2/3 are temperature independent and do not change even when T drops below the superconducting transition temperature. Similarly,  one finds no change in the power law fall-off on the gravity side, when one lowers the temperature of the black hole into the superconducting regime \cite{Horowitz:2013jaa}.

{\bf Entanglement entropy:} An important quantity in condensed matter is the entanglement entropy. Given a quantum state of a system and a subregion $A$, one can construct the density matrix $\rho_A$ by tracing over all degrees of freedom outside $A$. The entanglement entropy is then defined to be 
\be
S_\text{EE} = - Tr \rho_A \ln \rho_A
\,.
\ee
This is a measure of long range correlations and has proven useful in a variety of applications including 
identification of 
exotic ground states. Unfortunately, it is difficult to calculate in general interacting theories. A simple formula has been given for $S_\text{EE}$ using gauge/gravity duality \cite{Ryu:2006bv}. For a static spacetime, consider the minimal surface $\Sigma$ which ends on the boundary of $A$ at infinity. The conjecture is that
$S_\text{EE} = A_\Sigma/4G$, where $A_\Sigma$ is the area of $\Sigma$. In other words, the entanglement entropy is given by a formula which is very similar  to the Hawking-Bekenstein entropy of a black hole. Since $\Sigma$ extends out to infinity, its area is infinite. But $S_\text{EE}$ is also infinite if one includes arbitrarily short wavelength modes which cross the boundary of $A$. There is a prescription to regulate this divergence both in the gravity side and dual field theory. Using this prescription, this conjecture has passed a large number of nontrivial tests \cite{Nishioka:2009un}. A derivation using Euclidean gravity has recently been provided \cite{Lewkowycz:2013nqa}. It has even been suggested that quantum entanglement might be a key to reconstructing the bulk spacetime \cite{VanRaamsdonk:2010pw}.

\section{Conclusion}
\label{s:conclusion}
We have seen that combining supersymmetry with general relativity to form supergravity improves the behavior of perturbative graviton scattering amplitudes, but is not expected to provide a UV complete quantum theory. The situation is much improved in string theory, which not only has  perturbatively finite  scattering amplitudes, but also provides a complete  description of the microstates of certain black holes, 
 and gives a precise form of holography for certain boundary conditions.

In a review of such a large subject, several topics are inevitably 
 left out.  One is string phenomenology 
--- 
the attempt to find a choice of compactification and fluxes so that the low energy theory looks like some extension of  the standard model of particle physics (coupled to gravity).
Another area that we have not had space to describe are attempts to model de Sitter spacetime in string theory. This is useful for stringy models of both inflation and the current acceleration of the universe due to dark energy. 

There remain many directions for future research. One concerns spacetime
  singularities in string theory. It has been shown that string theory can resolve certain types of singularities, but these tend to be static timelike singularities. It is not known in detail how string theory resolves the naked singularities arising from the instability of black strings and black branes. More importantly, very little is known about the most significant singularities of general relativity, such as the big bang and the singularity inside black holes. It is not clear that string theory ``resolves" such singularities
 in the sense that there is another semiclassical spacetime on the other side. For example, in the case of the big bang, it is possible that time emerges from  a more fundamental description in much the same way that space emerges in our current theories of holography.
 In addition, we need to  better understand the dictionary relating quantum gravity with anti-de Sitter boundary conditions to the dual gauge theory. Some elements of this dictionary are known, but many more remain to be understood. A key test will be to understand how the information comes out of an evaporating black hole. Another area where progress is needed is to extend holography to asymptotically flat or de Sitter spacetimes.

String theory has had an amazing history. It began as a theory of hadrons and was reinvented as a theory of quantum gravity based on 10D strings. 
It can be understood as a weak-coupling limit of an 11D theory, M-theory,  which includes membranes and five-branes. And with anti-de Sitter boundary conditions, superstring theory is believed to be equivalent to a supersymmetric gauge theory. 
 Since string theory has both the ingredients of the standard model of particle physics and of quantum gravity, it unifies the fundamental forces and is therefore a candidate for a `theory of everything'. Thus, it was long hoped  that string theory would produce a single unique `vacuum' that  would unify our understanding of particle physics, gravity, and cosmology, i.e.~that it would explain all parameters in particle physics and beyond in a unified framework. It was later realized that string theory does not produce such unique predictions, and we will have to make choices among the various vacua. 
 
 However, string theory has recently surprised us again. Rather than ``just" being a unified framework for particle physics and quantum gravity, investigations of gauge/gravity duality have revealed totally unexpected connections with other areas of physics.  String theory is now expanding into the realm of nuclear physics, hydrodynamics and condensed matter.  We are clearly only beginning to explore the  depth and potential applications of this remarkable subject.

\vskip 1cm
\centerline{\bf Acknowledgements}
\vskip 3mm
It is a pleasure to thank N.~Engelhardt, G.~Hartnett, and T.~Olson for comments on the draft of this chapter. 
G.H.~is  supported in part by the National Science Foundation under Grant No. PHY12-05500. 
H.E.~is supported by NSF CAREER Grant PHY-0953232.


\begin{thebibliography}{99}

\bibitem{Tangherlini:1963bw}
  F.~R.~Tangherlini,
  ``Schwarzschild field in n dimensions and the dimensionality of space problem,''
  Nuovo Cim.\  {\bf 27} (1963) 636.
  
\bibitem{Banados:1992wn}
  M.~Banados, C.~Teitelboim and J.~Zanelli,
  ``The Black hole in three-dimensional space-time,''
  Phys.\ Rev.\ Lett.\  {\bf 69} (1992) 1849
  [hep-th/9204099].
  


\bibitem{Kaluza:1921tu}
  T.~Kaluza,
  ``On the Problem of Unity in Physics,''
  Sitzungsber.\ Preuss.\ Akad.\ Wiss.\ Berlin (Math.\ Phys.\ ) {\bf 1921} (1921) 966.
  
\bibitem{Klein:1926tv}
  O.~Klein,
  ``Quantum Theory and Five-Dimensional Theory of Relativity. (In German and English),''
  Z.\ Phys.\  {\bf 37} (1926) 895
   [Surveys High Energ.\ Phys.\  {\bf 5} (1986) 241].



\bibitem{Gregory:1993vy}
  R.~Gregory and R.~Laflamme,
  ``Black strings and p-branes are unstable,''
  Phys.\ Rev.\ Lett.\  {\bf 70} (1993) 2837
  [hep-th/9301052].
  
\bibitem{Horowitz:2011cq}
  G.~T.~Horowitz and T.~Wiseman,
  ``General black holes in Kaluza-Klein theory,'' in {\sl Black
Holes in Higher Dimensions} (G. Horowitz ed.), Cambridge University Press (2012);  arXiv:1107.5563 [gr-qc].
   
\bibitem{Hawking:1973uf}
  S.~W.~Hawking and G.~F.~R.~Ellis,
  ``The Large scale structure of space-time,''
  Cambridge University Press, Cambridge, 1973  
  
\bibitem{Horowitz:2001cz}
  G.~T.~Horowitz and K.~Maeda,
  ``Fate of the black string instability,''
  Phys.\ Rev.\ Lett.\  {\bf 87} (2001) 131301
  [hep-th/0105111].
  
\bibitem{Lehner:2011wc}
  L.~Lehner and F.~Pretorius,
  ``Final State of Gregory-Laflamme Instability,''
  arXiv:1106.5184 [gr-qc].
  
\bibitem{Cardoso:2006ks}
  V.~Cardoso and O.~J.~C.~Dias,
  ``Rayleigh-Plateau and Gregory-Laflamme instabilities of black strings,''
  Phys.\ Rev.\ Lett.\  {\bf 96} (2006) 181601
  [hep-th/0602017].
  


\bibitem{Wiseman:2002zc}
  T.~Wiseman,
  ``Static axisymmetric vacuum solutions and nonuniform black strings,''
  Class.\ Quant.\ Grav.\  {\bf 20} (2003) 1137
  [hep-th/0209051].
  
\bibitem{Kudoh:2003ki}
  H.~Kudoh and T.~Wiseman,
  ``Properties of Kaluza-Klein black holes,''
  Prog.\ Theor.\ Phys.\  {\bf 111} (2004) 475
  [hep-th/0310104].

\bibitem{Kleihaus:2006ee} 
  B.~Kleihaus, J.~Kunz and E.~Radu,
  ``New nonuniform black string solutions,''
  JHEP {\bf 0606}, 016 (2006)
  [hep-th/0603119].

\bibitem{Deser:1988fc}
  S.~Deser and M.~Soldate,
  ``Gravitational Energy in Spaces With Compactified Dimensions,''
  Nucl.\ Phys.\ B {\bf 311} (1989) 739.
  
\bibitem{Brill:1989di}
  D.~Brill and H.~Pfister,
  ``States of Negative Total Energy in {Kaluza-Klein} Theory,''
  Phys.\ Lett.\ B {\bf 228} (1989) 359.

\bibitem{Brill:1991qe}
  D.~Brill and G.~T.~Horowitz,
  ``Negative energy in string theory,''
  Phys.\ Lett.\ B {\bf 262} (1991) 437.


  
\bibitem{Witten:1981gj}
  E.~Witten,
  ``Instability of the Kaluza-Klein Vacuum,''
  Nucl.\ Phys.\ B {\bf 195} (1982) 481.
  
  \bibitem{Dai}
  X. Dai, ``A positive energy theorem for spaces with asymptotic SUSY compactification",
  Commun. Math. Phys. {\bf 244} (2004) 335.
  
\bibitem{Elvang:2002br}
  H.~Elvang and G.~T.~Horowitz,
  ``When black holes meet Kaluza-Klein bubbles,''
  Phys.\ Rev.\ D {\bf 67} (2003) 044015
  [hep-th/0210303].

\bibitem{Elvang:2004iz}
  H.~Elvang, T.~Harmark and N.~A.~Obers,
  ``Sequences of bubbles and holes: New phases of Kaluza-Klein black holes,''
  JHEP {\bf 0501} (2005) 003
  [hep-th/0407050].
  
\bibitem{Myers:1986un}
  R.~C.~Myers and M.~J.~Perry,
  ``Black Holes in Higher Dimensional Space-Times,''
  Annals Phys.\  {\bf 172} (1986) 304.
  
\bibitem{Emparan:2001wn}
  R.~Emparan and H.~S.~Reall,
  ``A Rotating black ring solution in five-dimensions,''
  Phys.\ Rev.\ Lett.\  {\bf 88} (2002) 101101
  [hep-th/0110260].
  
\bibitem{Emparan:2006mm}
  R.~Emparan and H.~S.~Reall,
  ``Black Rings,''
  Class.\ Quant.\ Grav.\  {\bf 23} (2006) R169
  [hep-th/0608012].
  
\bibitem{Elvang:2003yy}
  H.~Elvang,
  ``A Charged rotating black ring,''
  Phys.\ Rev.\ D {\bf 68} (2003) 124016
  [hep-th/0305247].
  
\bibitem{Elvang:2006dd}
  H.~Elvang, R.~Emparan and A.~Virmani,
  ``Dynamics and stability of black rings,''
  JHEP {\bf 0612} (2006) 074
  [hep-th/0608076].
  
\bibitem{Elvang:2007hg}
  H.~Elvang, R.~Emparan and P.~Figueras,
  ``Phases of five-dimensional black holes,''
  JHEP {\bf 0705} (2007) 056
  [hep-th/0702111].
  
\bibitem{Pomeransky:2006bd}
  A.~A.~Pomeransky and R.~A.~Sen'kov,
  ``Black ring with two angular momenta,''
  hep-th/0612005.
  
\bibitem{Belinsky:1971nt}
  V.~A.~Belinsky and V.~E.~Zakharov,
  ``Integration of the Einstein Equations by the Inverse Scattering Problem Technique and the Calculation of the Exact Soliton Solutions,''
  Sov.\ Phys.\ JETP {\bf 48} (1978) 985
   [Zh.\ Eksp.\ Teor.\ Fiz.\  {\bf 75} (1978) 1953].
  
\bibitem{Belinsky:1979mh}
  V.~A.~Belinsky and V.~E.~Sakharov,
  ``Stationary Gravitational Solitons with Axial Symmetry,''
  Sov.\ Phys.\ JETP {\bf 50} (1979) 1
   [Zh.\ Eksp.\ Teor.\ Fiz.\  {\bf 77} (1979) 3].
  
\bibitem{Belinski:2001ph}
  V.~Belinski and E.~Verdaguer,
  ``Gravitational solitons,''
  Cambridge, UK: Univ.~Pr.~(2001) 258 p  
  
  
\bibitem{Elvang:2007hs}
  H.~Elvang and M.~J.~Rodriguez,
  ``Bicycling Black Rings,''
  JHEP {\bf 0804} (2008) 045
  [arXiv:0712.2425 [hep-th]].
  
\bibitem{Elvang:2007rd}
  H.~Elvang and P.~Figueras,
  ``Black Saturn,''
  JHEP {\bf 0705} (2007) 050
  [hep-th/0701035].
  

  
\bibitem{Chrusciel:2010ix}
  P.~T.~Chrusciel, M.~Eckstein and S.~J.~Szybka,
  ``On smoothness of Black Saturns,''
  JHEP {\bf 1011} (2010) 048
  [arXiv:1007.3668 [hep-th]].

\bibitem{Iguchi:2007is} 
  H.~Iguchi and T.~Mishima,
  ``Black di-ring and infinite nonuniqueness,''
  Phys.\ Rev.\ D {\bf 75}, 064018 (2007)
  [Erratum-ibid.\ D {\bf 78}, 069903 (2008)]
  [hep-th/0701043].
  
\bibitem{Evslin:2007fv} 
  J.~Evslin and C.~Krishnan,
  ``The Black Di-Ring: An Inverse Scattering Construction,''
  Class.\ Quant.\ Grav.\  {\bf 26}, 125018 (2009)
  [arXiv:0706.1231 [hep-th]].
  
  
\bibitem{Izumi:2007qx}
  K.~Izumi,
  ``Orthogonal black di-ring solution,''
  Prog.\ Theor.\ Phys.\  {\bf 119} (2008) 757
  [arXiv:0712.0902 [hep-th]].
  
\bibitem{Reall:2002bh}
  H.~S.~Reall,
  ``Higher dimensional black holes and supersymmetry,''
  Phys.\ Rev.\ D {\bf 68} (2003) 024024
   [Erratum-ibid.\ D {\bf 70} (2004) 089902]
  [hep-th/0211290].
  
\bibitem{Hollands:2006rj}
  S.~Hollands, A.~Ishibashi and R.~M.~Wald,
  ``A Higher dimensional stationary rotating black hole must be axisymmetric,''
  Commun.\ Math.\ Phys.\  {\bf 271} (2007) 699
  [gr-qc/0605106].
  
\bibitem{Moncrief:2008mr}
  V.~Moncrief and J.~Isenberg,
  ``Symmetries of Higher Dimensional Black Holes,''
  Class.\ Quant.\ Grav.\  {\bf 25} (2008) 195015
  [arXiv:0805.1451 [gr-qc]].
  
\bibitem{Emparan:2009vd}
  R.~Emparan, T.~Harmark, V.~Niarchos and N.~A.~Obers,
  ``New Horizons for Black Holes and Branes,''
  JHEP {\bf 1004} (2010) 046
  [arXiv:0912.2352 [hep-th]].
  
\bibitem{Emparan:2003sy}
  R.~Emparan and R.~C.~Myers,
  ``Instability of ultra-spinning black holes,''
  JHEP {\bf 0309} (2003) 025
  [hep-th/0308056].
  
\bibitem{Emparan:2008eg}
  R.~Emparan and H.~S.~Reall,
  ``Black Holes in Higher Dimensions,''
  Living Rev.\ Rel.\  {\bf 11} (2008) 6
  [arXiv:0801.3471 [hep-th]].
 
\bibitem{Emparan:2007wm} 
  R.~Emparan, T.~Harmark, V.~Niarchos, N.~A.~Obers and M.~J.~Rodriguez,
  ``The Phase Structure of Higher-Dimensional Black Rings and Black Holes,''
  JHEP {\bf 0710}, 110 (2007)
  [arXiv:0708.2181 [hep-th]].
 
\bibitem{Emparan:2009at}
  R.~Emparan, T.~Harmark, V.~Niarchos and N.~A.~Obers,
  ``Essentials of Blackfold Dynamics,''
  JHEP {\bf 1003} (2010) 063
  [arXiv:0910.1601 [hep-th]].
    
\bibitem{Emparan:2009cs}
  R.~Emparan, T.~Harmark, V.~Niarchos and N.~A.~Obers,
  ``World-Volume Effective Theory for Higher-Dimensional Black Holes,''
  Phys.\ Rev.\ Lett.\  {\bf 102} (2009) 191301
  [arXiv:0902.0427 [hep-th]].
  
\bibitem{Camps:2012hw}
  J.~Camps and R.~Emparan,
  ``Derivation of the blackfold effective theory,''
  JHEP {\bf 1203} (2012) 038
   [Erratum-ibid.\  {\bf 1206} (2012) 155]
  [arXiv:1201.3506 [hep-th]].
  
\bibitem{Kleihaus:2012xh} 
  B.~Kleihaus, J.~Kunz and E.~Radu,
  ``Black rings in six dimensions,''
  Phys.\ Lett.\ B {\bf 718}, 1073 (2013)
  [arXiv:1205.5437 [hep-th]].
  
  
\bibitem{Coleman:1967ad}
  S.~R.~Coleman and J.~Mandula,
  ``All Possible Symmetries of the S Matrix,''
  Phys.\ Rev.\  {\bf 159} (1967) 1251.

\bibitem{Haag:1974qh}
  R.~Haag, J.~T.~Lopuszanski and M.~Sohnius,
  ``All Possible Generators of Supersymmetries of the s Matrix,''
  Nucl.\ Phys.\ B {\bf 88} (1975) 257.

\bibitem{Wess:1992cp}
  J.~Wess and J.~Bagger,
  ``Supersymmetry and supergravity,''
  Princeton, USA: Univ.~Pr.~(1992) 259 p  

\bibitem{Freedman:2012zz}
  D.~Z.~Freedman and A.~Van Proeyen,
  ``Supergravity,''
  Cambridge, UK: Cambridge Univ.~Pr.~(2012) 607 p


\bibitem{Freedman:1976xh}
  D.~Z.~Freedman, P.~van Nieuwenhuizen and S.~Ferrara,
  ``Progress Toward a Theory of Supergravity,''
  Phys.\ Rev.\ D {\bf 13} (1976) 3214.
  
\bibitem{Deser:1976eh}
  S.~Deser and B.~Zumino,
  ``Consistent Supergravity,''
  Phys.\ Lett.\ B {\bf 62} (1976) 335.
  
\bibitem{Ferrara:1976fu}
  S.~Ferrara and P.~van Nieuwenhuizen,
  ``Consistent Supergravity with Complex Spin 3/2 Gauge Fields,''
  Phys.\ Rev.\ Lett.\  {\bf 37} (1976) 1669.
 
\bibitem{Hawking:1981bu}
  E.~Cremmer, ``Supergravities in Five Dimensions," in 
  ``Superspace And Supergravity. Proceedings, Nuffield Workshop, Cambridge, Uk, June 16 - July 12, 1980,'' Eds.~S.~W.~Hawking and M.~Rocek, 
  Cambridge, Uk: Univ.~Pr.~(1981) 527p
 
\bibitem{Fronsdal:1978rb} 
  C.~Fronsdal,
  ``Massless Fields with Integer Spin,''
  Phys.\ Rev.\ D {\bf 18}, 3624 (1978).

\bibitem{Fradkin:1986qy} 
  E.~S.~Fradkin and M.~A.~Vasiliev,
  ``Cubic Interaction in Extended Theories of Massless Higher Spin Fields,''
  Nucl.\ Phys.\ B {\bf 291}, 141 (1987).
  
\bibitem{Vasiliev:1999ba} 
  M.~A.~Vasiliev,
  ``Higher spin gauge theories: Star product and AdS space,''
  In *Shifman, M.A. (ed.): The many faces of the superworld* 533-610
  [hep-th/9910096].
  
\bibitem{Klebanov:2002ja} 
  I.~R.~Klebanov and A.~M.~Polyakov,
  ``AdS dual of the critical O(N) vector model,''
  Phys.\ Lett.\ B {\bf 550}, 213 (2002)
  [hep-th/0210114].
  
\bibitem{Gaberdiel:2010pz} 
  M.~R.~Gaberdiel and R.~Gopakumar,
  ``An AdS$_3$ Dual for Minimal Model CFTs,''
  Phys.\ Rev.\ D {\bf 83}, 066007 (2011)
  [arXiv:1011.2986 [hep-th]].
 
 
\bibitem{Cremmer:1978km}
  E.~Cremmer, B.~Julia and J.~Scherk,
  ``Supergravity Theory in Eleven-Dimensions,''
  Phys.\ Lett.\ B {\bf 76} (1978) 409.


\bibitem{Gunaydin:1984qu}
  M.~Gunaydin, L.~J.~Romans and N.~P.~Warner,
  ``Gauged N=8 Supergravity in Five-Dimensions,''
  Phys.\ Lett.\ B {\bf 154} (1985) 268.
 
\bibitem{Bogomolny:1975de} 
  E.~B.~Bogomolny,
  ``Stability of Classical Solutions,''
  Sov.\ J.\ Nucl.\ Phys.\  {\bf 24}, 449 (1976)
  [Yad.\ Fiz.\  {\bf 24}, 861 (1976)].
 
\bibitem{Prasad:1975kr}
  M.~K.~Prasad and C.~M.~Sommerfield,
  ``An Exact Classical Solution for the 't Hooft Monopole and the Julia-Zee Dyon,''
  Phys.\ Rev.\ Lett.\  {\bf 35} (1975) 760.
 
\bibitem{Witten:1981mf}
  E.~Witten,
  ``A Simple Proof of the Positive Energy Theorem,''
  Commun.\ Math.\ Phys.\  {\bf 80} (1981) 381.
 
\bibitem{Gibbons:1982fy}
  G.~W.~Gibbons and C.~M.~Hull,
  ``A Bogomolny Bound for General Relativity and Solitons in N=2 Supergravity,''
  Phys.\ Lett.\ B {\bf 109} (1982) 190.
 
\bibitem{Gibbons:1993xt}
  G.~W.~Gibbons, D.~Kastor, L.~A.~J.~London, P.~K.~Townsend and J.~H.~Traschen,
  ``Supersymmetric selfgravitating solitons,''
  Nucl.\ Phys.\ B {\bf 416} (1994) 850
  [hep-th/9310118].
 
\bibitem{Cvetic:1996xz} 
  M.~Cvetic and D.~Youm,
  ``General rotating five-dimensional black holes of toroidally compactified heterotic string,''
  Nucl.\ Phys.\ B {\bf 476}, 118 (1996)
  [hep-th/9603100].
 
\bibitem{Breckenridge:1996is}
  J.~C.~Breckenridge, R.~C.~Myers, A.~W.~Peet and C.~Vafa,
  ``D-branes and spinning black holes,''
  Phys.\ Lett.\ B {\bf 391} (1997) 93
  [hep-th/9602065].
 
\bibitem{Elvang:2004rt}
  H.~Elvang, R.~Emparan, D.~Mateos and H.~S.~Reall,
  ``A Supersymmetric black ring,''
  Phys.\ Rev.\ Lett.\  {\bf 93} (2004) 211302
  [hep-th/0407065].

\bibitem{Bena:2004de}
  I.~Bena and N.~P.~Warner,
  ``One ring to rule them all ... and in the darkness bind them?,''
  Adv.\ Theor.\ Math.\ Phys.\  {\bf 9} (2005) 667
  [hep-th/0408106].

\bibitem{Elvang:2004ds}
  H.~Elvang, R.~Emparan, D.~Mateos and H.~S.~Reall,
  ``Supersymmetric black rings and three-charge supertubes,''
  Phys.\ Rev.\ D {\bf 71} (2005) 024033
  [hep-th/0408120].
 
\bibitem{Gauntlett:2004qy}
  J.~P.~Gauntlett and J.~B.~Gutowski,
  ``General concentric black rings,''
  Phys.\ Rev.\ D {\bf 71} (2005) 045002
  [hep-th/0408122].
  
\bibitem{Elvang:2013cua}
  H.~Elvang and Y.~-t.~Huang,
  ``Scattering Amplitudes,''
  arXiv:1308.1697 [hep-th].
  
\bibitem{ArkaniHamed:2012nw} 
  N.~Arkani-Hamed, J.~L.~Bourjaily, F.~Cachazo, A.~B.~Goncharov, A.~Postnikov and J.~Trnka,
  ``Scattering Amplitudes and the Positive Grassmannian,''
  arXiv:1212.5605 [hep-th].
 
\bibitem{Kawai:1985xq} 
  H.~Kawai, D.~C.~Lewellen and S.~H.~H.~Tye,
  ``A Relation Between Tree Amplitudes of Closed and Open Strings,''
  Nucl.\ Phys.\ B {\bf 269}, 1 (1986).

\bibitem{Bern:1998sv} 
  Z.~Bern, L.~J.~Dixon, M.~Perelstein and J.~S.~Rozowsky,
  ``Multileg one loop gravity amplitudes from gauge theory,''
  Nucl.\ Phys.\ B {\bf 546}, 423 (1999)
  [hep-th/9811140].


\bibitem{Bern:1999ji} 
  Z.~Bern and A.~K.~Grant,
  ``Perturbative gravity from QCD amplitudes,''
  Phys.\ Lett.\ B {\bf 457}, 23 (1999)
  [hep-th/9904026].

\bibitem{Bern:2000mf} 
  Z.~Bern, L.~J.~Dixon, D.~C.~Dunbar, A.~K.~Grant, M.~Perelstein and J.~S.~Rozowsky,
  ``On perturbative gravity and gauge theory,''
  Nucl.\ Phys.\ Proc.\ Suppl.\  {\bf 88}, 194 (2000)
  [hep-th/0002078].

\bibitem{Siegel:1993xq} 
  W.~Siegel,
  ``Two vierbein formalism for string inspired axionic gravity,''
  Phys.\ Rev.\ D {\bf 47}, 5453 (1993)
  [hep-th/9302036].

\bibitem{Bern:2002kj} 
  Z.~Bern,
  ``Perturbative quantum gravity and its relation to gauge theory,''
  Living Rev.\ Rel.\  {\bf 5}, 5 (2002)
  [gr-qc/0206071].
 
\bibitem{BCJ} 
  Z.~Bern, J.~J.~M.~Carrasco and H.~Johansson,
  ``New Relations for Gauge-Theory Amplitudes,''
  Phys.\ Rev.\ D {\bf 78}, 085011 (2008)
  [arXiv:0805.3993 [hep-ph]].
 
\bibitem{'tHooft:1974bx} 
  G.~'t Hooft and M.~J.~G.~Veltman,
  ``One loop divergencies in the theory of gravitation,''
  Annales Poincare Phys.\ Theor.\ A {\bf 20}, 69 (1974).

\bibitem{Goroff:1985sz}
  M.~H.~Goroff and A.~Sagnotti,
  ``Quantum Gravity At Two Loops,''
  Phys.\ Lett.\ B {\bf 160}, 81 (1985).

\bibitem{vandeVen:1991gw} 
  A.~E.~M.~van de Ven,
  ``Two loop quantum gravity,''
  Nucl.\ Phys.\ B {\bf 378}, 309 (1992).
 
 \bibitem{Deser:1974cz} 
  S.~Deser and P.~van Nieuwenhuizen,
  ``One Loop Divergences of Quantized Einstein-Maxwell Fields,''
  Phys.\ Rev.\ D {\bf 10}, 401 (1974).


\bibitem{Grisaru:1976ua} 
  M.~T.~Grisaru, P.~van Nieuwenhuizen and J.~A.~M.~Vermaseren,
  ``One Loop Renormalizability of Pure Supergravity and of Maxwell-Einstein Theory in Extended Supergravity,''
  Phys.\ Rev.\ Lett.\  {\bf 37}, 1662 (1976).

\bibitem{Grisaru:1976nn} 
  M.~T.~Grisaru,
  ``Two Loop Renormalizability of Supergravity,''
  Phys.\ Lett.\ B {\bf 66}, 75 (1977).


  
\bibitem{Tomboulis:1977wd} 
  E.~Tomboulis,
  ``On the Two Loop Divergences of Supersymmetric Gravitation,''
  Phys.\ Lett.\ B {\bf 67}, 417 (1977).

\bibitem{Deser:1977nt} 
  S.~Deser, J.~H.~Kay and K.~S.~Stelle,
  ``Renormalizability Properties of Supergravity,''
  Phys.\ Rev.\ Lett.\  {\bf 38}, 527 (1977).

 
\bibitem{Bern:2013uka}
  Z.~Bern, S.~Davies, T.~Dennen, A.~V.~Smirnov and V.~A.~Smirnov,
  ``The Ultraviolet Properties of N=4 Supergravity at Four Loops,''
  arXiv:1309.2498 [hep-th].  
  
\bibitem{Bern:2006kd} 
  Z.~Bern, L.~J.~Dixon and R.~Roiban,
  ``Is N = 8 supergravity ultraviolet finite?,''
  Phys.\ Lett.\ B {\bf 644}, 265 (2007)
  [hep-th/0611086].
  
\bibitem{Bern:2007hh} 
  Z.~Bern, J.~J.~Carrasco, L.~J.~Dixon, H.~Johansson, D.~A.~Kosower and R.~Roiban,
  ``Three-Loop Superfiniteness of N=8 Supergravity,''
  Phys.\ Rev.\ Lett.\  {\bf 98}, 161303 (2007)
  [hep-th/0702112].
  
\bibitem{Bern:2008pv} 
  Z.~Bern, J.~J.~M.~Carrasco, L.~J.~Dixon, H.~Johansson and R.~Roiban,
  ``Manifest Ultraviolet Behavior for the Three-Loop Four-Point Amplitude of N=8 Supergravity,''
  Phys.\ Rev.\ D {\bf 78}, 105019 (2008)
  [arXiv:0808.4112 [hep-th]].

\bibitem{Bern:2009kd} 
  Z.~Bern, J.~J.~Carrasco, L.~J.~Dixon, H.~Johansson and R.~Roiban,
  ``The Ultraviolet Behavior of N=8 Supergravity at Four Loops,''
  Phys.\ Rev.\ Lett.\  {\bf 103}, 081301 (2009)
  [arXiv:0905.2326 [hep-th]].
  
\bibitem{Bern:2010tq} 
  Z.~Bern, J.~J.~M.~Carrasco, L.~J.~Dixon, H.~Johansson and R.~Roiban,
  ``The Complete Four-Loop Four-Point Amplitude in N=4 Super-Yang-Mills Theory,''
  Phys.\ Rev.\ D {\bf 82}, 125040 (2010)
  [arXiv:1008.3327 [hep-th]].
  
\bibitem{Elvang:2010jv} 
  H.~Elvang, D.~Z.~Freedman and M.~Kiermaier,
  ``A simple approach to counterterms in N=8 supergravity,''
  JHEP {\bf 1011}, 016 (2010)
  [arXiv:1003.5018 [hep-th]].
  
\bibitem{Drummond:2010fp} 
  J.~M.~Drummond, P.~J.~Heslop and P.~S.~Howe,
  ``A Note on N=8 counterterms,''
  arXiv:1008.4939 [hep-th].
    
\bibitem{Elvang:2010kc} 
  H.~Elvang and M.~Kiermaier,
  ``Stringy KLT relations, global symmetries, and $E_{7(7)}$ violation,''
  JHEP {\bf 1010}, 108 (2010)
  [arXiv:1007.4813 [hep-th]].
    
\bibitem{Beisert:2010jx} 
  N.~Beisert, H.~Elvang, D.~Z.~Freedman, M.~Kiermaier, A.~Morales and S.~Stieberger,
  ``E7(7) constraints on counterterms in N=8 supergravity,''
  Phys.\ Lett.\ B {\bf 694}, 265 (2010)
  [arXiv:1009.1643 [hep-th]].
  
\bibitem{Green:2008bf}
  M.~B.~Green, J.~G.~Russo and P.~Vanhove,
  ``Modular properties of two-loop maximal supergravity and connections with
  string theory,''
  JHEP {\bf 0807}, 126 (2008)
  [arXiv:0807.0389 [hep-th]].
  
\bibitem{Bossard:2011tq} 
  G.~Bossard, P.~S.~Howe, K.~S.~Stelle and P.~Vanhove,
  ``The vanishing volume of D=4 superspace,''
  Class.\ Quant.\ Grav.\  {\bf 28}, 215005 (2011)
  [arXiv:1105.6087 [hep-th]].
  
\bibitem{Zwiebach:2004tj}
  B.~Zwiebach,
  ``A first course in string theory,''
  Cambridge, UK: Univ.~Pr.~(2009) 673 p

\bibitem{Polchinski:1998rq}
  J.~Polchinski,
  ``String theory. Vol. 1: An introduction to the bosonic string,''
  Cambridge, UK: Univ.~Pr.~(1998) 402 p;
  ``String theory. Vol. 2: Superstring theory and beyond,''
  Cambridge, UK: Univ.~Pr.~(1998) 531 p
  
\bibitem{Becker:2007zj}
  K.~Becker, M.~Becker and J.~H.~Schwarz,
  ``String theory and M-theory: A modern introduction,''
  Cambridge, UK: Cambridge Univ.~Pr.~(2007) 739 p
  
  \bibitem{GSO}
F. Gliozzi, J. Sherk, and D. Olive,
``Supersymmetry, supergravity, and the dual spinor model,"
Nucl.\ Phys.\ B {\bf 122} (1977)  253.

\bibitem{Callan:1985ia}
  C.~G.~Callan, Jr., E.~J.~Martinec, M.~J.~Perry and D.~Friedan,
  ``Strings in Background Fields,''
  Nucl.\ Phys.\ B {\bf 262} (1985) 593.
  
\bibitem{Witten:1985cc}
  E.~Witten,
  ``Noncommutative Geometry and String Field Theory,''
  Nucl.\ Phys.\ B {\bf 268} (1986) 253.
  
\bibitem{Sen:1999xm}
  A.~Sen,
  ``Universality of the tachyon potential,''
  JHEP {\bf 9912} (1999) 027
  [hep-th/9911116].
  
\bibitem{Moeller:2000xv}
  N.~Moeller and W.~Taylor,
  ``Level truncation and the tachyon in open bosonic string field theory,''
  Nucl.\ Phys.\ B {\bf 583} (2000) 105.
  [hep-th/0002237].
  
\bibitem{Witten:2012bh}
  E.~Witten,
  ``Superstring Perturbation Theory Revisited,''
  arXiv:1209.5461 [hep-th].
  
\bibitem{Witten:2013cia}
  E.~Witten,
  ``More On Superstring Perturbation Theory,''
  arXiv:1304.2832 [hep-th].
  
\bibitem{Candelas:1985en}
  P.~Candelas, G.~T.~Horowitz, A.~Strominger and E.~Witten,
  ``Vacuum Configurations for Superstrings,''
  Nucl.\ Phys.\ B {\bf 258} (1985) 46.
  
\bibitem{Giveon:1994fu}
  A.~Giveon, M.~Porrati and E.~Rabinovici,
  ``Target space duality in string theory,''
  Phys.\ Rept.\  {\bf 244} (1994) 77
  [hep-th/9401139].
  
\bibitem{Lerche:1989uy}
  W.~Lerche, C.~Vafa and N.~P.~Warner,
  ``Chiral Rings in N=2 Superconformal Theories,''
  Nucl.\ Phys.\ B {\bf 324} (1989) 427.
    
\bibitem{Greene:1990ud}
  B.~R.~Greene and M.~R.~Plesser,
  ``Duality in {Calabi-Yau} Moduli Space,''
  Nucl.\ Phys.\ B {\bf 338} (1990) 15.
  
\bibitem{Candelas:1990rm}
  P.~Candelas, X.~C.~De La Ossa, P.~S.~Green and L.~Parkes,
  ``A Pair of Calabi-Yau manifolds as an exactly soluble superconformal theory,''
  Nucl.\ Phys.\ B {\bf 359} (1991) 21.
  
\bibitem{Aspinwall:1993nu}
  P.~S.~Aspinwall, B.~R.~Greene and D.~R.~Morrison,
  ``Calabi-Yau moduli space, mirror manifolds and space-time topology change in string theory,''
  Nucl.\ Phys.\ B {\bf 416} (1994) 414
  [hep-th/9309097].
  
\bibitem{Schwarz:1982jn}
  J.~H.~Schwarz,
  ``Superstring Theory,''
  Phys.\ Rept.\  {\bf 89} (1982) 223.
  
\bibitem{Gross:1984dd}
  D.~J.~Gross, J.~A.~Harvey, E.~J.~Martinec and R.~Rohm,
  ``The Heterotic String,''
  Phys.\ Rev.\ Lett.\  {\bf 54} (1985) 502.
  
\bibitem{Green:1984sg}
  M.~B.~Green and J.~H.~Schwarz,
``Anomaly Cancellation in Supersymmetric D=10 Gauge Theory and Superstring Theory,''
  Phys.\ Lett.\ B {\bf 149} (1984) 117.
  
\bibitem{Polchinski:1995mt}
  J.~Polchinski,
  ``Dirichlet Branes and Ramond-Ramond charges,''
  Phys.\ Rev.\ Lett.\  {\bf 75} (1995) 4724
  [hep-th/9510017].
  
\bibitem{Kachru:2003aw}
  S.~Kachru, R.~Kallosh, A.~D.~Linde and S.~P.~Trivedi,
  ``De Sitter vacua in string theory,''
  Phys.\ Rev.\ D {\bf 68} (2003) 046005
  [hep-th/0301240].
  
\bibitem{Witten:1995ex}
  E.~Witten,
  ``String theory dynamics in various dimensions,''
  Nucl.\ Phys.\ B {\bf 443} (1995) 85
  [hep-th/9503124].
  
\bibitem{Banks:1996vh}
  T.~Banks, W.~Fischler, S.~H.~Shenker and L.~Susskind,
  ``M theory as a matrix model: A Conjecture,''
  Phys.\ Rev.\ D {\bf 55} (1997) 5112
  [hep-th/9610043].
  
\bibitem{Susskind:1993ws}
  L.~Susskind,
  ``Some speculations about black hole entropy in string theory,''
  In *Teitelboim, C. (ed.): The black hole* 118-131
  [hep-th/9309145].
  
\bibitem{Horowitz:1996nw}
  G.~T.~Horowitz and J.~Polchinski,
  ``A Correspondence principle for black holes and strings,''
  Phys.\ Rev.\ D {\bf 55} (1997) 6189
  [hep-th/9612146].
  
\bibitem{Strominger:1996sh}
  A.~Strominger and C.~Vafa,
  ``Microscopic origin of the Bekenstein-Hawking entropy,''
  Phys.\ Lett.\ B {\bf 379} (1996) 99
  [hep-th/9601029].
  
\bibitem{Callan:1996dv}
  C.~G.~Callan and J.~M.~Maldacena,
  ``D-brane approach to black hole quantum mechanics,''
  Nucl.\ Phys.\ B {\bf 472} (1996) 591
  [hep-th/9602043].
  
\bibitem{Horowitz:1996fn}
  G.~T.~Horowitz and A.~Strominger,
  ``Counting states of near extremal black holes,''
  Phys.\ Rev.\ Lett.\  {\bf 77} (1996) 2368
  [hep-th/9602051].
  
\bibitem{Maldacena:1996gb}
  J.~M.~Maldacena and A.~Strominger,
  ``Statistical entropy of four-dimensional extremal black holes,''
  Phys.\ Rev.\ Lett.\  {\bf 77} (1996) 428
  [hep-th/9603060].
  
\bibitem{Johnson:1996ga}
  C.~V.~Johnson, R.~R.~Khuri and R.~C.~Myers,
  ``Entropy of 4-D extremal black holes,''
  Phys.\ Lett.\ B {\bf 378} (1996) 78
  [hep-th/9603061].
  
\bibitem{Horowitz:1996ac}
  G.~T.~Horowitz, D.~A.~Lowe and J.~M.~Maldacena,
  ``Statistical entropy of nonextremal four-dimensional black holes and U duality,''
  Phys.\ Rev.\ Lett.\  {\bf 77} (1996) 430
  [hep-th/9603195].
  
  
\bibitem{Maldacena:1996ix}
  J.~M.~Maldacena and A.~Strominger,
  ``Black hole grey body factors and d-brane spectroscopy,''
  Phys.\ Rev.\ D {\bf 55} (1997) 861
  [hep-th/9609026].
  
\bibitem{Das:1996wn}
  S.~R.~Das and S.~D.~Mathur,
  ``Comparing decay rates for black holes and D-branes,''
  Nucl.\ Phys.\ B {\bf 478} (1996) 561
  [hep-th/9606185].

\bibitem{Cyrier:2004hj} 
  M.~Cyrier, M.~Guica, D.~Mateos and A.~Strominger,
  ``Microscopic entropy of the black ring,''
  Phys.\ Rev.\ Lett.\  {\bf 94}, 191601 (2005)
  [hep-th/0411187].


  
\bibitem{Bena:2004tk} 
  I.~Bena and P.~Kraus,
  ``Microscopic description of black rings in AdS / CFT,''
  JHEP {\bf 0412}, 070 (2004)
  [hep-th/0408186].
  
  
\bibitem{Bena:2005ay} 
  I.~Bena and P.~Kraus,
  ``Microstates of the D1-D5-KK system,''
  Phys.\ Rev.\ D {\bf 72}, 025007 (2005)
  [hep-th/0503053].
  
\bibitem{'tHooft:1993gx}
  G.~'t Hooft,
  ``Dimensional reduction in quantum gravity,''
  gr-qc/9310026.
  
\bibitem{Susskind:1994vu}
  L.~Susskind,
  ``The World as a hologram,''
  J.\ Math.\ Phys.\  {\bf 36} (1995) 6377
  [hep-th/9409089].
  
\bibitem{Avis:1977yn}
  S.~J.~Avis, C.~J.~Isham and D.~Storey,
  ``Quantum Field Theory in anti-De Sitter Space-Time,''
  Phys.\ Rev.\ D {\bf 18} (1978) 3565.
  
\bibitem{Hawking:1982dh}
  S.~W.~Hawking and D.~N.~Page,
  ``Thermodynamics of Black Holes in anti-De Sitter Space,''
  Commun.\ Math.\ Phys.\  {\bf 87} (1983) 577.

  
\bibitem{Marolf:2008mf}
  D.~Marolf,
  ``Unitarity and Holography in Gravitational Physics,''
  Phys.\ Rev.\ D {\bf 79} (2009) 044010
  [arXiv:0808.2842 [gr-qc]].
  
\bibitem{Maldacena:1997re}
  J.~M.~Maldacena,
  ``The Large N limit of superconformal field theories and supergravity,''
  Adv.\ Theor.\ Math.\ Phys.\  {\bf 2} (1998) 231
  [hep-th/9711200].
  
\bibitem{Gubser:1996de}
  S.~S.~Gubser, I.~R.~Klebanov and A.~W.~Peet,
  ``Entropy and temperature of black 3-branes,''
  Phys.\ Rev.\ D {\bf 54} (1996) 3915
  [hep-th/9602135].
  
\bibitem{Gubser:1998bc}
  S.~S.~Gubser, I.~R.~Klebanov and A.~M.~Polyakov,
  ``Gauge theory correlators from noncritical string theory,''
  Phys.\ Lett.\ B {\bf 428} (1998) 105
  [hep-th/9802109].
  
\bibitem{Witten:1998qj}
  E.~Witten,
  ``Anti-de Sitter space and holography,''
  Adv.\ Theor.\ Math.\ Phys.\  {\bf 2} (1998) 253
  [hep-th/9802150].
  
\bibitem{Drukker:2000rr}
  N.~Drukker and D.~J.~Gross,
  ``An Exact prediction of N=4 SUSYM theory for string theory,''
  J.\ Math.\ Phys.\  {\bf 42} (2001) 2896
  [hep-th/0010274].
  
\bibitem{Freedman:1999gp}
  D.~Z.~Freedman, S.~S.~Gubser, K.~Pilch and N.~P.~Warner,
  ``Renormalization group flows from holography supersymmetry and a c theorem,''
  Adv.\ Theor.\ Math.\ Phys.\  {\bf 3} (1999) 363
  [hep-th/9904017].
  
\bibitem{Berenstein:2002jq}
  D.~E.~Berenstein, J.~M.~Maldacena and H.~S.~Nastase,
  ``Strings in flat space and pp waves from N=4 superYang-Mills,''
  JHEP {\bf 0204} (2002) 013
  [hep-th/0202021].
  
\bibitem{Lin:2004nb}
  H.~Lin, O.~Lunin and J.~M.~Maldacena,
  ``Bubbling AdS space and 1/2 BPS geometries,''
  JHEP {\bf 0410} (2004) 025
  [hep-th/0409174].
  
\bibitem{Maldacena:1998bw}
  J.~M.~Maldacena and A.~Strominger,
  ``AdS(3) black holes and a stringy exclusion principle,''
  JHEP {\bf 9812} (1998) 005
  [hep-th/9804085].
  
\bibitem{McGreevy:2000cw}
  J.~McGreevy, L.~Susskind and N.~Toumbas,
  ``Invasion of the giant gravitons from Anti-de Sitter space,''
  JHEP {\bf 0006} (2000) 008
  [hep-th/0003075].
  
\bibitem{Myers:1999ps}
  R.~C.~Myers,
  ``Dielectric branes,''
  JHEP {\bf 9912} (1999) 022
  [hep-th/9910053].
  
\bibitem{Aharony:2008ug}
  O.~Aharony, O.~Bergman, D.~L.~Jafferis and J.~Maldacena,
  ``N=6 superconformal Chern-Simons-matter theories, M2-branes and their gravity duals,''
  JHEP {\bf 0810} (2008) 091
  [arXiv:0806.1218 [hep-th]].
  
\bibitem{Klebanov:1998hh}
  I.~R.~Klebanov and E.~Witten,
  ``Superconformal field theory on three-branes at a Calabi-Yau singularity,''
  Nucl.\ Phys.\ B {\bf 536} (1998) 199
  [hep-th/9807080].
  
\bibitem{Martelli:2011fw}
  D.~Martelli and J.~Sparks,
  ``The gravity dual of supersymmetric gauge theories on a biaxially squashed three-sphere,''
  Nucl.\ Phys.\ B {\bf 866} (2013) 72
  [arXiv:1111.6930 [hep-th]].
  

  
\bibitem{Freedman:2013oja} 
  D.~Z.~Freedman and S.~S.~Pufu,
  ``The Holography of F-maximization,''
  arXiv:1302.7310 [hep-th].
  
\bibitem{Hawking:1976ra}
  S.~W.~Hawking,
  ``Breakdown of Predictability in Gravitational Collapse,''
  Phys.\ Rev.\ D {\bf 14} (1976) 2460.
  
\bibitem{Mathur:2009hf}
  S.~D.~Mathur,
  ``The Information paradox: A Pedagogical introduction,''
  Class.\ Quant.\ Grav.\  {\bf 26} (2009) 224001
  [arXiv:0909.1038 [hep-th]].
  
\bibitem{Mathur:2008nj}
  S.~D.~Mathur,
  ``Fuzzballs and the information paradox: A Summary and conjectures,''
  arXiv:0810.4525 [hep-th].
  
\bibitem{Almheiri:2012rt}
  A.~Almheiri, D.~Marolf, J.~Polchinski and J.~Sully,
  ``Black Holes: Complementarity or Firewalls?,''
  JHEP {\bf 1302} (2013) 062
  [arXiv:1207.3123 [hep-th]];
\\[1mm]
  S.~L.~Braunstein, ``Black hole entropy as entropy of
entanglement, or it's curtains for the equivalence principle,"
[arXiv:0907.1190v1 [quant-ph]] published as S.~L.~Braunstein, S.~Pirandola and K.~Zyczkowski, ``Better Late than Never: Information
Retrieval from Black Holes," Physical Review Letters 110, 101301
(2013) for a similar prediction from different assumptions.
  
\bibitem{Balasubramanian:1999re}
  V.~Balasubramanian and P.~Kraus,
  ``A Stress tensor for Anti-de Sitter gravity,''
  Commun.\ Math.\ Phys.\  {\bf 208} (1999) 413
  [hep-th/9902121].
  
\bibitem{Horowitz:1998ha}
  G.~T.~Horowitz and R.~C.~Myers,
  ``The AdS / CFT correspondence and a new positive energy conjecture for general relativity,''
  Phys.\ Rev.\ D {\bf 59} (1998) 026005
  [hep-th/9808079].
  
\bibitem{de Haro:2000xn}
  S.~de Haro, S.~N.~Solodukhin and K.~Skenderis,
  ``Holographic reconstruction of space-time and renormalization in the AdS / CFT correspondence,''
  Commun.\ Math.\ Phys.\  {\bf 217} (2001) 595
  [hep-th/0002230].
  
\bibitem{Kovtun:2004de}
  P.~Kovtun, D.~T.~Son and A.~O.~Starinets,
  ``Viscosity in strongly interacting quantum field theories from black hole physics,''
  Phys.\ Rev.\ Lett.\  {\bf 94} (2005) 111601
  [hep-th/0405231].
  
\bibitem{Adams:2013vsa}
  A.~Adams, P.~M.~Chesler and H.~Liu,
  ``Holographic turbulence,''
  arXiv:1307.7267 [hep-th].
  
\bibitem{Gubser:2008px}
  S.~S.~Gubser,
  ``Breaking an Abelian gauge symmetry near a black hole horizon,''
  Phys.\ Rev.\ D {\bf 78} (2008) 065034
  [arXiv:0801.2977 [hep-th]].
  
\bibitem{Hartnoll:2008vx}
  S.~A.~Hartnoll, C.~P.~Herzog and G.~T.~Horowitz,
  ``Building a Holographic Superconductor,''
  Phys.\ Rev.\ Lett.\  {\bf 101} (2008) 031601
  [arXiv:0803.3295 [hep-th]].
  
\bibitem{Horowitz:2012ky}
  G.~T.~Horowitz, J.~E.~Santos and D.~Tong,
  ``Optical Conductivity with Holographic Lattices,''
  JHEP {\bf 1207} (2012) 168
  [arXiv:1204.0519 [hep-th]].
  
\bibitem{Horowitz:2013jaa}
  G.~T.~Horowitz and J.~E.~Santos,
  ``General Relativity and the Cuprates,''
  arXiv:1302.6586 [hep-th].
  
\bibitem{Ryu:2006bv}
  S.~Ryu and T.~Takayanagi,
  ``Holographic derivation of entanglement entropy from AdS/CFT,''
  Phys.\ Rev.\ Lett.\  {\bf 96} (2006) 181602
  [hep-th/0603001].
  
\bibitem{Nishioka:2009un}
  T.~Nishioka, S.~Ryu and T.~Takayanagi,
  ``Holographic Entanglement Entropy: An Overview,''
  J.\ Phys.\ A {\bf 42} (2009) 504008
  [arXiv:0905.0932 [hep-th]].
  
\bibitem{Lewkowycz:2013nqa}
  A.~Lewkowycz and J.~Maldacena,
  ``Generalized gravitational entropy,''
  JHEP {\bf 1308} (2013) 090
  [arXiv:1304.4926 [hep-th]].
  
\bibitem{VanRaamsdonk:2010pw}
  M.~Van Raamsdonk,
  ``Building up spacetime with quantum entanglement,''
  Gen.\ Rel.\ Grav.\  {\bf 42} (2010) 2323
   [Int.\ J.\ Mod.\ Phys.\ D {\bf 19} (2010) 2429]
  [arXiv:1005.3035 [hep-th]].


  
\end{thebibliography}
\end{document}